\documentclass[onecolumn]{sdl}

\usepackage{colortbl}

\def\0{\mbox{\tiny $0$}}
\def\1{\mbox{\tiny $1$}}
\def\2{\mbox{\tiny $2$}}
\def\3{\mbox{\tiny $3$}}
\def\4{\mbox{\tiny $4$}}
\def\5{\mbox{\tiny $5$}}
\def\6{\mbox{\tiny $6$}}
\def\7{\mbox{\tiny $7$}}
\def\8{\mbox{\tiny $8$}}
\def\9{\mbox{\tiny $9$}}
\def\sinfty{\mbox{\tiny $\infty$}}
\def\+infty{\mbox{\tiny $+\infty$}}
\def\-infty{\mbox{\tiny $-\infty$}}

\definecolor{navy}{rgb}{0,0,.6}
\definecolor{bg}{rgb}{0.90, 0.85, 0.40}

\newcommand{\cor}[1]{{\color{Black}{#1}}}

\logo{
\colorbox{DarkGoldenrod}{\color{white}$\mathbf{\Sigma\hspace*{0.06cm} \delta \hspace*{0.04cm} \Lambda}$}
}

\journal{\textbf{\color{DarkRed} Journal of Modern  Optics} \,\textbf{66}, 2142-2194 (2019).}

\titlelines{5}
\title{Lateral shifts and angular deviations\\ of Gaussian optical beams\\ reflected by and 
transmitted through\\ dielectric blocks: A tutorial review}

\imgbgabstract{In this work, we summarize the current state of understanding of lateral displacement and angular deviations of an optical beam propagating through dielectric blocks. In part I, the analytical formulas, found for critical incidence, are compared with numerical calculations and, when possible, extended from Gaussian to more general angular  distributions. Angular deviations are discussed both for the critical and for the Brewster incidence. The combined effect, known as the composite Goos-H\"anchen shift, shows an interesting axial dependence and, under particular circumstances, the intriguing  phenomenon of the light oscillation. In part II, the weak measurement technique is discussed in detail and compared with direct optical measurements.}

\author{
\names{Stefano De Leo$^{1,2,a}$ and Gabriel G. Maia$^{3,4}$}
\affiliation{$^{1}$Department of Applied Mathematics, State of University of Campinas, Brazil.}
\affiliation{$^{2}$Department of Mathematics and Physics, University of Salento, Italy}
\affiliation{$^{3}$Institute for Scientific and Industrial Research, University of Osaka, Japan}
\affiliation{$^{4}$Institute of Physics Gleb Wataghin, State University of Campinas, Brazil.}
\email{$^{a}$deleo@ime.unicamp.br}
}

\begin{document}

\sdlmaketitle

\section{Introduction}

In 1947\cor{,} the German physicists Hermann Goos and Hilda H\"anchen \cor{ observed and explained an anomalous beam shift that would grow into a long series of investigations in Optics. In seminal experiment \cite{GH1947}, they showed that, under the total internal reflection,} the origin point of a reflected ray would be shifted from the intersection point between the incident ray and the interface between the dielectric media, the size of the shift being proportional to the wavelength of the light being used. Up to that point the classical description of light's path through homogeneous media relied on Geometrical Optics \cite{saleh,born}, but their results showed that even classical light had more subtleties to it then previously thought. This phenomenon would be named in their honour the \emph{Goos-H\"anchen effect}.

In the following year\cor{,} the also German physicist Kurt Artmann presented a mathematical description of this experiment \cite{Art1948}, extending its analysis from the Transverse Electric polarisation, the only one experimentally verified up to that moment, to the Transverse Magnetic polarisation, which was then verified by Goos and H\"anchen in 1949 \cite{GH1949}. Artmann's approach consisted of considering that the multiple plane waves building up the resultant electromagnetic fields have rapidly varying phases that cancel each other \cor{out.} The stationary condition \cor{gives the main term of the phase which contributes to determine the optical path}. Upon total internal reflection, the Fresnel reflection coefficient becomes complex, giving origin to (what we will call here\cor{in}after) the \emph{Goos-H\"anchen phase}, the addition of which generates a lateral shift of the optical path. Although Artmanns's formula obtained by this method was generally successful, it has a troubling flaw inherent to it: it diverges for incidence angles near the critical angle when experimental data show that it should be finite.

The divergence problem of Artmann's formula was addressed by Wolter \cite{Wol1950}, and, independently, by Artmann himself \cite{Art1950}. Wolter's approach did not consider a bounded beam, but rather a wave composed of two slightly incoherent plane waves. This simplified model, however, proved itself to be overly simplified, since re-establishing coherence would bring the divergence back. Artmann was more successful in his enterprise, obtaining a formula for the Goos-H\"anchen shift \cor{for}  critical incidence  under the assumption of a large number of total internal reflections between two parallel interfaces.

Even though this problem remained unsolved for the decades to come, new insights helped its understanding. Brekhovskikh, in 1960 \cite{Brekh1960}, and Lotsch in the late 1960's \cite{Lots1968} and early 1970's \cite{Lots1970} demonstrated that the inconsistency between theory and experiment were due to a problem in the derivation of the analytical formula, which was not valid in the close vicinity of the critical angle. Meanwhile, in 1964, Renard presented a formula that was curiously different from \cor{the Artmann one far the critical incidence}, but would approach it as the incidence angle \cor{reaches the critical region} \cite{Ren1964}. The mistake in his derivations would be pointed out twenty years later by Lai, Cheng, and Tang \cite{Lai1986}, but the basis of his arguments would prove not only correct but very interesting. Renard's paper showed that the Goos-H\"anchen shift is a necessity imposed by the conservation of the energy flux for \cor{total internal reflection}.

In 1970, Horowitz and Tamir presented their famous attempt at solving this conundrum through a direct integration of the reflected electric field \cite{HoTa1970}. This was accomplished by a clever manipulation and expansion of the integrand, providing a complex formula in terms of the Weber function \cite{Mag1966}. Their solution, however, suffered two problems. The first one, as pointed out by Cowan and Ani\v{c}in \cite{Cow1977}, was that, even though it provided the correct value for the shift at the critical angle, in a small vicinity of it, the analytical curves presented an infinite slope possessing for some cases a cusp\cor{-}like structure. These features are \cor{neither} compatible with experimental results nor with the smooth continuity of Gaussian beams spectra of plane waves. \cor{The second one}, the mathematical rigour of the derivation was flawed since the resultant formula is inconsistent with the initial assumptions. Horowitz and Tamir assumed that the incidence angle was always close to the critical angle, but their formula also reproduced Artmann's results which are valid for angles \cor{far from the critical incidence}. This second problem was discovered by Lai \emph{et al.}, who, in 1986, presented a corrected version of Horowitz and Tamir's derivation where they assumed a beam that does not diverge as it propagates \cite{Lai1986}.

In the last decade, numerical analyses were used to study the regions of validity of the analytical formulas available in the literature, as well as the angles for which maximum shifts are obtained \cite{DeLeo2013}. Analytical investigations in the critical region were only resumed very recently. In 2016, a paper by \cor{De Leo} \emph{et al.} found analytical formulae for the Goos-H\"anchen shift of the maximum intensity point of a Gaussian beam as well as for the shift of its average intensity \cite{Maia2016}, using in \cor{the} derivation a different perspective than the one employed by Horowitz and Tamir, and by Lai \emph{et al.}. The shift of the maximum intensity point was obtained by considering the structure of the beam on the stationary condition employed originally by Artmann, while the shift of the average intensity was calculated by a mean value analysis of the electric field intensity in the direction perpendicular to the direction of propagation. Both methods return results that not only agree with each other away from the critical angle, but, more importantly, agree with Artmann's result in the same region. Near the critical angle the formulae disagree in magnitude only, which is expected, since in this region the Gaussian beam is not symmetrical \cite{DeLeo2014a, DeLeo2014b} and the maximum intensity point does not coincide with the average intensity. In 2017, a paper by the same authors studied more carefully the effects of this symmetry breaking on the Goos-H\"anchen shift, finding an oscillatory behaviour in the curves \cite{Maia2017}, depending on the position of the camera during measurements\cor{: A} phenomenon called \emph{Composite Goos-H\"anchen shift}. It is interesting to note that this new phenomenon was overlooked by previous works due to the assumption of non-diverging beams \cite{Lai1986}, which amounts to consider that measurements are carried out very close to the dielectric interface and \cor{to neglect} a portion of the beam outside the total internal reflection region \cite{Maia2016}.

Experimentally, the Goos-H\"anchen effect has been revisited several times since the original experiments. In 1973, Green, Kirkby, and Timsit re-measured the shift using a set-up similar to the one used by Goos and H\"anchen in 1947 and 1949, increasing the accuracy of the measurements \cite{Gree1973}. In 1977, Cowan and Ani\v{c}in measured it for the first time using microwaves \cite{Cow1977}, while Bretenaker, Le Floch, and Dutriaux, in 1992, presented the first measurement of the shift due to a single reflection using a \textbf{He-Ne} laser source \cite{Bret1992}.

Parallel to the study of the Goos-H\"anchen effect, angular deviations from the predictions of the Geometrical Optics were also discovered and studied throughout the 20th century up to \cor{present}, a phenomenon called \emph{angular Goos-H\"anchen shift}. In 1973, Ra, Bertoni, and Felsen identified this sort of shift from the analysis of the integrated reflected beam expression for the case of partial internal reflections, presenting the critical angle as a frontier between Goos-H\"anchen and angular shifts \cite{Ra1973}. In 1974, studying a similar system, but expressing the reflected beam as a superposition of beam modes, of which the Gaussian beam was the fundamental mode, Antar and Boerner, obtained an expression for the angular deviation at the Brewster angle \cite{Ant1974}. In their paper they describe how, for incidence at the polarisation angle, the fundamental mode is absent, being then angular deviations a higher order phenomenon at Brewster incidence. They also observed that, as the incidence angle moves from a value smaller than the Brewster angle to a value greater than it, a change in the sign of the shift occurs.

Up to that point, research on angular shifts was mostly driven by the mathematical properties of the electric field integrals. It was not until 1977 that a more physical interpretation of the effect was presented. White, Snyder, and Pask theorised that the angular shift was due to a change in the power distribution of the plane wave spectrum of the beam \cite{White1977}. Also considering internal reflections, they assumed that the propagation direction of the reflected beam was approximately the same as the propagation direction of the plane wave with the largest contribution to the beam power in the far field. With this assumption, they found that each part of the beam that was split in two at the Brewster angle had a different angular deviation. Their method corresponds to the maximum value analysis carried out for the Goos-H\"anchen shift in \cite{Maia2016}, while Antar and Boerner's method, which considered both peaks as parts of the same object, is analogous to the mean value calculation in the same reference. In 1985, Chan and Tamir analysed the angular shift in the region around the Brewster angle using a mathematical method resembling  the method used by Howoritz and Tamir for the analysis of lateral shifts in the critical region \cite{ChTa1985}. Their work offered an interesting perspective on the matter, arguing that, at the Brewster angle, the reflected beam is so deformed in comparison to the incident one that the concept of angular deviation lacks any meaning. Regarding angular deviations at the critical angle, Chan and Tamir, in a 1987's review work of beam phenomena in the critical region \cite{ChTa1987}, not only found the deviation value at critical incidence, but also reproduced results from Ra \emph{et al.}. In 2009, Aiello and Woerdman revisited the topic of angular deviations in the Brewster region \cite{AiWo2009}, calculating the shifts as the mean distance between the propagation direction according to Geometrical Optics and the center of the beam, and, in the same year Aiello, Merano, and Woerdman addressed beam deformation in the same region \cite{AiMeWo2009}.

Independently from the angular Goos-H\"anchen shift, researchers on microcavities studied a similar effect, mostly associated with transmissions instead of reflections. Tureci and Stone named this effect \emph{Fresnel Filtering} in 2002, in a paper where they showed that critical incidence does not actually originate a tangent transmission, \cor{with} large deviations occurring from this expectation \cite{TuSt2002}. Their explanation for the phenomenon was the same presented by White \emph{et al.} regarding the angular Goos-H\"anchen shift \cite{White1977}, that is, the shift is due to a change in the power distribution of the (in this case) transmitted beam, induced by the interface. In 2013, G\"otte, Shinohara, and Hentschel demonstrated that both phenomena, angular Goos-H\"anchen shift and Fresnel Filtering, were in fact the same in nature \cite{GoShiHen2013}.

Curiously, while the experimental results of Gmachl \emph{et al.} with microcavities \cite{Gma1998} stimulated the theoretical investigations of Tureci and Stone, there is a time interval of more than thirty years between the first theoretical studies on the angular Goos-H\"anchen shift and its experimental verification in 2006 by M\"uller \emph{et al.} for microwaves \cite{Mull2006}. Three years later, Merano \emph{et al.} measured the effect using a superluminescent light emiting diode \cite{Mera2009}.

The Goos-H\"anchen shift and the angular deviations have in common the minute nature of their manifestations, which poses a practical problem for the experimentalist. In their 1992 paper, Bretenaker \emph{et al.} even defended the importance of their experimental work measuring the Goos-H\"anchen shift of lasers for a single reflection by stating that, up to that point, all measurements had made use of one out of two techniques: they either employed a system with multiple reflections or made use of microwaves. The goal of both methods being the amplification of the shift. In the history of angular deviations' experiments a similar pattern can be observed. The Fresnel Filtering naturally involves several interactions with the interfaces of the microcavity, while the angular Goos-H\"anchen shift was measured \cor{first} for microwaves in 2006 and then for a single reflection of a laser beam in 2009.

A third route to amplification, however, \cor{was been} found in a technique originally designed for quantum mechanical systems. In 1988, Aharonov, Albert, and Vaidman presented their famous paper, introducing what they called \emph{Weak Measurements} \cite{AAV1988}. The details of their quantum theory are out of the scope of our work (though the reader interested may refer to the excellent review of the subject by Svensson \cite{Sve2013}), but its general idea was that particular choices of final states and a weak interaction between system and meter could provide a trade-off between the final state's probability and the eigenvalue characterising it. By selecting an event with a very low probability, it was, consequently, possible to greatly increase the measured value associated to it (in their original paper they discuss spin measurements). One year later, Duck and Stevenson published a paper addressing some inconsistencies of Aharonov \emph{et al.}'s work, but acknowledging the worth of their results \cite{DuSt1989}. In the same paper they also adapted the theory to an optical system, initiating the \emph{Optical Weak Measurements} field of research. Under this classical point of view, the trade-off happens between the electromagnetic field's intensity and an induced deviation of its path. In 2012, Dennis and G\"otte developed the full correspondence theory between Quantum and Optical weak measurements \cite{DeGo2012}, and in 2013, Jayaswal, Mistura, and Merano made the first weak measurement of the Goos-H\"anchen shift \cite{JaMisMera2013}, while, in the following year, they \cite{JaMisMera2014} and Goswami \emph{et al.} \cite{Gos2014} employed the technique, independently, to observe angular deviations. In 2016, Santana \emph{et al.} made the first weak measurement experiment in order to investigate the composite Goos-H\"anchen shift \cite{San2016}. In 2015, Ara\'{u}jo, De Leo, and Maia presented their study on how weak measurements of the Goos-H\"anchen shift in the critical region could suffer axial deformations \cite{DeLeo2015}, due to the symmetry breaking of the beam in the region, and, in 2017, a paper by the same authors made a comparative analysis of weak measurements versus direct measurements of angular deviations near the Brewster and critical regions, evaluating the efficiency of the amplification technique \cite{DeLeo2017}. Finally, in an accepted, but yet unpublished paper, Maia \emph{et al.} investigated the effect of the Goos-H\"anchen phase in weak measurements \cite{Maia2018}. Such a phase is usually discounted from theoretical works as an unnecessary complication, and is removed from experiments with the aid of waveplates. The paper fills the gap in the literature concerning the formal description of its effects, describing its destructive influence on measurements. The understanding of the precise nature of this influence, the authors argue, \cor{is} relevant in preventing discrepancies between theoretical expectations and experimental evaluations.

This brief history of beam shifts is by no means an exhaustive account, but \cor{it focuses} only on the most relevant aspects of \cor{the paste}. It does not consider, for instance, other kinds of shifts such as the \emph{Imbert-Fedorov} effect \cite{GHIF2008,GHIF2012,GHIF2013}, which is a transversal shift occurring for circularly polarised light, nor does it consider shifts for metallic interfaces \cite{GHmetal2007} and waveguides \cite{GHwaveguide1974}, shifts occurring for different beam modes \cite{HermiteGHIF2012}, or their seismic counterpart \cite{GHseismic2012,DK2018}. Hopefully, however, these \cor{references} will help to illustrate the large range of applicability of such phenomena.

The present work is divided \cor{into} two parts. The first \cor{one contains an analysis  of the two-dimensional beam shift phenomena}, meaning that all shifts considered occur in the plane of incidence, and the second one \cor{refers to} the use of optical weak measurements in the study of \cor{lateral shifts and angular deviations}. It is structured as follows: the first Section \cor{was}  an introductory review of \cor{optical} beam shifts. In the \cor{sequential}  Sections, \cor{the formalism followed and the notation used is introduced} by a brief study of the electromagnetic waves propagation in dielectric media, the calculation of their optical paths, and \cor{by showing} how the use of Gaussian beams change such calculations. 
\cor{In Section \ref{sec2}, the Fresnel coefficients, for Transverse Electric and Magnetic waves, are obtained, by using the continuity equations. The optical system is discussed in Section \ref{sec3}. In Section \ref{sec4}, the Gaussian beam formalism, the proper coordinate systems for each interface, and the Brewster and critical regions  are introduced. Section \ref{sec5} contains the Artmann's results  and a discussion on the divergence problem solved by an analytical solution for the maximum intensity point of the beam and for the average intensity as well. In Section \ref{sec6}, the analytical expressions for the angular deviations are obtained, and in Section \ref{sec7}  the composite Goos-H\"anchen shift and the oscillatory behaviour is investigated. This} concludes Part I. \cor{In Section \ref{sec8}}, the first Section of Part II, the axial \cor{dependence} of weak measurements of the Goos-H\"anchen effect in the critical region is studied as well as the effects of the Goos-H\"anchen phase on such measurements. Section \ref{sec9} \cor{contains an analysis} of the efficiency of optical weak measurements versus direct measurements for angular deviations. \cor{Conclusions and outlooks are given in the last Section.} 

\section{Waves propagation in dielectric media}
\label{sec2}

The behaviour of electromagnetic fields is described by Maxwell's equations \cite{saleh}. In dielectric media, which are the ones we are interested in, there are no free charges nor current densities.  \cor{These} equations can be written in their differential form as

\begin{subequations}
\label{eq:MaxEqD}
\begin{equation}
\label{eq:MaxEqD1}
\mathbf{\nabla}\cdot(\epsilon_{_{j}}\,\mathbf{E}) = 0,
\end{equation}
\begin{equation}
\label{eq:MaxEqD2}
\mathbf{\nabla}\cdot\mathbf{B} = 0,
\end{equation}
\begin{equation}
\label{eq:MaxEqD3}
\mathbf{\nabla}\times\mathbf{E} = -\frac{\partial \mathbf{B}}{\partial t},
\end{equation}
and
\begin{equation}
\label{eq:MaxEqD4}
\mathbf{\nabla}\times\frac{\mathbf{B}}{\mu_{_{j}}} = -\frac{\partial \mathbf{(\epsilon_{_{j}}\,E})}{\partial t},
\end{equation}
\end{subequations}
being $\mathbf{E}$ and $\mathbf{B}$ the electric and magnetic fields, respectively, $\epsilon_{_{j}}$ the permittivity of the propagating medium $j$ and $\mu_{_{j}}$ its permeability. The fields are implied to be a function of spatial coordinates and of time. From these equations it is possible to write differential wave equations \cite{saleh},

\begin{equation}
\label{eq:EMwaves}
\mathbf{\nabla}^{^2}\mathbf{F}-\frac{\displaystyle n_{\mbox{\tiny $j$}}^{\2}}{\displaystyle c^{\2}}\,\frac{\displaystyle \partial^{\2} \mathbf{F}}{\displaystyle \partial \,t^{\2}} = 0,
\end{equation}
the solutions of which describe the propagation of electromagnetic waves. In the equation above, $\mathbf{F}$ can be thought of as representing $\mathbf{E}$ or $\mathbf{B}$, since both equations have the same form. In its derivation we have used  $\mu_{_{j}}\,\epsilon_{_{j}} = n_{\mbox{\tiny $j$}}^{\2}\,\mu_{\0}\,\epsilon_{\0}$, being $n_{\mbox{\tiny $j$}}$ the refractive index of the medium $j$, and  $\mu_{\0}\,\epsilon_{\0} = 1/c^{\2}$, where $c$ is the speed of light in vacuum. The solutions to the Eq. (\ref{eq:EMwaves}) can be obtained by \cor{the} separation of spatial and time variables. Assuming $\mathbf{F} = \mathbf{A}(\mathbf{r})\,T(t)$ we obtain the equations

\begin{subequations}
\begin{equation}
\label{eq:EMwaves1}
\left(\mathbf{\nabla}^{^{2}}+k^{\2}\right)\mathbf{A}(\mathbf{r}) = 0,
\end{equation}
and
\begin{equation}
\label{eq:EMwaves2}
\left(\frac{\displaystyle \partial^{\2}}{\displaystyle \partial t^{\2}} + \frac{\displaystyle c^{\2}\,k^{\2}}{\displaystyle n_{\mbox{\tiny $j$}}^{\2}}\right)T(t) = 0.
\end{equation}
\end{subequations}
\cor{The plane wave solutions are}
\begin{subequations}
\begin{equation}
\label{eq:EMwavesSol1}
\mathbf{A}(\mathbf{r}) = \mathbf{u}(\mathbf{k})\,\exp\left(i\,\mathbf{k}\cdot\mathbf{r}\right)
\end{equation}
and
\begin{equation}
\label{eq:EMwavesSol2}
T(t) = \exp\left( -i\,\omega_{\mbox{\tiny $j$}}\,t \right),
\end{equation}
\end{subequations}
respectively, \cor{where} $\omega_{\mbox{\tiny $j$}} = c\,k/n_{\mbox{\tiny $j$}}$ the angular frequency of the wave and $k=2\pi/\lambda$ its wavenumber, \cor{with} $\lambda$ is its wavelength. The vector amplitude $\mathbf{u(\mathbf{k})}$ determines not only the amplitude of the field's oscillation, but its direction as well. Since electromagnetic waves are transversal waves, this vector amplitude depends on the vector $\mathbf{k}$, which provides us with the propagation direction of the wave and is such that $\left|\mathbf{k}\right| = k = n_{\mbox{\tiny $j$}}\,k_{\0}$, where $k_{\0}$ is the wavenumber in vacuum. The components of this vector are $\mathbf{k} = (k_{x},k_{y},k_{z})$, and they can be written as a function of spherical coordinates as
\begin{equation}
\label{eq:kvec}
\mathbf{k} = k\,(\sin\theta_{z}\cos\theta_{x},\sin\theta_{z}\sin\theta_{x},\cos\theta_{z}),
\end{equation}
where $\theta_{z}$ is the polar angle and $\theta_{x}$ the azimuthal angle.

The plane wave solutions obtained describe waves propagating through an uniform medium. Upon interaction with an interface between media, however, these waves are split in reflected and refracted portions. \cor{The ratio at which this division occurs is determined by the Fresnel coefficients. These} are obtained from the boundary conditions' analysis of the Maxwell's equations \cite{born}. From (\ref{eq:MaxEqD1}) and (\ref{eq:MaxEqD2}) we have that
\begin{subequations}
\label{eq:BC}
\begin{equation}
\label{eq:BC1}
\epsilon_{\1}\,E_{{\1\bot}} = \epsilon_{\2}\,E_{{\2\bot}}
\end{equation}
and
\begin{equation}
\label{eq:BC2}
B_{{\1\bot}} = B_{{\2\bot}},
\end{equation}
and from Eqs (\ref{eq:MaxEqD3}) and (\ref{eq:MaxEqD4}),
\begin{equation}
\label{eq:BC3}
E_{{\1\parallel}} = E_{{\2\parallel}}
\end{equation}
and
\begin{equation}
\label{eq:BC4}
\frac{\displaystyle B_{{\1\parallel}}}{\displaystyle \mu_{\1}} = \frac{\displaystyle B_{{\2\parallel}}}{\displaystyle \mu_{\2}},
\end{equation}
\end{subequations}
where the subscripts $1$ and $2$ are a reference to the \cor{first} and \cor{second} medium with refractive indices $n_{\1}$ and $n_{\2}$, respectively. When we introduced the vector amplitude $\mathbf{u}(\mathbf{k})$ no details about the oscillation direction were presented, because for the propagation in a uniform medium this information is of no relevance. An interface, however, breaks this uniformity, and the oscillation direction becomes important. The normal to the interface and the incidence direction, $\mathbf{k}/k$, define the plane of incidence, which is the reference in the definition of the polarisation state of light. The component of the electric field orthogonal to the plane of incidence characterises the Transverse Electric (TE) polarisation, while the component of the magnetic field orthogonal to the plane of incidence characterises the Transverse Magnetic (TM) polarisation, see Figure 1. In Eqs. (\ref{eq:BC}) the subscript `$\parallel$' indicates the component of the field that is parallel to the interface and `$\bot$' the component perpendicular to it. Since we are considering a plane interface, which is, consequently, perpendicular to the plane of incidence, these symbols indicate field components that are perpendicular and parallel to the plane of incidence, respectively.

\WideFigureSideCaption{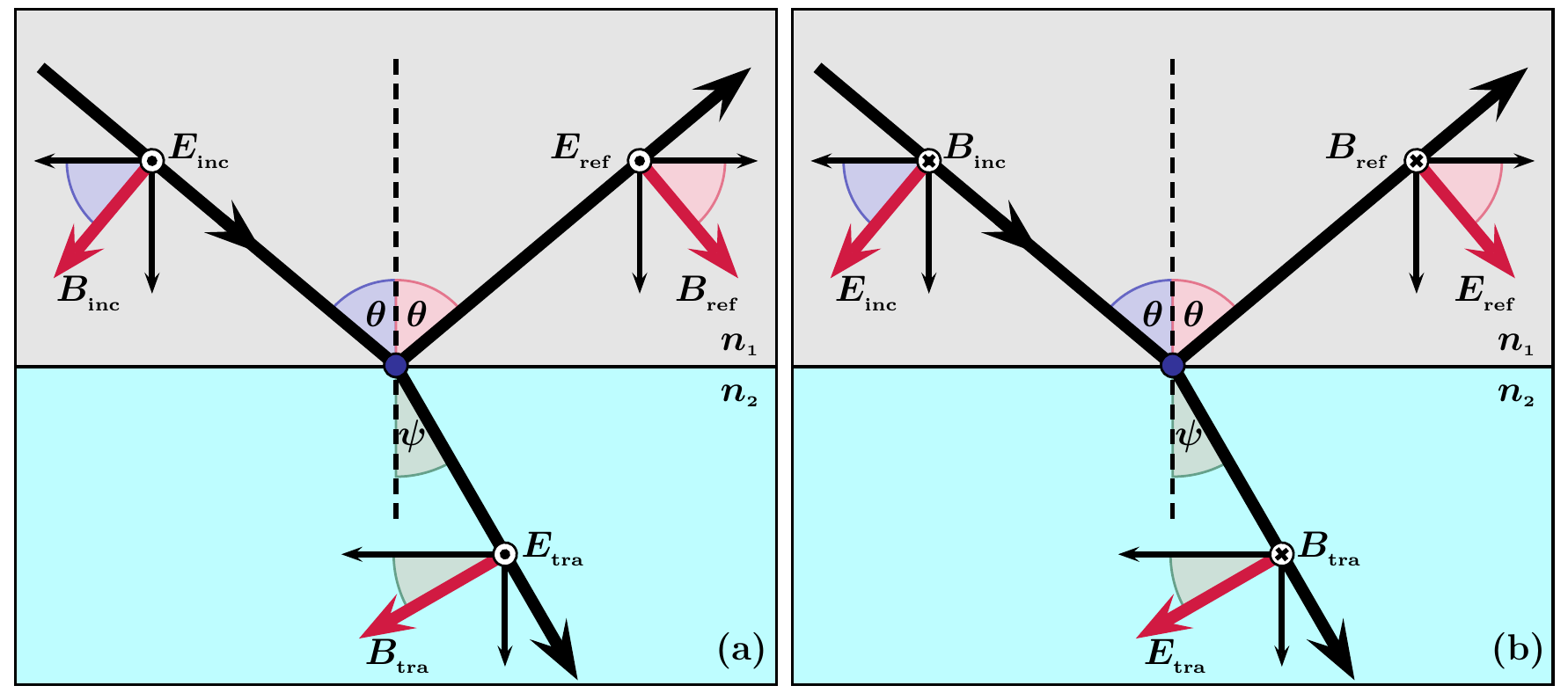}{The incidence plane defined by the interaction between an electromagnetic wave and an interface between two dielectric media of refractive index $n_{\1}$ (incident medium) and $n_{\2}$ (refractive medium). The incident wave hits the interface \cor{creating} an angle $\theta$ with its normal, being then partially reflected with the same angle and partially refracted with an angle $\psi$. The electric field orthogonal to the incidence plane in (a) defines the Transverse Electric (TE) polarisation and the magnetic field orthogonal to it in (b) defines the Transverse Magnetic (TM) polarisation.}

The four Eqs. (\ref{eq:BC}) are redundant, and we can focus only on Eqs. (\ref{eq:BC3}) and (\ref{eq:BC4}) to find Fresnel's coefficients. To do so, let us consider a coordinate system \cor{with the interface being}  the plane $z=z_{\0}$, and the plane of incidence the $x-z$ plane. This perfectly valid choice is equivalent to orient the coordinate system in such a way that $\theta_{x} = 0$ in Eq. (\ref{eq:kvec}), which becomes then
\begin{equation}
\mathbf{k} = k\,(\sin\theta,0,\cos\theta),
\end{equation}
where the notation was simplified by making $\theta_{z} = \theta$. The $z-$axis is parallel to the normal of the interface, which makes $\theta$ the incidence angle. The propagation direction of the reflected wave also makes an angle $\theta$ with the $z-$axis, but upon reflection the $z$-component of its phase acquires a minus sign, according to the law of reflection. The transmitted wave makes an angle $\psi$ with the normal, which is given by the Snell's law
\begin{equation}
\label{eq:Snell}
n_{\1}\,\sin\theta = n_{\2}\,\sin\psi,
\end{equation}
see Figure 1.

For the TE polarisation, Figure 1(a), Eqs. (\ref{eq:BC3}) and (\ref{eq:BC4}) can be written as
\begin{subequations}
\begin{equation}
\label{eq:FresnelTE1}
E_{_{\mathrm{inc}}} + E_{_{\mathrm{ref}}} = E_{_{\mathrm{tra}}}
\end{equation}
and
\begin{equation}
\label{eq:FresnelTE2}
B_{_{\mathrm{inc}}}\,\cos\theta - B_{_{\mathrm{ref}}}\cos\theta = B_{_{\mathrm{tra}}}\cos\psi,
\end{equation}
\end{subequations}
where ``inc'', ``ref'', and ``tra'' denote the incident, reflected, and transmitted fields, respectively, and where we considered that the permeability $\mu_{_{j}}$ does not differ appreciably from one dielectric to another \cite{born}. Besides, the subscript `$\parallel$' was suppressed. Noticing that $B=n_{\mbox{\tiny $j$}}\,E/c$ \cite{jackson}, and defining the reflection and transmission coefficients to be the ratio between electric field amplitudes at the interface, that is,
\begin{equation}
r = \frac{\displaystyle E_{_{\mathrm{ref}}}}{\displaystyle E_{_{\mathrm{inc}}}}\,e^{2\,i\,k\,z_{\0}\cos\theta}\,\,\,\,\,\,\,\,\,\,\mathrm{and}\,\,\,\,\,\,\,\,\,\, t = \frac{\displaystyle E_{_{\mathrm{tra}}}}{\displaystyle E_{_{\mathrm{inc}}}}\,e^{i\,k\,(\cos\theta-n\cos\psi)\,z_{\0}},
\end{equation}
respectively, where we have defined the relative refractive index $n=n_{\2}/n_{\1}$, we have that
\begin{subequations}
\label{eq:TE0}
\begin{equation}
\label{eq:RTE}
r^{^{[\mathrm{TE}]}}(\theta)  = \frac{\displaystyle \cos\theta - n\,\cos\psi}{\displaystyle \cos\theta + n\,\cos\psi}\,e^{2\,i\,k\,z_{\0}\cos\theta}
\end{equation}
and
\begin{equation}
\label{eq:TTE}
t^{^{[\mathrm{TE}]}}(\theta)  = \frac{\displaystyle 2\,\cos\theta}{\displaystyle \cos\theta + n\,\cos\psi}\,e^{i\,k\,(\cos\theta-n\cos\psi)\,z_{\0}}.
\end{equation}
\end{subequations}

For the TM polarisation, Figure 1(b), Eqs. (\ref{eq:BC3}) and (\ref{eq:BC4}) assume the form
\begin{subequations}
\begin{equation}
\label{eq:FresnelTM1}
B_{_{\mathrm{inc}}} + B_{_{\mathrm{ref}}} = B_{_{\mathrm{tra}}}
\end{equation}
and
\begin{equation}
\label{eq:FresnelTM2}
E_{_{\mathrm{inc}}}\cos\theta_{i} - E_{_{\mathrm{ref}}}\cos\theta_{i} = E_{_{\mathrm{tra}}}\cos\psi_{i},
\end{equation}
\end{subequations}
and a similar analysis as the one carried out for the TE polarisation provides us with
\begin{subequations}
\label{eq:TM0}
\begin{equation}
\label{eq:RTM}
r^{^{[\mathrm{TM}]}}(\theta)  = \frac{\displaystyle n\,\cos\theta - \cos\psi}{\displaystyle n\,\cos\theta + \cos\psi}\,e^{2\,i\,k\,z_{\0}\cos\theta}
\end{equation}
and
\begin{equation}
\label{eq:TTM}
t^{^{[\mathrm{TM}]}}(\theta)  = \frac{\displaystyle 2\,\cos\theta}{\displaystyle n\,\cos\theta + \cos\psi}\,e^{i\,k\,(\cos\theta-n\cos\psi)\,z_{\0}}.
\end{equation}
\end{subequations}

Optics textbooks usually do not present these complex exponentials as part of the Fresnel's coefficients and at first they do seem like an unnecessary complication since we could have chosen the interface to be the plane $z=0$. However, in more complex structures, composed of several interfaces, such as a prism, it is not ideal to avoid these exponentials at every interface, which is why they were presented here, since they carry important information regarding the light's path inside the structure, and its analysis provides a smooth introduction to lateral shifts, as will be seen in the next Section.

\section{The optical system}
\label{sec3}

With the basic notions of electromagnetic waves propagation established in the last Section, let us now turn to the description of the optical system we will be studying. In order to bring our results closer to possible experimental implementations, we will consider as optical system a dielectric right angle triangular prism of vertices $A$, $B$, and $C$, as depicted in Figure 2(a). The interaction of light with it \cor{occurs} as follows: light hits the left face of the prism \cor{forming} an angle $\theta$ with its normal. Part of it is reflected with the same angle of incidence and \cor{another} part is transmitted into the prism with an angle $\psi$ with the left face's normal, according to the Snell's law, see Eq. (\ref{eq:Snell}). The portion of light transmitted into the structure hits then its lower face with an angle $\varphi$, which is determined by the geometry of the system, being, in this case,
\begin{equation}
\label{eq:varphi}
\varphi = \frac{\displaystyle \pi}{\displaystyle 4} + \psi.
\end{equation}
Here again part of the light is transmitted to the outside of the prism with an angle $\phi$, and part is reflected with an angle $\varphi$, hitting then the right face of the prism with an angle $\psi$ and being finally transmitted with an angle $\theta$.

In the process described above, to each face of the prism there is a set of associated Fresnel's coefficients which modify the plane waves interacting with \cor{those} faces. These coefficients are obtained from the same procedure carried out in the last Section with a few modifications regarding the coordinate systems used in their derivation. Let us define two coordinate systems, both with origin on the left face, at a distance $d$ from the vertex $A$ of the prism, as depicted in Figure 2(b). The $x-z$ system has its $z-$axis orthogonal to the left face, while the $x_{_{*}}-z_{_{*}}$ system has its $z_{_{*}}-$axis orthogonal to the lower face of the prism. Using the $x-z$ system, the Fresnel's coefficients associated to the left face are simply the ones given by Eqs. (\ref{eq:TE0}) and (\ref{eq:TM0}) with $z_{\0}=0$,

\WideFigureSideCaption{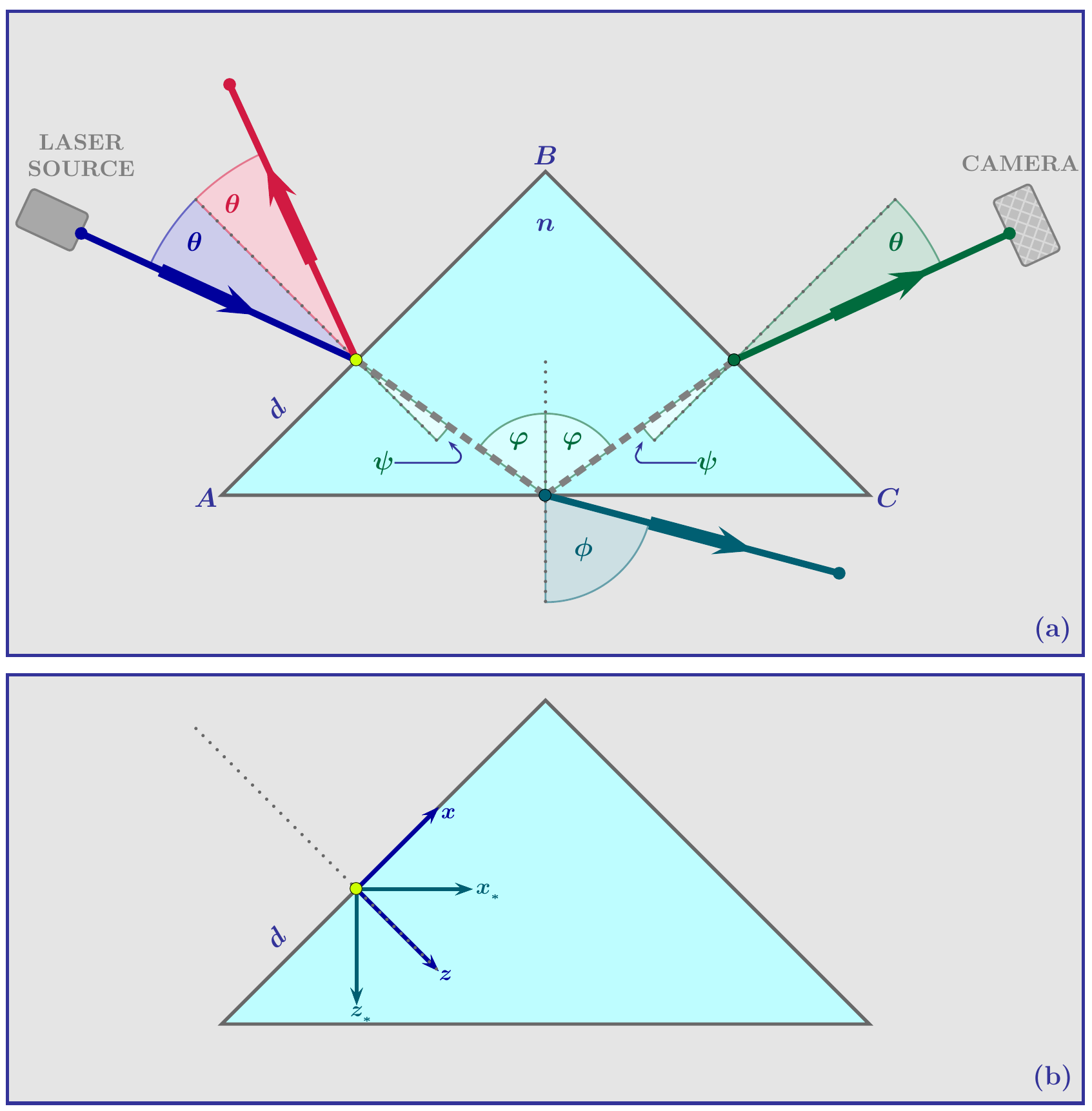}{(a) The optical system of interest: a right angle triangular prism of vertices $A$, $B$, and $C$, and relative refractive index $n$. Light \cor{is emitted by} laser source and hits the left face ($\overline{AB}$) of the prism \cor{forming} an angle $\theta$ with its normal. Part of this beam is reflected with the same angle of incidence and part is transmitted with an angle $\psi$ given by the Snell's law. The transmitted beam hits then the lower interface ($\overline{AC}$) of the prism \cor{at} an angle $\varphi = \pi/4 + \psi$, being then partially transmitted to the outside with an angle $\phi$ given by $n\,\sin\varphi = \sin\phi$, and partially reflected with the same angle $\varphi$. This reflected portion then hits the right interface ($\overline{BC}$) of the prism with an angle $\psi$ and is finally transmitted with an angle $\theta$, being \cor{recorded} by a camera. (b) The coordinate-systems of interest, sharing a common origin at a distance $d$ from the vertex $A$. The $x-z$ system has its $z-$component perpendicular to the left face of the prism while the $x_{_{*}}-z_{_{*}}$ system has its $z_{_{*}}-$component perpendicular to the lower face.}

\begin{subequations}
\label{eq:FcLeft}
\begin{equation}
\left\{r^{^{[\mathrm{TE}]}}_{_\mathrm{left}}(\theta),r^{^{[\mathrm{TM}]}}_{_\mathrm{left}}(\theta) \right\}= \left\{\frac{\displaystyle \cos\theta - n\cos\psi}{\displaystyle \cos\theta + n\cos\psi}, \frac{\displaystyle n\cos\theta - \cos\psi}{\displaystyle n\cos\theta + \cos\psi}\right\}
\end{equation}
and
\begin{equation}
\left\{ t^{^{[\mathrm{TE}]}}_{_\mathrm{left}}(\theta) ,t^{^{[\mathrm{TM}]}}_{_\mathrm{left}}(\theta) \right\} = \left\{ \frac{\displaystyle 2\,\cos\theta}{\displaystyle \cos\theta + n\cos\psi}, \frac{\displaystyle 2\,n\,\cos\theta}{\displaystyle n\cos\theta + \cos\psi}\right\}.
\end{equation}
\end{subequations}
In the $x_{_{*}}-z_{_{*}}$ system, the lower interface is placed at $z_{_{*}}=d/\sqrt{2}$, and the correspondent coefficients are
\begin{subequations}
\label{eq:FcLower}
\begin{equation}
\left\{r^{^{[\mathrm{TE}]}}_{_\mathrm{lower}}(\theta) , r^{^{[\mathrm{TM}]}}_{_\mathrm{lower}}(\theta) \right\} = \left\{ \frac{\displaystyle n\cos\varphi - \cos\phi}{\displaystyle n\cos\varphi + \cos\phi},  \frac{\displaystyle \cos\varphi - n\cos\phi}{\displaystyle \cos\varphi + n\cos\phi} \right\}\,e^{\sqrt{\2}\,i\,n\,k\,d\cos\varphi}
\end{equation}
and
\begin{equation}
\left\{ t^{^{[\mathrm{TE}]}}_{_\mathrm{lower}}(\theta) ,  t^{^{[\mathrm{TM}]}}_{_\mathrm{lower}}(\theta)   \right\} = \left\{ \frac{\displaystyle 2\,n\cos\varphi}{\displaystyle n\cos\varphi + \cos\phi} , \frac{\displaystyle 2\,\cos\varphi}{\displaystyle \cos\varphi + n\cos\phi} \right\}\,e^{i\,k\,(n\cos\varphi-\cos\phi)\,d/\sqrt{\2}},
\end{equation}
\end{subequations}
where attention must be paid to fact that now the incoming medium has a refractive index $n_{\2}$, while the refractive index of the refracting medium is $n_{\1}$. Finally, for the right face of the prism we can use the $x-z$ system again, noticing that now the discontinuity is in the $x-$axis, in the plane $x = \overline{AB}-d$, obtaining

\begin{subequations}
\label{eq:FcRight}
\begin{equation}
\left\{r^{^{[\mathrm{TE}]}}_{_\mathrm{right}}(\theta) ,r^{^{[\mathrm{TM}]}}_{_\mathrm{right}}(\theta)  \right\}= \left\{\frac{\displaystyle n\cos\psi -\cos\theta}{\displaystyle n\cos\psi + \cos\theta}, \frac{\displaystyle \cos\psi - n\cos\theta}{\displaystyle \cos\psi + n\cos\theta}\right\}\,e^{2\,i\,n\,k(\overline{AB}-d)\cos\psi}
\end{equation}
and
\begin{equation}
\left\{ t^{^{[\mathrm{TE}]}}_{_\mathrm{right}}(\theta) ,t^{^{[\mathrm{TM}]}}_{_\mathrm{right}}(\theta) \right\} = \left\{ \frac{\displaystyle 2\,n\cos\psi}{\displaystyle \cos\theta + n\cos\psi}, \frac{\displaystyle 2\,\cos\psi}{\displaystyle n\cos\theta + \cos\psi}\right\}\,e^{i\,k(n\cos\psi-\cos\theta)(\overline{AB}-d)}.
\end{equation}
\end{subequations}

The reflectivity ($R=|r|^{^2}$) and transmissivity ($T = 1-|r|^{^2}$) associated to Fresnel's coefficients above are represented as a function of the incidence angle $\theta$ in Figures 3 and 4, respectively, for a borosilicate ($n=1.515$) prism. \cor{Note} that the coefficients for the right interface are not plotted since they would reproduce the plots 3(a) and 4(a). From these graphics we can see that some incidence angles present particularly interesting effects. The TM reflection at the left face, Figure 3(a), for instance, which is an external reflection, meaning that the refractive index of the transmitting medium is greater than the incident medium's, becomes null for an incidence angle

\WideFigureSideCaption{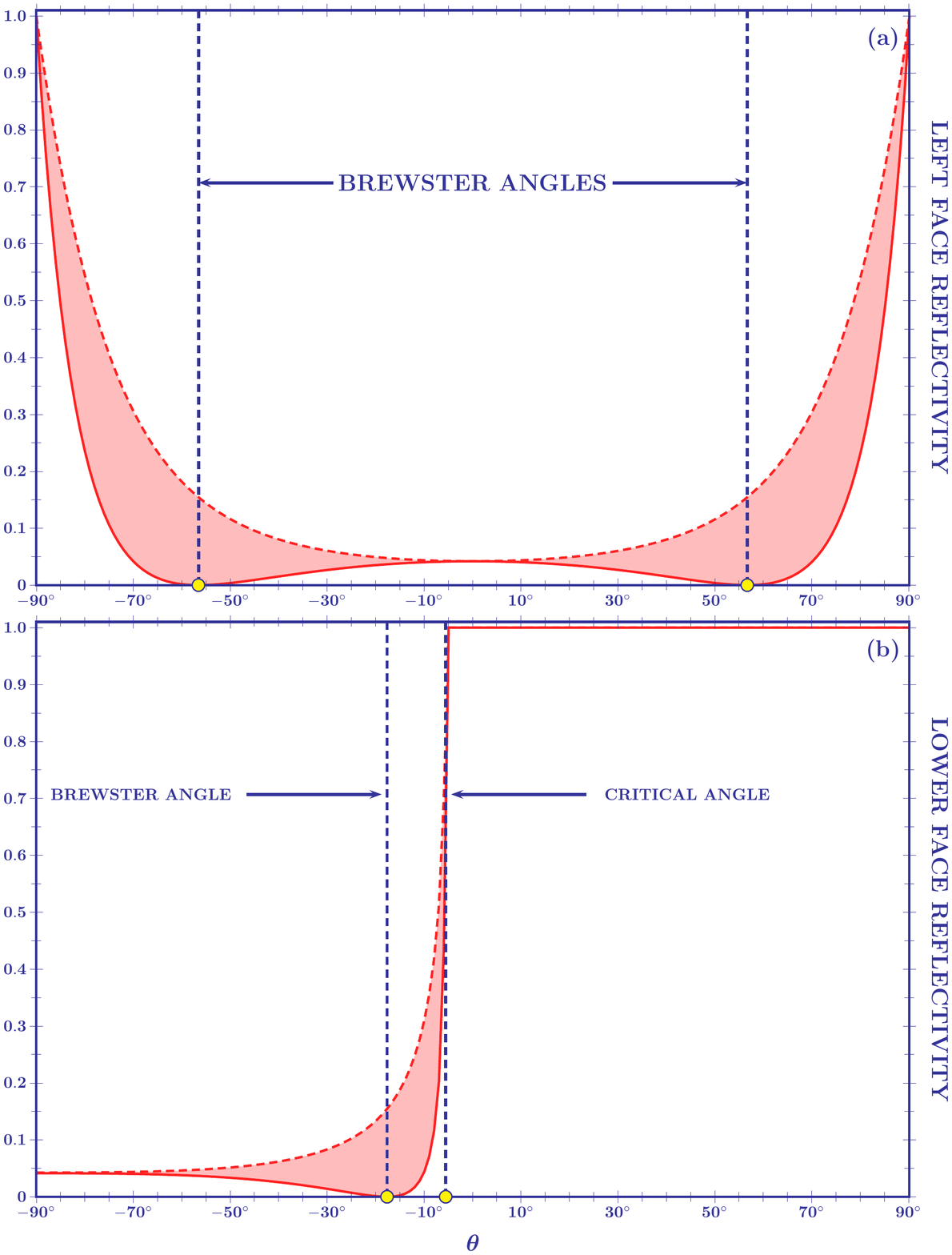}{The reflectivity of the left (a) and lower (b) faces of a borosilicate ($n=1.515$) prism as a function of the incident angle $\theta$. The solid lines stand for the TM-polarisation, while the dashed lines stand for the TE-polarisation. For the left face of the prism there are two Brewster angles, that is, angles for which there is no reflection of TM-polarised waves, located at $\theta_{_{\mathrm{B(ext)}}}=\pm 56.57^{\circ}$. For the the lower face of the prism there is one Brewster angle ($\theta_{_{\mathrm{B(int)}}}=-14.38^{\circ}$) and one critical angle ($\theta_{_{\mathrm{cri}}}=-5.603^{\circ}$). Critical incidence makes the reflection coefficient complex and we enter in the so called Total Internal Reflection regime, where the wave's energy is reflected in its entirety.}

\WideFigureSideCaption{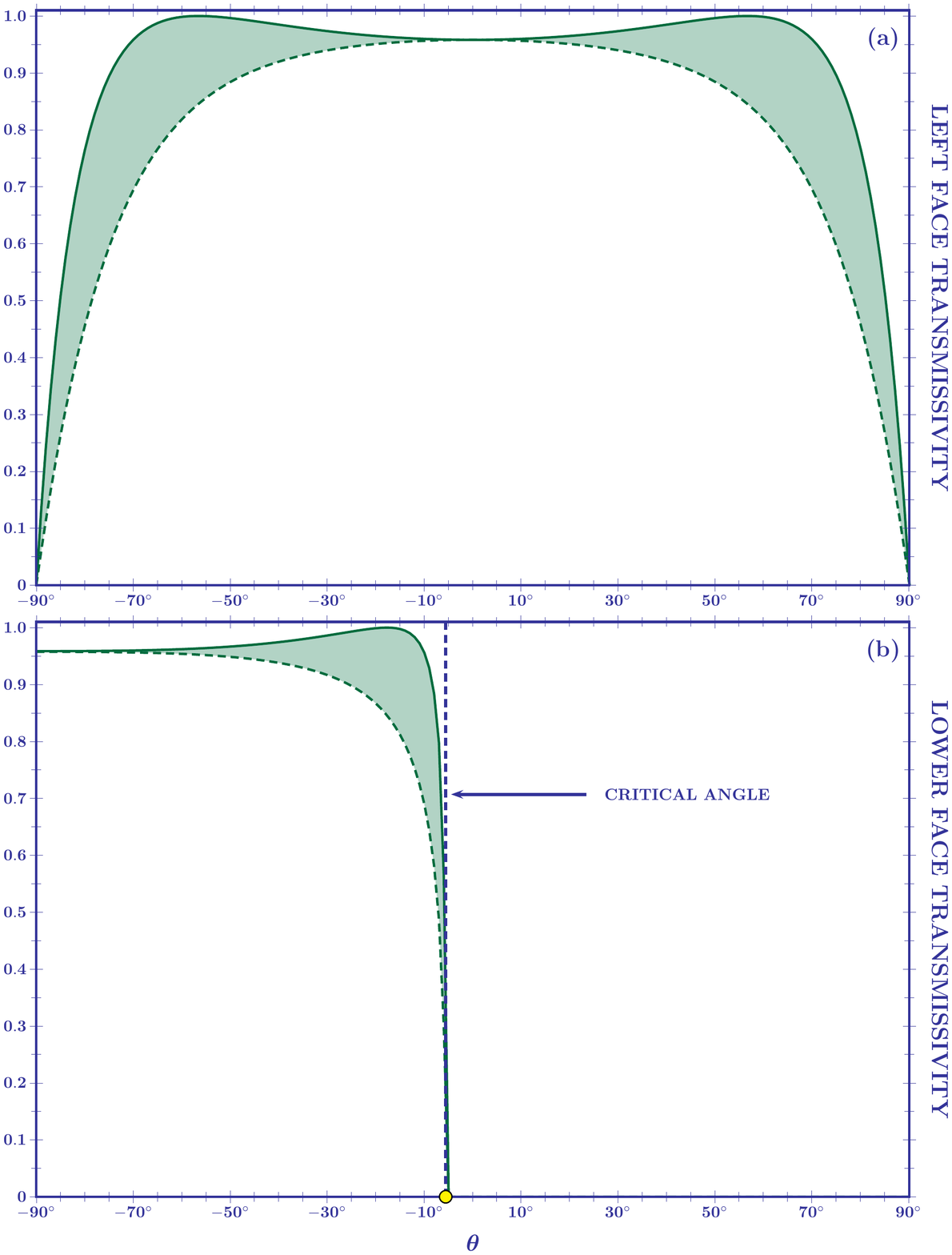}{The transmissivity of the left (a) and lower (b) faces of a borosilicate ($n=1.515$) prism as a function of the incident angle $\theta$. The solid lines stand for the TM-polarisation, while the dashed lines stand for the TE-polarisation. For the lower face, where the incident medium is denser than the refracting one, the transmissivity is characterised by the presence of evanescent waves, hence, the null transmission after the critical angle ($\theta_{_{\mathrm{cri}}}=-5.603^{\circ}$).}

\begin{equation}
\theta_{_\mathrm{B(ext)}}^{^{\pm}} = \pm \arcsin\left[ \frac{n}{\sqrt{n^{\2}+1}} \right].
\end{equation}
This is known as the external Brewster angle, hence the subscript in the equation above, $\mathrm{B(ext)}$, and it is a polarisation angle: a plane wave reflected at this angle will have its TM polarisation component filtered out. The same effect occurs for the reflection at the bottom of the prism, Figure 3(b), this time an internal reflection, with the refractive index of the incident medium being greater. For the angle
\begin{equation}
\varphi_{_{\mathrm{B(int)}}} = \arcsin\left[\frac{1}{\sqrt{n^{\2}-1}}\right],
\end{equation}
no TM-polarised light is reflected. This can be written for the incidence angle $\theta$, the angle the experimentalist has direct control over, as

\begin{equation}
\theta_{_{\mathrm{B(int)}}} = \arcsin\left[ \frac{n\,(1-n)}{\sqrt{2}\,\sqrt{n^{\2}+1}} \right].
\end{equation}
\cor{Note} that there are two symmetrical external Brewster angles and only one internal. The reason is that for the external reflection we can choose the incident angle symmetrically around the normal to the interface, being the `$+$' sign associated with an anticlockwise rotation from the normal and the `$-$' sign with a clockwise rotation. The geometry of the system, however, limits the angle $\varphi$ of the internal reflection to an anticlockwise rotation only. As we will see later, the Brewster angles play an important role in angular deviations from geometrical optics.

Besides the Brewster angles, there is another angle of interest called the \emph{critical} angle. It is given by $\varphi_{_{\mathrm{cri}}} = \arcsin\left[1/n\right]$, or, expressing it for $\theta$, by

\begin{equation}
\theta_{_{\mathrm{cri}}} = \arcsin\left[ \frac{1-\sqrt{n^{\2}-1}}{\sqrt{2}} \right].
\end{equation}
This angle marks a threshold. For incidence angles greater than $\theta_{_{\mathrm{cri}}}$ the reflection coefficient of the lower interface becomes complex and its reflectivity becomes 1, which characterises the phenomenon known as Total Internal Reflection. In this regime we have that
\begin{equation}
\label{eq:rGH}
r_{_{\mathrm{lower}}}^{^{[\mathrm{TE,TM}]}}(\theta)  = e^{\sqrt{\2}\,i\,n\,k\,d\cos\varphi+i\,\Phi_{_{\mathrm{GH}}}^{^{[\mathrm{TE,TM}]}}},
\end{equation}
where $\Phi_{_{\mathrm{GH}}}^{^{[\mathrm{TE,TM}]}}$ are the Goos-H\"anchen phases
\begin{equation}
\label{eq:GHphs}
\left\{ \Phi_{_{\mathrm{GH}}}^{^{[\mathrm{TE}]}}, \Phi_{_{\mathrm{GH}}}^{^{[\mathrm{TM}]}} \right\} = -2 \, \left\{ \arctan\left[\frac{\displaystyle \sqrt{n^{\2}\sin^{\2}\varphi-1}}{n\cos\varphi}\right],\arctan\left[\frac{\displaystyle n\,\sqrt{n^{\2}\sin^{\2}\varphi-1}}{\cos\varphi}\right] \right\},
\end{equation}
which, as the name suggests, are intrinsically related to the Goos-H\"anchen shift, as we will see in the next Section.

At the beginning of the Section, we described the light's path through the optical system in terms of the law of reflection and the Snell's law. This information is encoded in the mathematical description of these electromagnetic plane waves, being hold by their phase. The light's trajectory is obtained from the condition
\begin{equation}
\frac{\partial\Phi(\Theta)}{\partial\Theta} = 0,
\end{equation}
where $\Phi(\Theta)$ is the wave's phase and $\Theta$ is the angle of the trajectory's inclination in the \cor{proper} coordinate system. For the incident wave on the lower face of the prism, for instance, $\Phi(\varphi) = \Phi_{_{\mathrm{LowInc}}}(\varphi) = n\,k\,(x_{_{*}}\sin\varphi+z_{_{*}}\cos\varphi)$. It travels then along the line $x_{_{*}} = \tan\varphi \, z_{_{*}}$, meeting the lower face at $x_{_{*}} = \tan\varphi \, d/\sqrt{2}$. Here we can see the role played by the exponentials in Fresnel's coefficients, see Eqs. (\ref{eq:FcLeft}) to (\ref{eq:FcRight}). They provide the intersection points between wave vectors and the prism. For the beam reflected at the lower interface (in the Partial Reflection Regime) we have that
\begin{equation}
\label{eq:lowREFPHS}
\Phi_{_{\mathrm{LowRef}}}(\varphi) = n\,k\left(x_{_{*}}\sin\varphi - z_{_{*}}\cos\varphi + \sqrt{2}\,d\cos\varphi\right),
\end{equation}

\WideFigureSideCaption{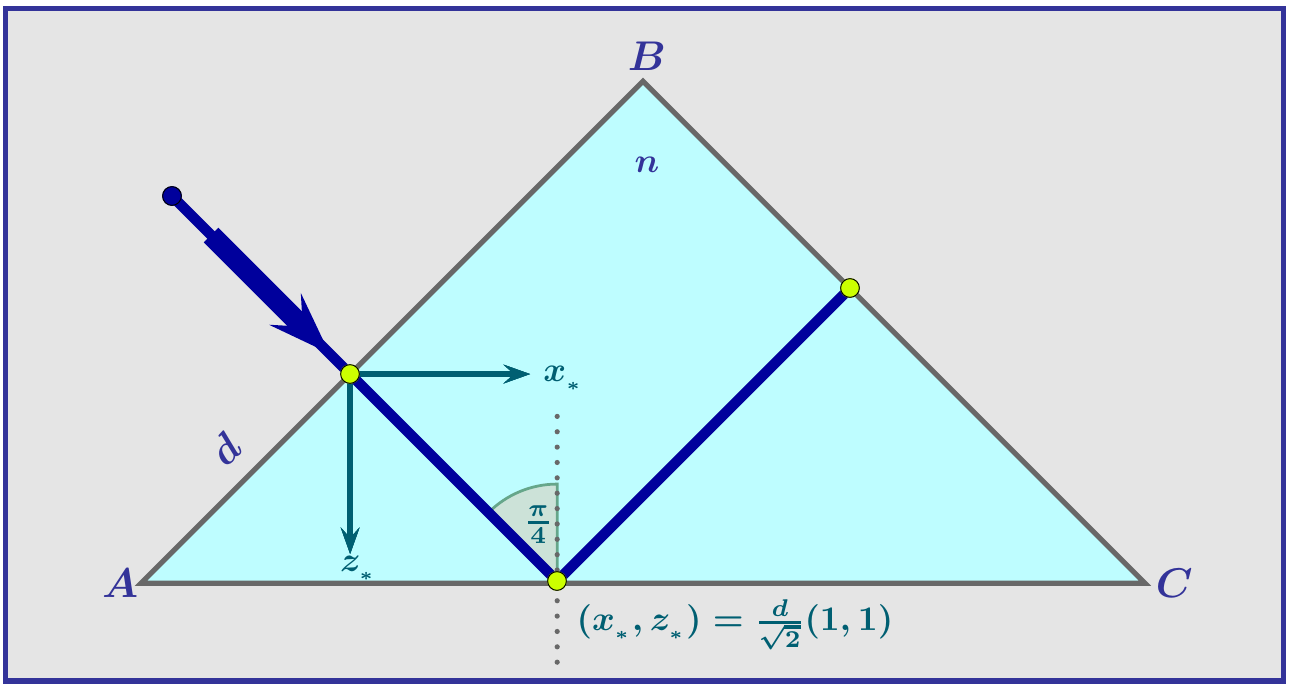}{For a plane wave with an incidence angle $\theta=0$ the coordinates of the intersection point between the wavevector refracted by the left face of the prism and its lower face in the $x_{_{*}}-z_{_{*}}$ system can be easily obtained geometrically, which is confirmed by the phase analysis yielding Eq. (\ref{eq:xstar}). In the Partial Reflection Regime, this intersection point is the same for the incident, reflected, and refracted wave vectors.}

and the condition
\begin{equation}
\frac{\partial \Phi_{_{\mathrm{LowRef}}}(\varphi)}{\partial\varphi} = 0
\end{equation}
yields
\begin{equation}
\label{eq:xstar}
x_{_{*}} = \tan\varphi\,\left(\sqrt{2}\,d-z_{_{*}}\right).
\end{equation}
\cor{So}, the reflection occurs at $x_{_{*}} = \tan\varphi\,d/\sqrt{2}$. The same result is obtained for the wave transmitted through the lower interface. In this case we have that
\begin{equation}
\frac{\partial}{\partial\phi}\left(k\left[x_{_{*}}\sin\phi + z_{_{*}}\cos\phi + \left(n\cos\varphi - \cos\phi\right)d/\sqrt{2}\right]\right) = 0,
\end{equation}
which provides us with the line equation
\begin{equation}
x_{_{*}} = \tan\phi\,z_{_{*}} + \left( \tan\varphi - \tan\phi \right)\frac{d}{\sqrt{2}}.
\end{equation}
For $z_{_{*}} = d/\sqrt{2}$ the equation above gives $x_{_{*}} = \tan\varphi\,d/\sqrt{2}$, showing that, in the Partial Reflection Regime, the incoming, transmitted, and reflected waves at the lower interface meet at the same point. This can be easily geometrically verified for $\theta=0$, hence $\varphi=\pi/4$, as can be seen in Figure 5. The waves' phases that come from the complex exponentials in Fresnel's coefficients are called \emph{geometrical phases} because they hold information about the light's path according to Geometrial Optics. From the cases analysed above, we can see that the displacement from the origin due to the geometrical phase $\Phi_{_{\mathrm{geo}}}$ is given by
\begin{equation}
x_{_{\mathrm{geo}}} = -\frac{1}{k\cos\Theta}\frac{\partial\Phi_{_{\mathrm{geo}}}}{\partial\Theta}.
\end{equation}

This connection between phase and trajectory tells us that adding incident-angle-dependent phases to the wave displaces the intersection between its path and a particular face of the prism. This is the mechanism behind the Goos-H\"anchen shift. As we saw in Eqs. (\ref{eq:rGH}) and (\ref{eq:GHphs}), in Total Internal Reflection,  the wave acquires an additional phase, the Goos-H\"anchen phase, which will generate a displacement of the reflected \cor{beam} from the point where the incident and the transmitted \cor{beams} meet. In the next Section, we will study the Gaussian beam formalism, and in \cor{Section \ref{sec5}} the Goos-H\"anchen effect in greater detail.

\section{The Gaussian beam formalism}
\label{sec4}

Plane waves are a straightforward solution to the electromagnetic wave equation and their simplicity makes them easy to work with. They are not, however, physical solutions, carrying an infinite amount of energy and spreading throughout the whole space. The plane wave limit is an useful approximation, valid when the region with an appreciable electric field amplitude is greater than the characteristic dimensions of the optical system, but a more precise description of light used in experiments, for example, will relay on the concept of bounded beams.

Mathematically, beams are a collection of plane waves with amplitudes following a given distribution of their propagation direction. For this reason, contrary to plane waves, which have a well-defined direction of incidence, they present an angular spreading around their incidence angle. In the following, we will study  four Gaussian beams. The first one is the beam incident upon our optical system and the other three \cor{are} the beams resulting from such interaction, being one the beam reflected by the left interface, one the beam transmitted through the lower interface, and the last one the beam transmitted through the right interface.

Let us consider a Gaussian beam hitting the left face of the optical system discussed in section \ref{sec3}. Its electric field is given by
\begin{equation}
\label{eq:Ei}
E_{_{\mathrm{inc}}} = E_{\0}\,\int_{\mbox{\tiny $-\pi/\2$}}^{\mbox{\tiny $+\pi/\2$}}{\hspace{-0.25cm}}\mathrm{d}\theta\,g(\theta-\theta_{\0})\,e^{i\,k\,\left(x\,\sin\theta + z\,\cos\theta\right)},
\end{equation}
where $E_{\0}$ is the electric field's amplitude in the center of the beam, and $g(\theta-\theta_{\0})$ is the Gaussian angular distribution, given by
\begin{equation}
\label{eq:gaussianDist}
g(\theta-\theta_{\0}) = \frac{\displaystyle k\,\mathrm{w}_{\0}}{\displaystyle 2\,\sqrt{\pi}}\,\exp\left[ -(\theta-\theta_{\0})^{^2}\,\frac{\displaystyle (k\,\mathrm{w}_{\0})^{^2}}{4}\right].
\end{equation}
In the expression above, $\mathrm{w}_{\0}$ is the distribution's waist and $\theta_{\0}$ the position of its center, which is the incidence angle of the beam. \cor{Note} that as $\mathrm{w}_{\0}$ becomes greater\cor{,} the Gaussian function becomes more strongly centred around $\theta_{\0}$, taking the beam to the plane wave limit. In order for our analysis of beam shifts to be more approachable\cor{,} we will consider the paraxial limit, which lies on the path to the plane wave limit, without being as drastic. The paraxial limit considers strongly collimated beams, with $k\,\mathrm{w}_{\0} \gg 2\pi$, which in turn allows the expansion of the trigonometric functions in the phase of the electric field given by Eq. (\ref{eq:Ei}) up to second order around $\theta_{\0}$:
\begin{eqnarray*}
\sin\theta &\approx & \sin\theta_{\0} + \cos\theta_{\0}(\theta-\theta_{\0})-\frac{1}{2}\sin\theta_{\0}(\theta-\theta_{\0})^{\2},
\\
\cos\theta & \approx & \cos\theta_{\0} - \sin\theta_{\0}(\theta-\theta_{\0})-\frac{1}{2}\cos\theta_{\0}(\theta-\theta_{\0})^{\2}.
\end{eqnarray*}
By defining a coordinate system $x_{_{\mathrm{inc}}}-z_{_{\mathrm{inc}}}$, see Figure 6(b), which is parallel to the incidence direction of the beam, we have that
\begin{equation}
\left[\begin{array}{cc}
x_{_{\mathrm{inc}}} \\ z_{_{\mathrm{inc}}}
\end{array}\right] = \left( \begin{array}{cc}
\cos\theta_{\0} & -\sin\theta_{\0} \\
\sin\theta_{\0} & \cos\theta_{\0}
\end{array} \right) \, \left[ \begin{array}{cc}
x \\ z
\end{array}\right],
\end{equation}
and we can then write Eq. (\ref{eq:Ei}) as
\begin{equation}
\label{eq:IncF}
E_{_\mathrm{inc}} = E_{\0}\,e^{i\,k\,z_{\mathrm{inc}}}\,\int_{\-infty}^{\+infty}{\hspace{-0.25cm}}\mathrm{d}\theta\,g(\theta-\theta_{\0})\,e^{i\,k\,\left[x_{_{\mathrm{inc}}}\,(\theta-\theta_{\0}) - z_{_{\mathrm{inc}}}\,(\theta-\theta_{\0})^{\2}/2\right]},
\end{equation}
\WideFigureSideCaption{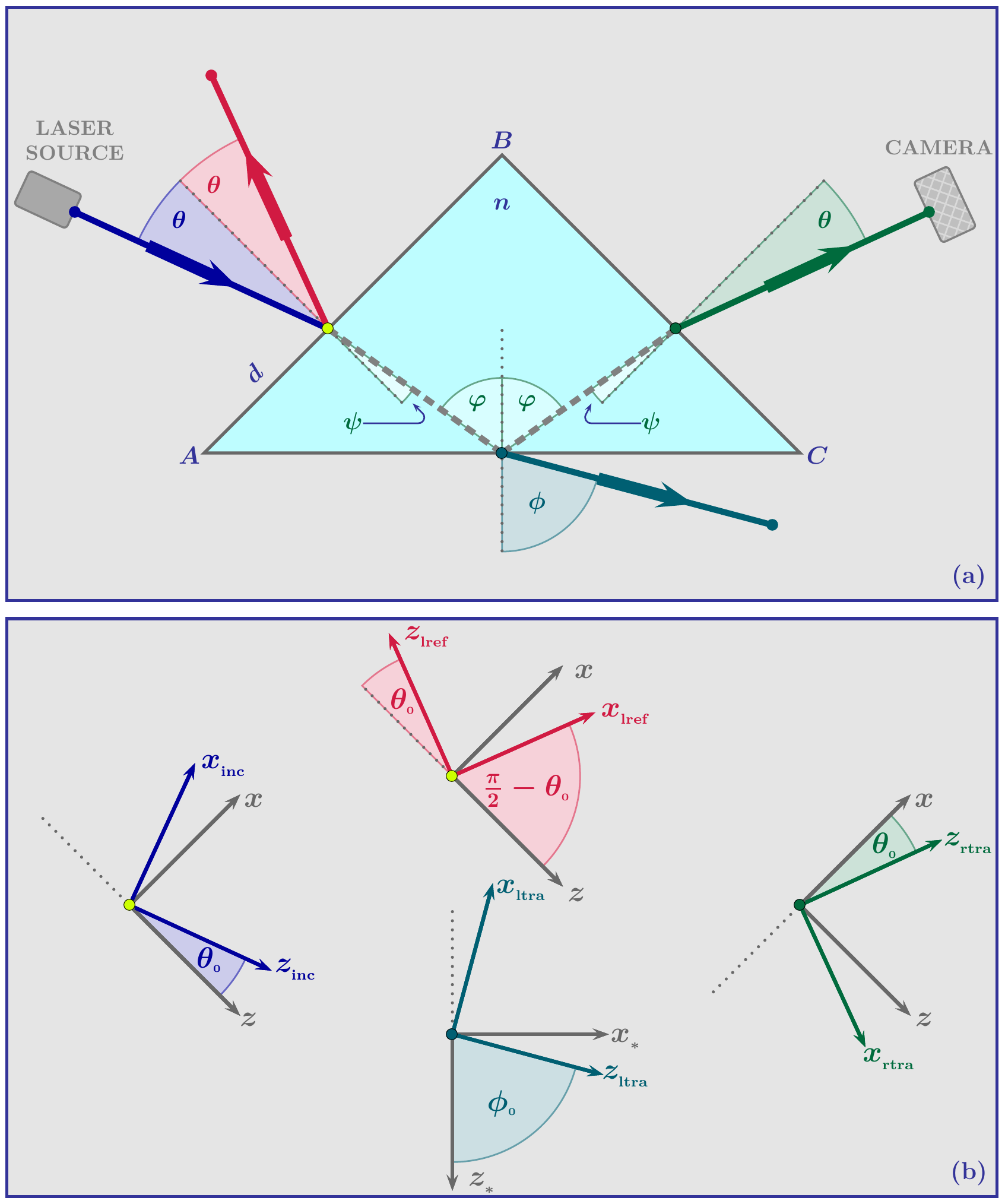}{The optical system of interest (a) and the convenient coordinate systems (b) defined in order to better study it. There are two coordinate systems associated to the prism, the $x-z$ and the $x_{_{*}}-z_{_{*}}$ systems. The $z-$coordinate is perpendicular to the left face of the prism, while the $z_{_{*}}-$coordinate is perpendicular to its lower face. In addition, we define four other systems, with $z-$components following the propagation direction of the beams we are interested in. The $x_{_{\mathrm{inc}}}-z_{_{\mathrm{inc}}}$ system is parallel to the incoming beam, the $x_{_{\mathrm{lref}}}-z_{_{\mathrm{lref}}}$ system to the beam reflected by the prism's left face, the $x_{_{\mathrm{ltra}}}-z_{_{\mathrm{ltra}}}$ system to the beam transmitted through the lower face of the prism, and the $x_{_{\mathrm{rtra}}}-z_{_{\mathrm{rtra}}}$ system is parallel to the beam transmitted through the right face of the prism.}

where we have used the paraxial approximation to make the integration limits infinite. This integral is integrable, returning
\begin{equation}
\label{eq:IncIntegrated}
E_{_{\mathrm{inc}}} = E_{\0}\,e^{i\,k\,z_{_{\mathrm{inc}}}}\frac{\displaystyle \mathrm{w}_{\0}}{\displaystyle \mathrm{w}(z_{_{\mathrm{inc}}})}\,\exp\left[-\frac{\displaystyle x_{_{\mathrm{inc}}}^{\2}}{\displaystyle \mathrm{w}^{\2}(z_{_{\mathrm{inc}}})}-i\,\frac{\Psi_{\0}}{2}+i\,\frac{x_{_{\mathrm{inc}}}^{\2}}{\mathrm{w}^{\2}(z_{_{\mathrm{inc}}})}\,\zeta\right],
\end{equation}
where $\zeta=z_{_{\mathrm{inc}}}/z_{_{\mathrm{R}}}$, being $z_{_{\mathrm{R}}}=k\,\mathrm{w}_{\0}^{\2}/2$ the Rayleigh length, which gives the distance from the point of minimum waist to where the area of the cross section of the beam doubles. Besides,
\begin{equation}
\Psi_{\0} = \arctan\zeta
\end{equation}
is the Gouy's phase, which describes the phase change of the beam after the point of minimal beam waist, that is, $\zeta=0$, and
\begin{equation}
\label{eq:wz}
\mathrm{w}(z_{_{\mathrm{inc}}}) = \mathrm{w_{\0}}\sqrt{1+\zeta^{^2}},
\end{equation}
defines the diameter of the Gaussian beam, giving the radius where the electric field intensity falls to $1/2e^{\2}$ of its peak value \cite{PamEnd2004}.

It is interesting to analyse the power associated to this Gaussian beam, and how the interaction with the prism changes it. This investigation is analogous to the reflectivity and transmissivity study carried out in the last Section for plane waves. The incident power is given by
\begin{equation}
P_{_{\mathrm{inc}}} = \int_{\-infty}^{\+infty}{\hspace{-0.25cm}}\mathrm{d}x_{_{\mathrm{inc}}} \, |E_{_{\mathrm{inc}}}|^{^2},
\end{equation}
which is a straightforward integration using Eq. (\ref{eq:IncIntegrated}). However, in order to prepare for future, more complicated calculations, let us use the integral form of the incident electric field given by Eq. (\ref{eq:IncF}). Using the relation
\begin{equation}
\label{eq:DiracD}
\int_{\-infty}^{\+infty}{\hspace{-0.25cm}}\mathrm{d}x_{_{\mathrm{inc}}}\,e^{i\,k\,(\theta-\tilde{\theta})\,x_{_{\mathrm{inc}}}} = \frac{2\pi}{k}\,\delta(\theta-\tilde{\theta}),
\end{equation}
where $\delta(\theta-\tilde{\theta})$ is the Dirac's delta function, we can write the power integral in its angular form:
\begin{equation}
\label{eq:PowerInc}
P_{_{\mathrm{inc}}} = \frac{2\pi}{k}\,|E_{\0}|^{^2}\,\int_{\-infty}^{\+infty}{\hspace{-0.25cm}}\mathrm{d}\theta\,g^{\2}(\theta-\theta_{\0}) = \sqrt{\frac{\pi}{2}}\,\mathrm{w}_{\0}\,|E_{\0}|^{^{2}}.
\end{equation}

As for the electric field of the beam reflected at the left face of the prism, it has its angular distribution modified by the reflection coefficient of that interface:
\begin{equation}
\label{eq:Elref}
E_{_{\mathrm{lref}}}^{^{\mathrm{[TE,TM]}}} = E_{\0}\,\int_{\-infty}^{\+infty}{\hspace{-0.25cm}}\mathrm{d}\theta\,r^{^{\mathrm{[TE,TM]}}}_{_{\mathrm{left}}}(\theta)\,g(\theta-\theta_{\0})\,e^{i\,k\,\left(x\,\sin\theta - z\,\cos\theta\right)}.
\end{equation}
Expanding the sine and cosine functions up to \cor{the} second order as done for the incident beam, and defining the reflected coordinate system $x_{_{\mathrm{lref}}}-z_{_{\mathrm{lref}}}$, see Figure 6(b), which respects the following relation,
\begin{equation}
\left[\begin{array}{cc}
z_{_{\mathrm{lref}}} \\ x_{_{\mathrm{lref}}}
\end{array}\right] = \left( \begin{array}{cc}
\sin\theta_{\0} & -\cos\theta_{\0} \\
\cos\theta_{\0} & \sin\theta_{\0}
\end{array} \right) \, \left[ \begin{array}{cc}
x \\ z
\end{array}\right],
\end{equation}
we can rewrite Eq. (\ref{eq:Elref}) as
\begin{equation}
\label{eq:ElrefFinal}
E_{_\mathrm{lref}}^{^{\mathrm{[TE,TM]}}} = E_{\0}\,e^{i\,k\,z_{\mathrm{lref}}}\,\int_{\-infty}^{\+infty}\mathrm{d}\theta\,r^{^{\mathrm{[TE,TM]}}}_{_{\mathrm{left}}}(\theta)\,g(\theta-\theta_{\0})\,e^{i\,k\,\left[x_{_{\mathrm{lref}}}\,(\theta-\theta_{\0}) - z_{_{\mathrm{lref}}}\,(\theta-\theta_{\0})^{\2}/2\right]}.
\end{equation}

Following \cor{the procedure} for the incident power we see that the reflected power is simply Eq. (\ref{eq:PowerInc}) modified by the reflection coefficient $r_{_{\mathrm{left}}}^{^{\mathrm{[TE,TM]}}}(\theta)$,
\begin{equation}
\label{eq:PowerRef}
P_{_{\mathrm{lref}}}^{^{\mathrm{[TE,TM]}}} = \frac{2\pi}{k}\,|E_{\0}|^{^2}\,\int_{\-infty}^{\+infty}{\hspace{-0.25cm}}\mathrm{d}\theta\,\left[g(\theta-\theta_{\0})r_{_{\mathrm{left}}}^{^{\mathrm{[TE,TM]}}}(\theta)\right]^{^2}.
\end{equation}
Expanding $[r_{_{\mathrm{left}}}^{^{\mathrm{[TE,TM]}}}(\theta)]^{^2}$ up to \cor{the} first order we can integrate the equation above to obtain the normalised reflected power as

\begin{equation}
\mathcal{P}_{_{\mathrm{lref}}}^{^{\mathrm{[TE,TM]}}} = \frac{P_{_{\mathrm{lref}}}^{^{\mathrm{[TE,TM]}}}}{P_{_{\mathrm{inc}}}} = |r_{_{\mathrm{left}}}^{^{\mathrm{[TE,TM]}}}(\theta_{\0})|^{^{2}}.
\end{equation}
The power reflected by the left face of the prism is depicted in Figure 7(a).

The beam transmitted through the right face of the prism is the result of three interactions with the interfaces, being modified by the transmission coefficient of the left face, by the reflection coefficient of the lower face, and, finally, by the transmission coefficient of the right face, being written as
\begin{equation}
\label{eq:Er}
E_{_{\mathrm{rtra}}}^{^{\mathrm{[TE,TM]}}} = E_{\0}\,\int_{\-infty}^{\+infty}{\hspace{-0.25cm}}\mathrm{d}\theta\,t^{^{\mathrm{[TE,TM]}}}_{_{\mathrm{left}}}(\theta)r^{^{\mathrm{[TE,TM]}}}_{_{\mathrm{lower}}}(\theta)t^{^{\mathrm{[TE,TM]}}}_{_{\mathrm{right}}}(\theta)\,g(\theta-\theta_{\0})\,e^{i\,k\,\left(z\,\sin\theta + x\,\cos\theta\right)}.
\end{equation}
Following the same expansion step as before, and defining the right transmission coordinate system $y_{_{\mathrm{rtra}}}-z_{_{\mathrm{rtra}}}$, see Figure 6(b),

\WideFigureSideCaption{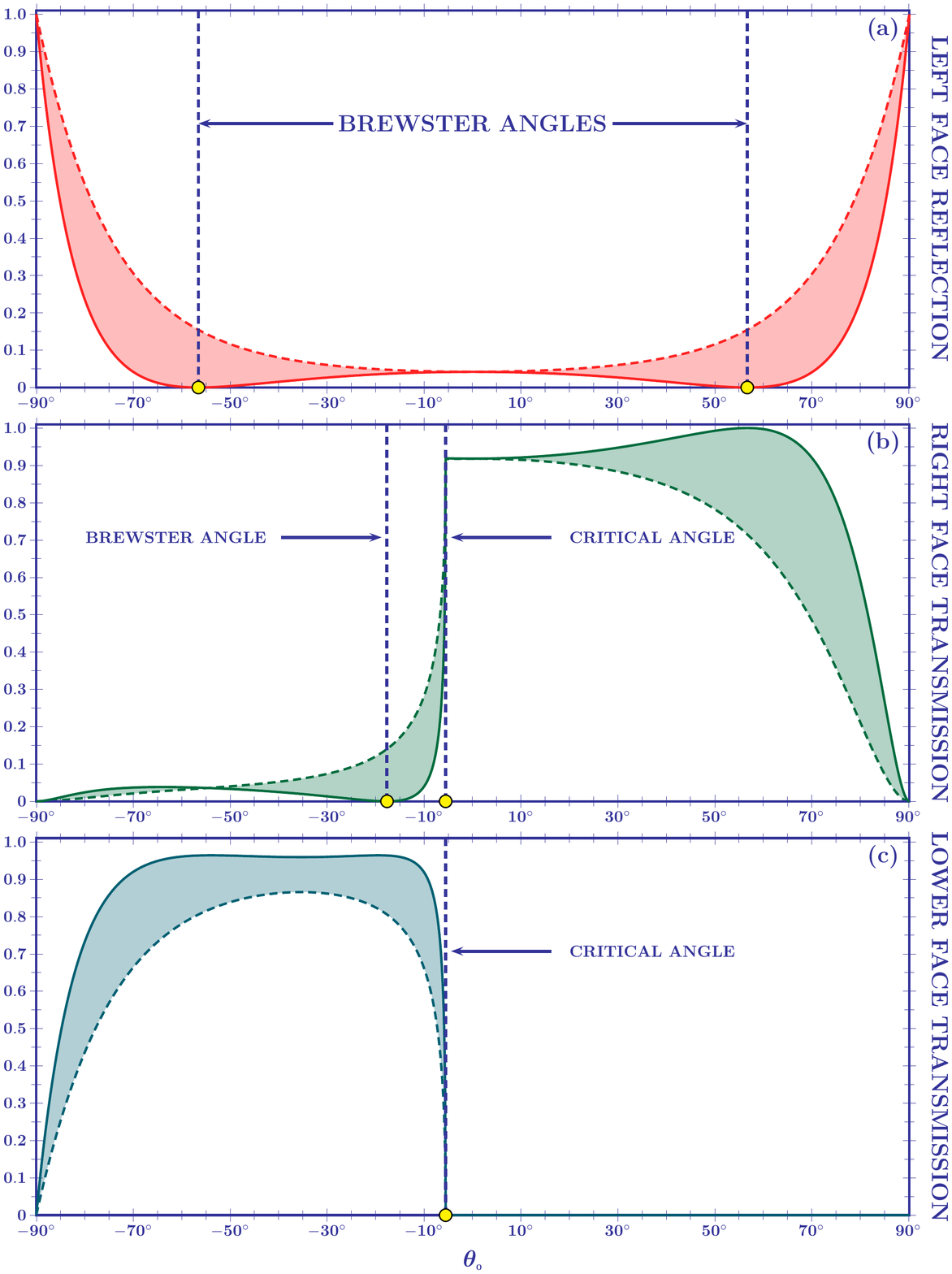}{The relative power reflected by the left face of a borosilicate ($n=1.515$) prism (a), transmitted through its right face (b) and transmitted through its lower face (c) as a function of the incidence angle $\theta_{\0}$. The solid and dashed lines represent the TM and TE-polarised light, respectively. For incidence angles greater than the critical angle ($\theta_{_{\mathrm{cri}}}=-5.60^{\circ}$) the waves across the lower interface are evanescent, and consequently no power is transmitted through that face. As a result, the transmission through the right face greatly increases after the critical angle.}

\begin{equation}
\left[\begin{array}{cc}
z_{_{\mathrm{rtra}}} \\ x_{_{\mathrm{rtra}}}
\end{array}\right] = \left( \begin{array}{cc}
\cos\theta_{\0} & \sin\theta_{\0} \\
-\sin\theta_{\0} & \cos\theta_{\0}
\end{array} \right) \, \left[ \begin{array}{cc}
x \\ z
\end{array}\right],
\end{equation}
we have
\begin{eqnarray}
E_{_\mathrm{rtra}}^{^{\mathrm{[TE,TM]}}} = E_{\0}\,e^{i\,k\,z_{\mathrm{rtra}}}\,\int_{\-infty}^{\+infty}{\hspace{-0.25cm}}\mathrm{d}\theta\,t^{^{\mathrm{[TE,TM]}}}_{_{\mathrm{left}}}(\theta)r^{^{\mathrm{[TE,TM]}}}_{_{\mathrm{lower}}}(\theta)t^{^{\mathrm{[TE,TM]}}}_{_{\mathrm{right}}}(\theta)\,g(\theta-\theta_{\0}){\nonumber}
\\
\times \, e^{i\,k\,\left[x_{_{\mathrm{rtra}}}\,(\theta-\theta_{\0}) - z_{_{\mathrm{rtra}}}\,(\theta-\theta_{\0})^{\2}/2\right]}.
\end{eqnarray}
The Fresnel's coefficients in the integral above have the geometrical phases discussed in \cor{the} last Section embedded in them. In order to make things clearer, let us detach these phases from the coefficients,
\[ r^{^{\mathrm{[TE,TM]}}}_{_{\mathrm{lower}}} \rightarrow r^{^{\mathrm{[TE,TM]}}}_{_{\mathrm{lower}}} e^{\sqrt{\2} i\,n\,k\,d\cos\varphi}\]
\[ t^{^{\mathrm{[TE,TM]}}}_{_{\mathrm{right}}} \rightarrow t^{^{\mathrm{[TE,TM]}}}_{_{\mathrm{right}}} e^{i\,k(n\cos\psi-\cos\theta)(\overline{AB}-d)},\]
and group them together in the geometrical phase of the right face transmission:
\begin{eqnarray}
\label{eq:rgeophs}
\Phi_{_{\mathrm{rgeo}}}(\theta) &=& \sqrt{2}\,n\,k\,d\cos\varphi + k(n\cos\psi-\cos\theta)(\overline{AB}-d){\nonumber}
\\
&=& k \left[\, (\cos\theta-\sin\theta)\,d + (n\cos\psi - \cos\theta)\,\overline{AB}\, \right].
\end{eqnarray}
Expanding this phase up to second order we have that the first order derivative shifts the $x_{_{\mathrm{rtra}}}$ component, giving the exit point of the beam along the $x_{_{\mathrm{rtra}}}$ direction,
\begin{eqnarray}
\label{eq:xrgeo}
x_{_{\mathrm{rgeo}}} &=& -\frac{1}{k}\left.\frac{\partial \Phi_{_{\mathrm{rgeo}}}(\theta)}{\partial\theta}\right|_{\0}{\nonumber}
\\
&=&(\sin\theta_{\0}+\cos\theta_{\0})d+\left(\frac{\cos\theta_{\0}}{n\,\cos\psi_{\0}}-1\right)\sin\theta_{\0}\,\overline{AB},
\end{eqnarray}
while the second order derivative acts as a beam profile modifier, as has been recently suggested \cite{Lima2016} and experimentally verified \cite{Car2016}. By \cor{then} defining the variables
\begin{subequations}
\label{eq:ytildertra}
\begin{equation}
\tilde{x}_{_{\mathrm{rtra}}} = x_{_{\mathrm{rtra}}} + \Phi_{_{\mathrm{rgeo}}}^{\prime}(\theta_{\0})/k = x_{_{\mathrm{rtra}}} - x_{_{\mathrm{rgeo}}},
\end{equation}
and
\begin{equation}
\tilde{z}_{_{\mathrm{rtra}}} = z_{_{\mathrm{rtra}}} + \Phi_{_{\mathrm{rgeo}}}^{\prime\prime}({\theta_{\0}})/k = z_{_{\mathrm{rtra}}} - z_{_{\mathrm{rgeo}}},
\end{equation}
\end{subequations}
we \cor{can write}  the electric field integral as
\begin{eqnarray}
\label{eq:Ertrafinal}
E_{_\mathrm{rtra}}^{^{\mathrm{[TE,TM]}}} = E_{\0}\,e^{i\,k\,z_{\mathrm{rtra}}+i\,\Phi_{_{\mathrm{rgeo}}}(\theta_{\0})}\,\int_{\-infty}^{\+infty}{\hspace{-0.25cm}}\mathrm{d}\theta\,t^{^{\mathrm{[TE,TM]}}}_{_{\mathrm{left}}}(\theta)r^{^{\mathrm{[TE,TM]}}}_{_{\mathrm{lower}}}(\theta)t^{^{\mathrm{[TE,TM]}}}_{_{\mathrm{right}}}(\theta)\,g(\theta-\theta_{\0}){\nonumber}
\\
\times e^{i\,k\,\left[\tilde{x}_{_{\mathrm{rtra}}}\,(\theta-\theta_{\0}) - \tilde{z}_{_{\mathrm{rtra}}}\,(\theta-\theta_{\0})^{\2}/2\right]}.
\end{eqnarray}

The same procedure as carried out before gives us the normalised power transmitted through the right face of the prism as
\begin{equation}
\label{eq:PowerRTRA}
\mathcal{P}_{_{\mathrm{rtra}}}^{^{\mathrm{[TE,TM]}}} = |t_{_{\mathrm{left}}}^{^{\mathrm{[TE,TM]}}}(\theta_{\0})r_{_{\mathrm{lower}}}^{^{\mathrm{[TE,TM]}}}(\theta_{\0})t_{_{\mathrm{right}}}^{^{\mathrm{[TE,TM]}}}(\theta_{\0})|^{^2},
\end{equation}
Note that the power integral is taken along the direction perpendicular to the propagation direction of the beam, which, in this case, is $z_{_{\mathrm{rtra}}}$. The variable $\tilde{x}_{_{\mathrm{rtra}}}$, however, is simply shifted by a constant value from $x_{_{\mathrm{rtra}}}$, and, therefore, $\mathrm{d} x_{_{\mathrm{rtra}}} = \mathrm{d}\tilde{x}_{_{\mathrm{rtra}}}$, not affecting the integral. The power transmitted through the right face of the prism is depicted in Figure 7(b). We can see in this plot that the transmitted power is greatly increased after the critical angle. This is due to the fact that in this regime no power is lost through the lower face.

Finally, the beam transmitted through the lower face of the prism\footnote{\cor{Note} that this electric field is represented by $E_{_{\mathrm{ltra}}}$, where the sub-index stands for ``lower transmission'', not to be confused with the sub-index ``lref'', which stands for ``left reflection''. Since there is no camera collecting the beam transmitted through the left interface, this choice \cor{of} indices should not cause any confusion.} has its angular distribution modified by the transmission coefficients of the left and lower interfaces,
\begin{equation}
\label{eq:Er}
E_{_{\mathrm{ltra}}}^{^{\mathrm{[TE,TM]}}} = E_{\0}\,\int_{\-infty}^{\+infty}{\hspace{-0.25cm}}\mathrm{d}\theta\,t^{^{\mathrm{[TE,TM]}}}_{_{\mathrm{left}}}(\theta)t^{^{\mathrm{[TE,TM]}}}_{_{\mathrm{lower}}}(\theta)\,g(\theta-\theta_{\0})\,e^{i\,k\,\left(x_{_{*}}\,\sin\phi + z_{_{*}}\,\cos\phi\right)+i\,\Phi_{_{\mathrm{lgeo}}}(\theta)},
\end{equation}
where the lower transmission geometrical phase has already been detached from the Fresnel's coefficients and is given by
\begin{equation}
\Phi_{_{\mathrm{lgeo}}}(\theta) = \frac{k \, d}{\sqrt{2}}\,(n\cos\varphi-\cos\phi).
\end{equation}
\cor{Noting} that
\begin{eqnarray*}
\sin\phi &\approx & \sin\phi_{\0} + \cos\phi_{\0}\,(\theta-\theta_{\0})\phi^{\prime}_{\0} - \left[ \sin\phi_{\0}\,(\phi^{\prime})^{\2}  +\cos\phi_{\0}\, \phi_{\0}^{\prime\prime}\right](\theta-\theta_{\0})^{\2}/2
\\
\cos\phi & \approx & \cos\phi_{\0} - \sin\phi_{\0}\,(\theta-\theta_{\0})\phi^{\prime}_{\0} - \left[ \cos\phi_{\0}\,(\phi^{\prime})^{\2} + \sin\phi_{\0}\, \phi_{\0}^{\prime\prime}\right](\theta-\theta_{\0})^{\2}/2 \,,
\end{eqnarray*}
and defining the coordinate system $y_{_{\mathrm{ltra}}} - z_{_{\mathrm{ltra}}}$, see Figure 6(b),
\begin{equation}
\left[\begin{array}{cc}
x_{_{\mathrm{ltra}}} \\ z_{_{\mathrm{ltra}}}
\end{array}\right] = \left( \begin{array}{cc}
\cos\phi_{\0} & -\sin\phi_{\0} \\
\sin\phi_{\0} & \cos\phi_{\0}
\end{array} \right) \, \left[ \begin{array}{cc}
x_{_{*}} \\ z_{_{*}}
\end{array}\right],
\end{equation}
we have
\begin{eqnarray}
E_{_\mathrm{ltra}}^{^{\mathrm{[TE,TM]}}} = E_{\0}\,e^{i\,k\,z_{\mathrm{ltra}}}\,\int_{\-infty}^{\+infty}{\hspace{-0.25cm}}\mathrm{d}\theta\,t^{^{\mathrm{[TE,TM]}}}_{_{\mathrm{left}}}(\theta)t^{^{\mathrm{[TE,TM]}}}_{_{\mathrm{lower}}}(\theta)\,g(\theta-\theta_{\0}){\nonumber}
\\
\times \, e^{i\,k\,\{x_{_{\mathrm{ltra}}}\,\phi_{\0}^{\prime}(\theta-\theta_{\0}) - z_{_{\mathrm{ltra}}}\,[\phi_{\0}^{\prime}(\theta-\theta_{\0})]^{\2}/2\} + i\,\Phi_{_{\mathrm{lgeo}}}(\theta)},
\end{eqnarray}
where we have neglected the term dependent on the second order derivative of $\phi$. This is a valid approximation since $\phi_{\0}^{\prime} \sim \phi_{\0}^{\prime\prime}$ and $z_{_{\mathrm{ltra}}}\gg x_{_{\mathrm{ltra}}}$, meaning that measurements are carried out at a distance far greater than the
characteristic dimensions of the beam. Expanding now the geometrical phase up to \cor{the} second order we can define the new spacial coordinates
\begin{subequations}
\begin{equation}
\label{eq:TILyltra}
\tilde{x}_{_{\mathrm{ltra}}} = \phi^{\prime}_{\0}x_{_{\mathrm{ltra}}} + \Phi^{\prime}_{_{\mathrm{lgeo}}}(\theta_{\0})/k  =  \phi^{\prime}_{\0}x_{_{\mathrm{ltra}}}  -  x_{_{\mathrm{lgeo}}},
\end{equation}
and
\begin{equation}
\tilde{z}_{_{\mathrm{ltra}}} = (\phi^{\prime}_{\0})^{\2} z_{_{\mathrm{ltra}}} + \Phi^{\prime\prime}_{_{\mathrm{lgeo}}}(\theta_{\0})/k  =  (\phi^{\prime}_{\0})^{\2}\,z_{_{\mathrm{ltra}}}  -  z_{_{\mathrm{lgeo}}},
\end{equation}
\end{subequations}
being
\begin{equation}
x_{_{\mathrm{lgeo}}} = \frac{d}{\sqrt{2}}\,(\tan\varphi_{\0}\cos\phi_{\0}-\sin\phi_{\0})
\end{equation}
the shift of the beam's trajectory on the $x_{_{\mathrm{ltra}}}-$axis, and
\begin{equation}
\phi_{\0}^{\prime} = \frac{\cos\varphi_{\0}\,\cos\theta_{\0}}{\cos\psi_{\0}\,\cos\phi_{\0}}.
\end{equation}
The electric field of the lower transmitted beam is then
\begin{eqnarray}
E_{_\mathrm{ltra}}^{^{\mathrm{[TE,TM]}}} = E_{\0}\,e^{i\,k\,z_{\mathrm{ltra}}+i\,\Phi_{_{\mathrm{lgeo}}}(\theta_{\0})}\,\int_{\-infty}^{\+infty}{\hspace{-0.25cm}}\mathrm{d}\theta\,t^{^{\mathrm{[TE,TM]}}}_{_{\mathrm{left}}}(\theta)t^{^{\mathrm{[TE,TM]}}}_{_{\mathrm{lower}}}(\theta)\,g(\theta-\theta_{\0}){\nonumber}
\\
\times \, e^{i\,k\,\left[\tilde{x}_{_{\mathrm{ltra}}}\,(\theta-\theta_{\0}) - \tilde{z}_{_{\mathrm{ltra}}}\,(\theta-\theta_{\0})^{\2}/2\right]},
\end{eqnarray}
with an associated relative power
\begin{equation}
\label{eq:PowerLTRA}
\mathcal{P}_{_{\mathrm{ltra}}}^{^{\mathrm{[TE,TM]}}} = \frac{1}{\phi^{\prime}_{\0}}\,|t_{_{\mathrm{left}}}^{^{\mathrm{[TE,TM]}}}(\theta_{\0})t{_{_\mathrm{lower}}}^{^{\mathrm{[TE,TM]}}}(\theta_{\0})|^{^2},
\end{equation}
the factor $1/\phi^{\prime}_{\0}$ coming from the Dirac's delta in Eq. (\ref{eq:DiracD}), since $\tilde{x}_{_{\mathrm{ltra}}}$ has a $\phi^{\prime}_{\0}$ factor multiplying $x_{_{\mathrm{ltra}}}$, see Eq. (\ref{eq:TILyltra}). The power transmitted through the lower face of the prism is plotted in Figure 7(c). \cor{Note} that for incidence angles greater than the critical angle, no power is transmitted.

With the expressions for the relevant electric fields and their associated powers found, it remains to be discussed how to determine the light's trajectory under the Gaussian beams formalism. The first method considers a tool of asymptotic analysis called \emph{Stationary Phase Method} \cite{SPM2015}, which considers that rapidly varying oscillatory functions in the integral will cancel each other \cor{out}, the stationary condition \cor{making} then the most important contribution to the trajectory. This condition is

\begin{equation}
\left.\frac{\partial \Phi(\Theta)}{\partial\Theta}\right|_{\0}= 0,
\end{equation}
where $\Phi(\Theta)$ is the integrand's oscillatory phase. This method can be thought of as a generalisation of the method employed in Section \ref{sec3}, since a plane wave can be regarded as a Gaussian beam in the limit where $\mathrm{w}_{\0}\rightarrow \infty$. In this case the beam's width encompasses a single incidence angle $\theta=\theta_{\0}$.

Another approach is to consider the mean path of the beam. This is accomplished by evaluating the mean value of the electric field's intensity along the direction perpendicular to its propagation direction. For a beam propagating along the $z$ direction, we have that

\begin{equation}
\langle x^{^{\mathrm{[TE,TM]}}} \rangle = \frac{\displaystyle \int_{\-infty}^{\+infty}{\hspace{-0.25cm}}\mathrm{d}x\,x\,|E^{^{\mathrm{[TE,TM]}}}|^{^2}}{\displaystyle \int_{\-infty}^{\+infty}{\hspace{-0.25cm}}\mathrm{d}x\,|E^{^{\mathrm{[TE,TM]}}}|^{^2}},
\end{equation}
which is akin to \cor{the} mean value calculations in Quantum Mechanics \cite{sakurai}. The particularities of this integration depend on the incidence region, that is, if the beam's center is in the Partial or Total Internal Reflection regime, and it will be carried out individually for each case in the following sections.

Before concluding this Section, however, it is important to \cor{note} that,  \cor{under the Gaussian beams formalism}, the special angles, the Brewster and critical angles, studied in section \ref{sec3} for plane waves become special regions, called Brewster and critical regions. This happens\cor{,} because even though a beam may not be centred at such angles, it may still be centred at an angle close enough for it to be affected by them. Under the paraxial approximation, the region where the Gaussian distribution has an appreciable magnitude is given by
\begin{equation}
\label{eq:GaussianWidth}
\theta_{\0} - \frac{\lambda}{\mathrm{w}_{\0}} \, < \, \theta \,  < \, \theta_{\0} + \frac{\lambda}{\mathrm{w}_{\0}}.
\end{equation}
This allows us to define the special regions in the following manner: incidence angles in the interval
\[\theta_{_{\mathrm{B(ext)}}} - \frac{\lambda}{\mathrm{w}_{\0}} \, < \, \theta_{\0} \,  < \, \theta_{_{\mathrm{B(ext)}}} + \frac{\lambda}{\mathrm{w}_{\0}}\]
are said to be in the external Brewster region, while incidence angles in the interval
\[\theta_{_{\mathrm{B(int)}}} - \frac{\lambda}{\mathrm{w}_{\0}} \, < \, \theta_{\0} \,  < \, \theta_{_{\mathrm{B(int)}}} + \frac{\lambda}{\mathrm{w}_{\0}},\]
are in the internal Brewster region. The critical region is defined by
\[\theta_{_{\mathrm{cri}}} - \frac{\lambda}{\mathrm{w}_{\0}} \, < \, \theta_{\0} \,  < \, \theta_{_{\mathrm{cri}}} + \frac{\lambda}{\mathrm{w}_{\0}},\]
and, after this region, that is, for
\[\theta_{\0} \,  > \, \theta_{_{\mathrm{cri}}} + \frac{\lambda}{\mathrm{w}_{\0}},\]
is the so-called Artmann region, or Artmann zone, since this is the region where Artmann's results for the Goos-H\"anchen shift are valid.

\section{Closed formula of the Goos-H\"anchen shift}
\label{sec5}

\subsection{The Artmann's formula}

Upon total internal reflection of light at an interface between two media, evanescent waves appear in the less dense medium \cite{saleh,born} while interference occurs between the incident and the reflected waves in the denser one. As a result of such an interference, the origin point of the reflected electromagnetic radiation appears to be displaced from the point where the incident wave met the interface. This effect was experimentally verified in 1947 \cite{GH1947} by Hermann Goos and Hilda H\"anchen for TE-polarised light, and \cor{it} has been named in their honour as the Goos-H\"anchen effect. The mathematical description of the phenomenon, however, was only provided one year later by Kurt Artmann, who also presented the analysis \cite{Art1948}, later confirmed by Goos and H\"anchen \cite{GH1949}, for TM-polarised light. \cor{The} core of Artmann's analysis is the relation between the light's path and its phase and the Stationary Phase Method presented in section \ref{sec4}. Artmann  considers  the incident light \cor{to be} a composition of plane waves like the one in Eq. (\ref{eq:Ei}), but without specifying the distribution $g(\theta-\theta_{\0})$. In our system, the stationary condition \cor{applied to} the beam reflected at the lower interface of the prism \cor{gives}

\begin{equation}
\left\{\frac{\mathrm{d}}{\mathrm{d}\varphi}\left[n\,k\left(x_{_{*}}\sin\varphi - z_{_{*}}\cos\varphi + \sqrt{2}\,d\cos\varphi\right) + \Phi^{^{\mathrm{[TE,TM]}}}_{_{\mathrm{GH}}}\right]\right\}_{\0} = 0,
\end{equation}
which is the derivative of the phase given in Eq. (\ref{eq:lowREFPHS}) with the addition of the Goos-H\"{a}nchen phase. This condition gives us the maximum intensity of the beam as moving along the line
\begin{equation}
x_{_{*}} = \tan\varphi\,\left(\sqrt{2}\,d-z_{_{*}}\right) + \delta^{^{\mathrm{[TE,TM]}}}_{_{\mathrm{GH(Art)}}},
\end{equation}
being
\begin{eqnarray}
\label{eq:deltaArt}
\left\{\, \delta_{_{{\mathrm{GH(Art)}}}}^{^{\mathrm{[TE]}}}\,,\,\,\delta_{_{{\mathrm{GH(Art)}}}}^{^{\mathrm{[TM]}}} \,\right\}
& = & -\, \frac{1}{n\,k\,\cos\varphi_{\0}}\,\,\left\{\,\frac{\partial \Phi_{_{{\mathrm{GH}}}}^{^{\mathrm{[TE]}}}  }{\partial\varphi}\,,\,\frac{\partial \Phi_{_{{\mathrm{GH}}}}^{^{\mathrm{[TM]}}}  }{\partial\varphi}
\,\right\}_{\0}
\nonumber \\
 & = &
\frac{2\tan\varphi_{\0}}{k\,\sqrt{n^{\2}\sin^{\2}\varphi_{\0}-1}}
\left\{\,1\,,\,\frac{1}{n^{\2}\sin^{\2}\varphi_{\0}-\cos^{\2}\varphi_{\0}}\,\right\},
\end{eqnarray}
the Artmann's formulae for the Goos-H\"{a}nchen shift of TE- and TM-polarised light, respectively. \cor{Note} that the shift is associated with the reflection coefficient exclusively, \cor{and it is not}
present \cor{for} the  refracted light. Nevertheless, in our system, the displacement is measured only after the transmission through the right face of the prism, which, due to the geometry of the system, is given by
\begin{equation}
d_{_{\mathrm{GH(Art)}}}^{^{[\mathrm{TE,TM}]}} =  \frac{\cos\varphi_{\0}\,\cos \theta_{\0}}{\cos\psi_{\0}}\,\,\delta_{_{\mathrm{GH(Art)}}}^{^{\mathrm{[TE,TM]}}}\,\,,
\end{equation}
see Figure 8(a). This geometrical factor can be obtained directly from the Goos-H\"anchen phase by taking its derivation with respect to $\theta$, since $\theta$, $\varphi$, and $\psi$ are connected through the geometry of the prism, see Eq. (\ref{eq:varphi}). We have then that

\begin{equation}
\label{eq:PhsShft}
d_{_{\mathrm{GH(Art)}}}^{^{[\mathrm{TE,TM}]}} = -\frac{1}{k}\left.\frac{\displaystyle \partial\, \Phi_{_{\mathrm{GH}}}^{^{\mathrm{[TE,TM]}}}}{\partial\theta}\right|_{\0}.
\end{equation}

Two points must be made regarding Artmann's formulae given in Eq. (\ref{eq:deltaArt}). \cor{First: As it is}  proportional to $k^{-\1}$, it is \cor{also} proportional to the wavelength of the light being used, making the displacement inaccessible to naked eyes. The way Goos and H\"anchen dealt with the minute nature of the effect was by employing a structure that allowed multiple internal reflections, as the ones showed in Figures 8(b-c), built from our system. Every time light reflects at an interface it gains a new Goos-H\"anchen phase. Consequently, for $N_{_{\mathrm{r}}}$ total internal reflections the total displacement is simply $N_{_{\mathrm{r}}}\,d_{_{\mathrm{GH(Art)}}}^{^{\mathrm{[TE,TM]}}}$.

The second point \cor{concerns} the validity of Artmann's result. We can see that as the incidence angle approaches the critical angle, $n\sin\varphi_{\0}$ becomes closer to 1 and Eq. (\ref{eq:deltaArt}) diverges, suggesting an infinitely great displacement. Experimental data shows us that the shift is finite around the critical angle \cite{GH1947,GH1949,Bret1992}, as does numerical calculations \cite{DeLeo2013}, see Figure 9. In the next sections we will derive an analytical expression for the Goos-H\"anchen shift valid in the vicinity of the critical angle.

\subsection{Analytical solution to the critical divergence}

The stationary condition gives the main contribution among all the phases composing the beam and so it amounts to an analysis of the trajectory of the beam's maximum intensity. Artmann's formula, however is a limit case. The Stationary Phase Method employed does not take into consideration the structure of the beam and it is only valid while the derivative of the Goos-H\"anchen phase can be evaluated at $\theta_{\0}$ and factorised from the electric field integral, see Figure 9. To overcome the divergence problem we have to analyse the stationary condition under the angular distribution $g(\theta-\theta_{\0})$. Let us do this analysis for the TE-polarisation first. The extension to the TM case, as will be seen later, is straightforward. The integral we have to solve is

\begin{equation}
\label{eq:dgh}
d_{_{\mathrm{GH}}}^{^{\mathrm{[TE]}}} = -\,\frac{1}{k}\,
\frac{\displaystyle \int_{\theta_{_\mathrm{cri}}}^{\mbox{\tiny $+\pi/\2$}}
\hspace{-0.25cm}\mbox{d}\theta\,g(\theta-\theta_{\0})\,\frac{\partial \Phi_{_{\mathrm{GH}}}^{^{\mathrm{[TE]}}}}{\partial \theta\,\,}}
{\displaystyle \int_{\mbox{\tiny $-\pi/\2$}}^{\mbox{\tiny $+\pi/\2$}}
\hspace{-0.25cm}\mbox{d}\theta\,g(\theta-\theta_{\0})}\,.
\end{equation}

\WideFigureSideCaption{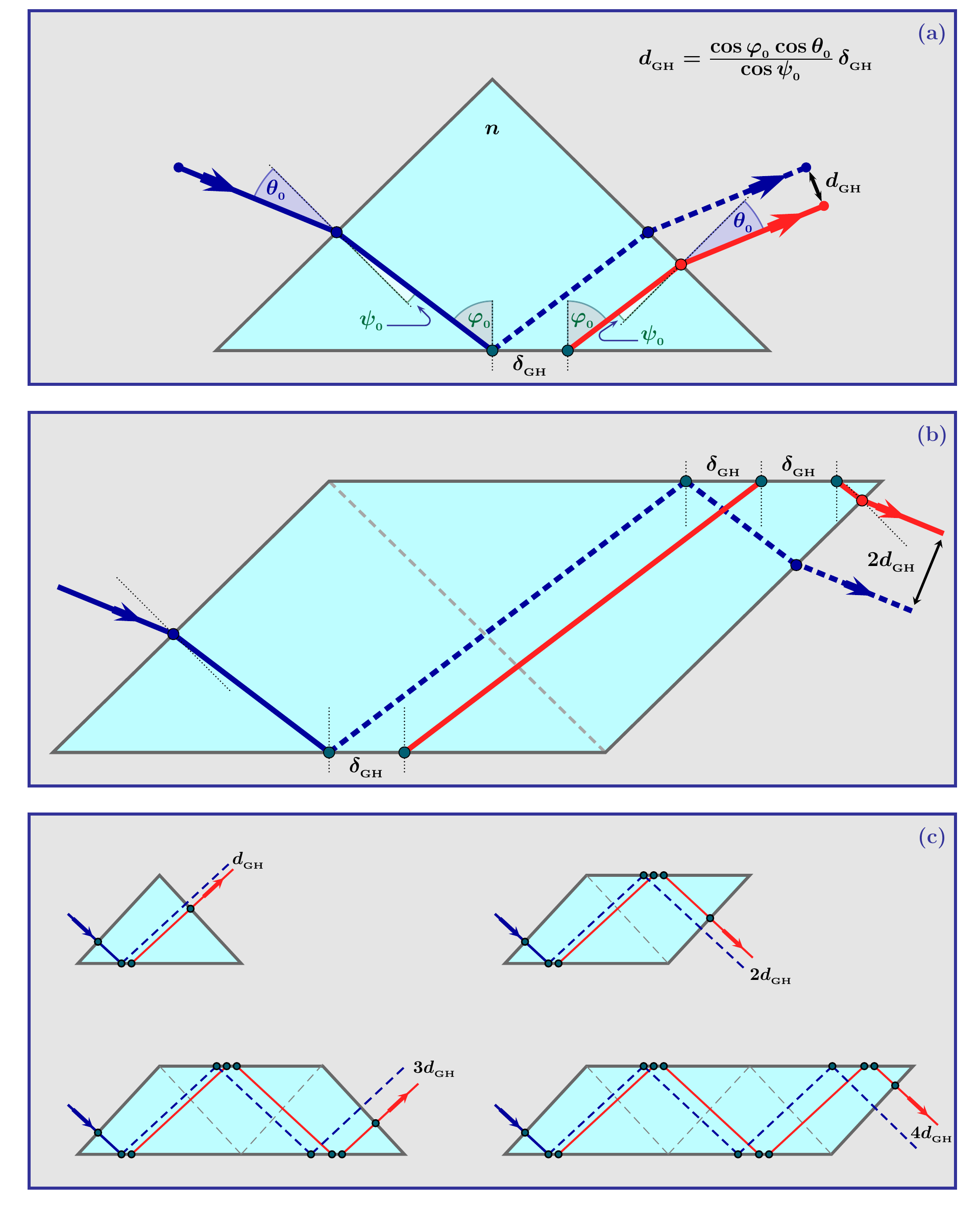}{(a) In the Total Internal Reflection regime the reflection coefficient of the lower interface of the prism acquires an additional phase, which prompts a displacement of the reflected beam, known as the Goos-H\"anchen shift. Here, $\delta_{_{\mathrm{GH}}}$ is the shift occurring at the reflecting interface, but since the light is only collected after it leaves the prism, a geometrical factor must be taken into account, yielding the measured shift $d_{_{\mathrm{GH}}}$. (b) This shift is associated with the reflection coefficient and so, for every total internal reflection inside a dielectric structure, there is an additional shift $\delta_{_{\mathrm{GH}}}$. (c) Using our original structure, a right angle triangular prism, it is possible to build a multiple reflection system, \cor{like} the one originally employed by Goos and H\"anchen.}

\WideFigureSideCaption{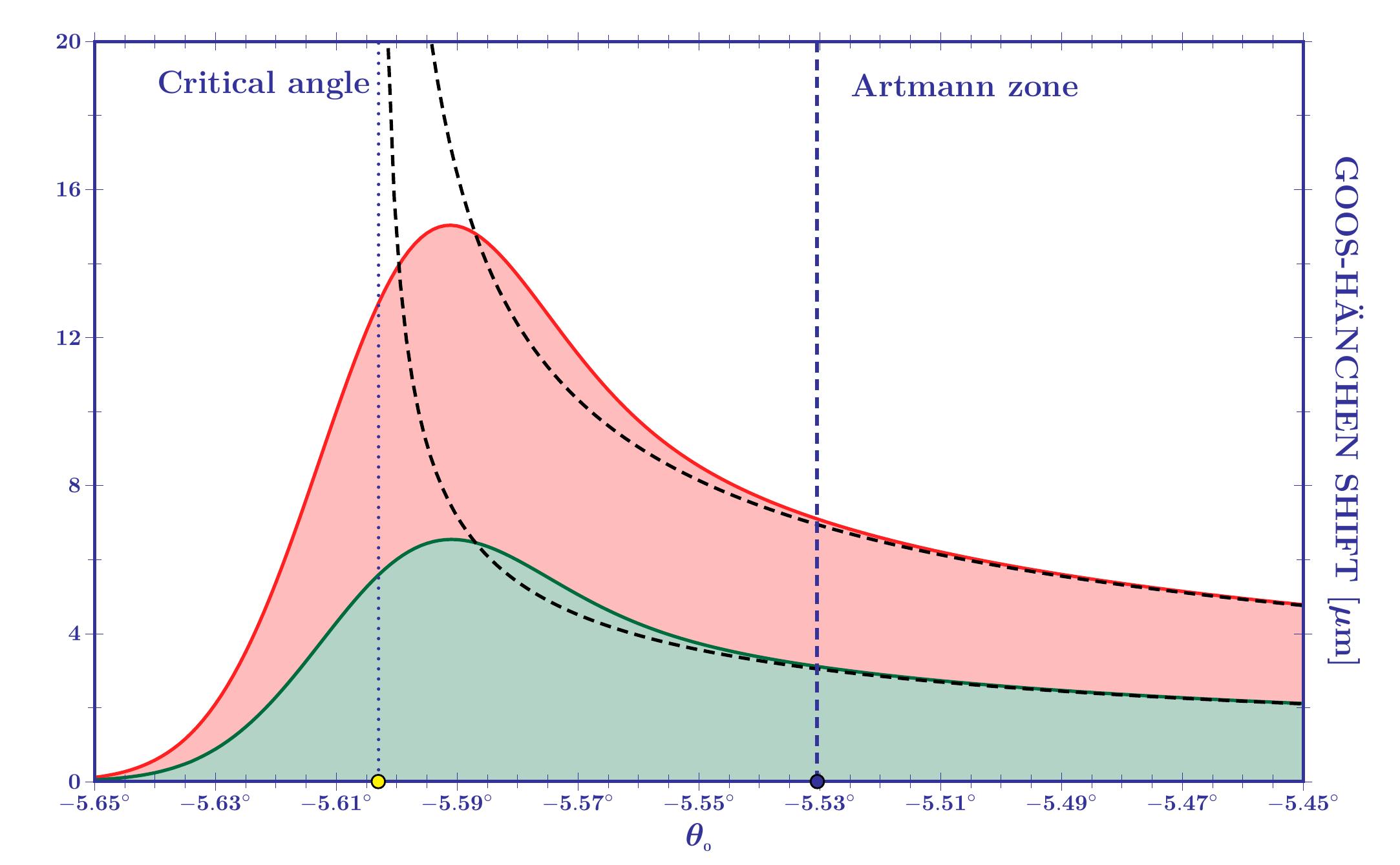}
{The Goos-H\"anchen shift as a function of the incidence angle $\theta_{\0}$ for a borosilicate ($n=1.515$) prism and a laser with $\lambda=0.633\,\mu\mathrm{m}$. The red and green curves are the shifts for the TM- and TE-polarisations, respectively, calculated numerically by evaluating the point of maximum intensity along the direction perpendicular to the propagation direction of the beam transmitted through the right face of a right angle triangular prism. The beams employed have a $\mathrm{w}_{\0}=150\,\mu\mathrm{m}$ minimum waist. The associated dashed black lines are Artmann's analytical curves. Note that as the critical angle ($\theta_{\0} = -5.603^{\circ}$) is approached, these curves go to infinity, contradicting the numerical analysis. In the so-called Artmann zone both results are in agreement. The reason for this is that in this region the Goos-H\"anchen shift is nearly constant and can be factored out of integral (\ref{eq:dgh}) while in the critical region ($\theta_{_\mathrm{cri}}-\mathrm{w}_{\0}/\lambda<\theta_{\0}<\theta_{_\mathrm{cri}}+\mathrm{w}_{\0}/\lambda$) the structure of the beam must be taken into account.}

\noindent
\cor{Note} that the integral in the numerator has its lower integration limit in $\theta_{_{\mathrm{cri}}}$ because before this angle the Goos-H\"anchen phase is null. Besides, its upper limit and both limits of the integral in the denominator can be made to infinity since we are considering the paraxial limit. The divergence of the Goos-H\"anchen shift near the critical angle comes from the term $(n^{\2}\sin^{\2}\varphi-1)^{-\1/\2}$ in the derivative of the Goos-H\"anchen phase, see Eq. (\ref{eq:deltaArt}). So, we can simplify Eq. (\ref{eq:dgh}) to

\begin{equation}
\label{eq:dgh2}
d_{_{\mathrm{GH}}}^{^{\mathrm{[TE]}}} = \frac{2\,\sin\varphi_{\0}\cos\theta_{\0}}{k\,\cos\psi_{\0}}\,
\int_{\theta_{\mathrm{cri}}}^{\+infty}
\hspace{-0.25cm}\mbox{d}\theta\,\frac{g(\theta-\theta_{\0})}{\sqrt{n^{^2}\sin^{\2}\varphi-1}}.
\end{equation}
Expanding the term inside the square root in the denominator around the incident angle $\theta_{\0}$,
\begin{eqnarray}
\label{eq:expansion}
n^{\2}\sin^{\2}\varphi\,-\,1 & \approx & n^{\2}\sin^{\2}\varphi_{\0}\,-1\, +\,   \,\frac{n\,\sin(2\,\varphi_{\0})\cos{\theta_{\0}}}{\cos\psi_{\0}}\,\,\,(\theta -\theta_{\0}) \nonumber\\
 & = &  \frac{n\,\sin(2\,\varphi_{\0})\cos{\theta_{\0}}}{\cos\psi_{\0}}\,\,\left(\,\theta - \theta_{\0} - \sigma_{\0}\,\right),
\end{eqnarray}
with
\begin{equation}
\label{eq:sigma0}
\sigma_{\0} = \frac{\cos\psi_{\0}}{n\,\sin(2\,\varphi_{\0})\cos{\theta_{\0}}}\,\left(1- n^{\2}\sin^{\2}\varphi_{\0} \right),
\end{equation}
we can rewrite Eq. (\ref{eq:dgh2}) as
\begin{eqnarray}
\label{eq:dghfI}
d_{_{\mathrm{GH}}}^{^{\mathrm{[TE]}}} = {\mathrm{w}}_{\0}\,\,\sqrt{\frac{\tan\varphi_{\0}\cos\theta_{\0}}{2\,\,n\,\pi\,\cos\psi_{\0}}}\,\,
  \int_{{\mbox{\tiny $ \theta_{\mbox{\tiny $_0$}}+\sigma_{\mbox{\tiny $_0$}}$}}}^{\+infty}
\hspace*{-0.25cm}\mbox{d}\theta\,\frac{\exp \left[-\,(\,k\,\mbox{w}_{\0}\,)^{^{2}} (\theta -\theta_{\0})^{^2}/\,4\,\right]}{\sqrt{ \theta-\theta_{\0}-\sigma_{\0}}}\,.
\end{eqnarray}
The condition for total internal reflection is, with the expansion (\ref{eq:expansion}), now given by $\theta\geq\theta_{\0}+\sigma_{\0}$. Let us introduce the new integration variable $\rho = k\,{\mathrm{w}}_{\0}\,\left(\,\theta - \theta_{\0} - \sigma_{\0}\right)/2$. The displacement $d^{^{\mathrm{[TE]}}}_{_{\mathrm{GH}}}$ can then be written as
\begin{equation}
d_{_{\mathrm{GH}}}^{^{\mathrm{[TE]}}}
 =  \sqrt{\frac{{\mathrm{w}}_{\0}}{k}\,\,\frac{\tan\varphi_{\0}\cos\theta_{\0}}{n\,\pi\,\cos\psi_{\0}}}\,\, \mathcal{I}(k\,\mathrm{w}_{\0}\,\sigma_{\0}),
\end{equation}
where
\begin{equation}
\mathcal{I}(k\,\mathrm{w}_{\0}\,\sigma_{\0}) = \int_{\0}^{\sinfty}
\hspace{-0.25cm}\mbox{d}\rho\,\exp\left[-\,\left(\rho \,+\,
\frac{k\,{\mathrm{w}}_{\0}\sigma_{\0}}{2}\right)^{^{2}}\right]\,\rho^{-\1/\2}.
\end{equation}
Opening the squared argument of the exponential and expressing the term linear in $\rho$ as a summation we obtain
\begin{eqnarray}
\label{eq:GHIntexp}
\nonumber
\mathcal{I}(k\,\mathrm{w}_{\0}\,\sigma_{\0}) &=& \exp\left[-\,\left(\frac{k\,{\mathrm{w}}_{\0}\sigma_{\0}}{2}\,\right)^{\2}\right]\,\sum_{_{m=0}}^{\sinfty}\frac{\,\,
(-\,k\,{\mathrm{w}}_{\0}\sigma_{\0})^{^{m}}}{m!}\,
\int_{\0}^{\+infty}\hspace*{-0.25cm}\mbox{d}\rho\,\,\,e^{-\rho^{\2}}\,\rho^{m-\1/\2} \nonumber
\\
&=& \frac{1}{2}\,\exp\left[-\,\left(\frac{k\,{\mathrm{w}}_{\0}\sigma_{\0}}{2}\,\right)^{\2}\right] \,\sum_{_{m=0}}^{\sinfty}\frac{( -\,k\,{\mathrm{w}}_{\0}\sigma_{\0}  )^{^{m}}}{m!}\,\Gamma\left[\frac{1\,+\,2\,m}{4}\right].
\end{eqnarray}
By defining then the variable $x = k\mathrm{w}_{\0}\sigma_{\0}/2\sqrt{2}$ we have that
\begin{equation}
\sum_{_{m=0}}^{\sinfty}\frac{( -\,2\,\sqrt{2}\,x)^{^{m}}}{m!}\,\Gamma\left[\frac{1\,+\,2\,m}{4}\right] = 2^{^{1/4}} \pi\,  \, \sqrt{|x|}\,\,\,e^{x^{\2}}\,\left[I_{_{-1/4}}\left(x^{\2}\right) - {\mathrm{sgn}}(x)\, I_{_{1/4}}\left(x^{\2}\right) \right],
\end{equation}
where the functions $I_{_{\alpha}}(x^{\2})$ are modified Bessel functions of the first kind and $\mathrm{sgn}(x)$ is the sign function. Defining then the shift function
\begin{equation}
\mathcal{S}(x) = e^{-x^{\2}}\sqrt{|x|}\,\left[ I_{_{-1/4}}(x^{\2}) - \mathrm{sgn}(x)I_{_{1/4}}(x^{\2}) \right],
\end{equation}
we arrive at the closed-form expression for the Goos-H\"anchen shift as
\begin{equation}
\label{eq:dghMAX}
d_{_{\mathrm{GH}}}^{^{\mathrm{[TE]}}} = \sqrt{\frac{\,\,\pi \,\tan\varphi_{\0} \cos\theta_{\0}}{2\,\sqrt{2}\,\,n\,\cos\psi_{\0}}}\,\,\mathcal{S}\left[\,\frac{k\,{\mathrm{w}}_{\0}\sigma_{\0}}{2\,\sqrt{2}}\,\right]\,\,\sqrt{\frac{{\mathrm{w}}_{\0}}{k}}.
\end{equation}
Let us now check the behaviour of the function at critical incidence. For $\theta_{\0} = \theta_{_\mathrm{cri}}$, Eq. (\ref{eq:sigma0}) tells us that $\sigma_{\0} = 0$. We have then that
\begin{equation}
\lim_{x \rightarrow 0}\mathcal{S}(x) = \frac{\Gamma\left[\frac{1}{4}\right]}{2^{^{1/4}}\pi},
\end{equation}
being the Goos-H\"anchen shift at the critical angle
\begin{eqnarray}
\label{eq:dghcriTE}
d^{^{\mathrm{[TE]}}}_{_{\mathrm{GH}(\mathrm{cri})}} &=& \sqrt{\frac{\,\,\,\tan\varphi_{_{\mathrm{cri}}} \cos\theta_{_{\mathrm{cri}}}}{n\,\pi\,\cos\psi_{_\mathrm{cri}}}}\,\frac{\Gamma\left[\frac{1}{4}\right]}{2}\,\sqrt{\frac{{\mathrm{w}}_{\0}}{k}}{\nonumber}\\
&\approx & \frac{\Gamma\left[\frac{1}{4}\right]}{2\sqrt{n\pi}(n^{\2}-1)^{\1/\4}}\,\sqrt{\frac{{\mathrm{w}}_{\0}}{k}}.
\end{eqnarray}

Finally, it is interesting to show that our formula restores $d^{^{\mathrm{[TE]}}}_{_{\mathrm{GH(Art)}}}$ for incidence angles far from the critical angle. By requesting that $k\mathrm{w}_{\0}\sigma_{\0}<-2\pi$, which amounts to say that we are considering a collimated beam with an incidence angle greater than $\theta_{_{\mathrm{cri}}}$, we can extend the lower limit of the integral (\ref{eq:dgh2}) to $-\infty$ (since the Gaussian distribution $g(\theta-\theta_{\0})$ will be close to zero before the critical angle) and take the limit
\begin{equation}
\lim_{_{x\rightarrow -\infty }}\mathcal{S}(x) = \sqrt{\frac{2}{\pi\,|x|}},
\end{equation}
which will then provide us with $d^{^{\mathrm{[TE]}}}_{_{\mathrm{GH}}(k\mathrm{w}_{\0}\sigma_{\0}<-2\pi)} \rightarrow d^{^{\mathrm{[TE]}}}_{_{\mathrm{GH(Art)}}}$. This analysis also allows us to determine the frontier of the critical region where Artmann's formula becomes valid,
\[\theta_{\0_{\mathrm{(Art)}}} \geq \theta_{_{\mathrm{cri}}} + \frac{2\pi}{k\mathrm{w}_{\0}} = \theta_{_\mathrm{cri}} + \frac{\lambda}{\mathrm{w_{\0}}}, \]
as expected. It is interesting to note that this result does not depend on the relative refractive index between media, but only on beam's parameters.

A remarkable characteristic of numerical and experimental data on the Goos-H\"anchen shift is that, contrary to what might be expected from the divergent result of Eq. (\ref{eq:deltaArt}), the maximum shift is not found at critical incidence, but for an incidence angle slightly greater. The function $\mathcal{S}(x)$ has a maximum at $x_{_{\mathrm{max}}} = -0.38$, so we have that
\[\frac{k\mathrm{w}_{\0}\sigma_{\0}}{2\sqrt{2}} = -0.38.  \]
Now, let us consider an angle $\varphi_{\0} = \varphi_{_{\mathrm{cri}}} + \delta\varphi_{_{\mathrm{max}}}$, which is the critical angle $\varphi_{_{\mathrm{cri}}}$ plus an increment that will leave us at the angle for which the Goos-H\"anchen shift is maximum. For such an angle, we have that $-\sigma_{\0} \approx n\delta\varphi_{_{\mathrm{max}}} = \delta\theta_{_{\mathrm{max}}}.$ By placing this approximation in the equation above we obtain
\[ \frac{k\mathrm{w}_{\0}\,\delta\theta_{_{\mathrm{max}}}}{2\sqrt{2}} = 0.38, \]
which gives us that $\delta\theta_{_{\mathrm{max}}} \approx 1/k\mathrm{w}_{\0}$. So, the incidence angle which returns the maximum shift is, approximately,

\begin{equation}
\theta_{\0} \approx \theta_{_{\mathrm{cri}}} + \frac{1}{k\mathrm{w}_{\0}}.
\end{equation}

The results obtained for the TE polarisation are immediately extended to the TM case:
\begin{equation}
d_{_{\mathrm{GH}}}^{^{\mathrm{[TM]}}} = \frac{d_{_{\mathrm{GH}}}^{^{\mathrm{[TE]}}}}{n^{\2}\sin^{\2}\varphi_{\0}-\cos^{\2}\varphi_{\0}},
\end{equation}

\WideFigureSideCaption{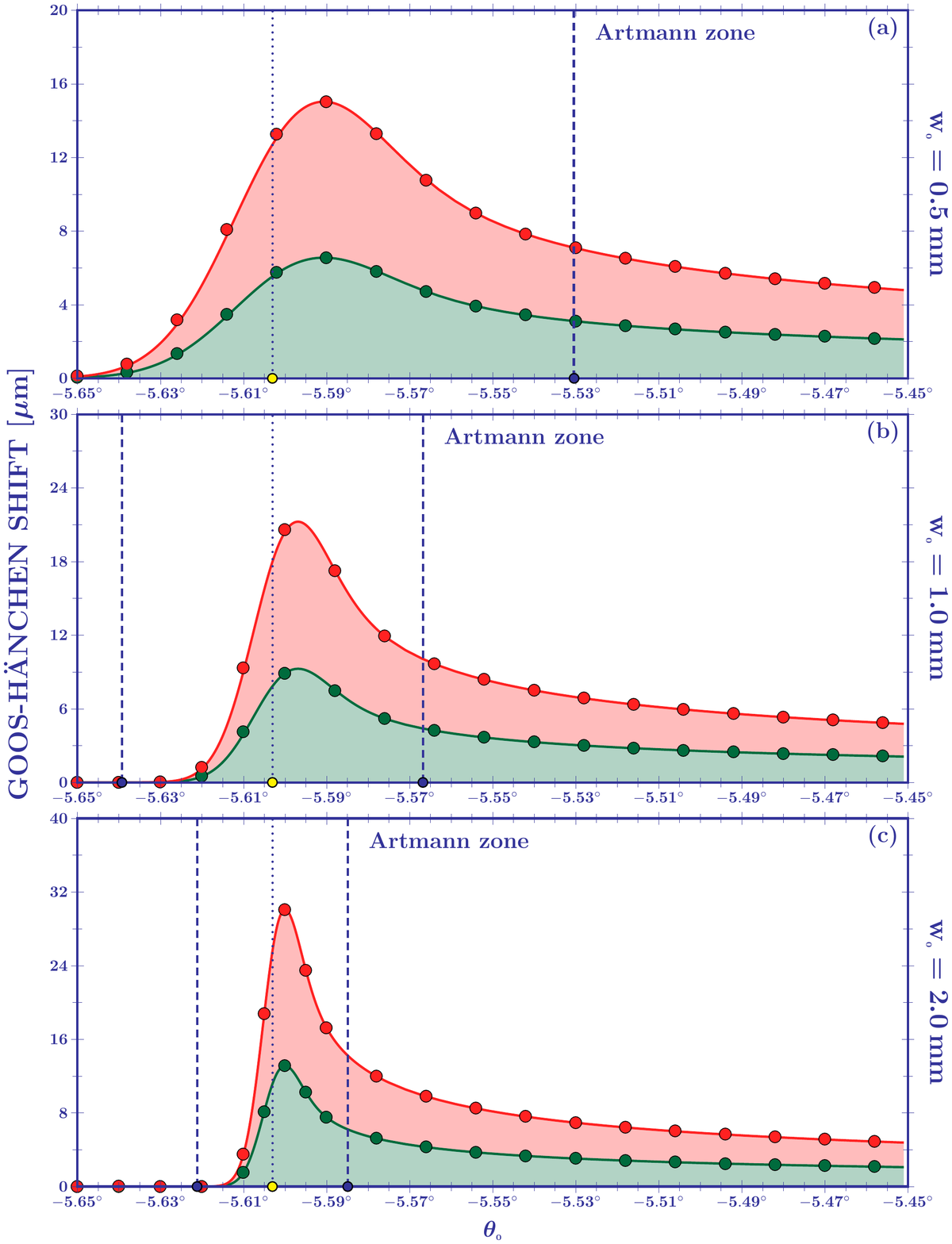}{The analytical curves for Goos-H\"anchen shift of the maximum intensity point of a beam with $\lambda=0.633\,\mu\mathrm{m}$ and (a) $\mathrm{w}_{\0}=0.5\,\mathrm{mm}$, (b) $\mathrm{w}_{\0}=1\,\mathrm{mm}$, and (c) $\mathrm{w}_{\0}=2\,\mathrm{mm}$, as a function of the incidence angle $\theta_{\0}$. The prism considered is made of borosilicate $(n=1.515)$. The red curves denote the TM-polarisation and the green curves the TE-polarisation. The dots show numerical calculations, which are in good agreement with our analytical results.}

\noindent
which returns at the critical angle
\begin{equation}
\label{eq:dghcriTM}
d_{_{\mathrm{GH(cri)}}}^{^{\mathrm{[TM]}}}  = n^{\2}\,\, d_{_{\mathrm{GH(cri)}}}^{^{\mathrm{[TE]}}}.
\end{equation}
The curves for $d_{_{\mathrm{GH}}}^{^{\mathrm{[TE,TM]}}}$ as a function of the incidence angle $\theta_{\0}$ for a laser source with $\lambda=0.633\,\mu$m and different values of $\mathrm{w}_{\0}$ are found in Figure 10.

\subsection{Mean value analysis}

In the last Section we obtained an analytical expression for the Goos-H\"anchen shift valid at and around the critical angle by taking into account the Gaussian structure of the light beam when analysing the stationary condition presented by Artmann \cite{Art1948}. A second possible approach considers the shift of the average intensity of the beam. In the coordinate system parallel to the propagation direction of the outgoing beam, $x_{_{\mathrm{rtra}}}-z_{_{\mathrm{rtra}}}$, after the transmission through the right face of the prism, light travels along the $z_{_{\mathrm{rtra}}}$ direction, and its average intensity is at

\begin{equation}
\label{eq:zmean}
\langle x_{_{\mathrm{rtra}}}^{^{\mathrm{[TE,TM]}}} \rangle = \langle \tilde{x}_{_{\mathrm{rtra}}}^{^{\mathrm{[TE,TM]}}} \rangle - x_{_{\mathrm{rgeo}}} = \frac{\displaystyle \int_{\-infty}^{\+infty}{\hspace{-0.25cm}}\mathrm{d}x_{_{\mathrm{rtra}}}\,x_{_{\mathrm{rtra}}}\,\left|E_{_{\mathrm{rtra}}}^{^{\mathrm{[TE,TM]}}}\right|^{\2} }{\displaystyle  \int_{\-infty}^{\+infty}{\hspace{-0.25cm}}\mathrm{d}x_{_{\mathrm{rtra}}}\,\left|E_{_{\mathrm{rtra}}}^{^{\mathrm{[TE,TM]}}}\right|^{\2}},
\end{equation}
see Eqs. (\ref{eq:ytildertra}) and (\ref{eq:Ertrafinal}). By removing the geometrical shift, the expression above gives us the Goos-H\"anchen shift  $\langle d^{^{\mathrm{[TE,TM]}}}_{_{\mathrm{GH}}}\rangle$ directly. Let us focus firstly on the denominator of Eq. (\ref{eq:zmean}). The electric field has the form presented in Eq. (\ref{eq:Ertrafinal}), but, since the transmission coefficients of the left and right faces are smoothly varying functions of the incidence angle, they can be evaluated at $\theta_{\0}$ and factored out of the integrals. The integral in $x_{_{\mathrm{rtra}}}$ has the same form of Eq. (\ref{eq:DiracD}),
\begin{eqnarray*}
\int_{\-infty}^{\+infty}{\hspace{-0.25cm}}\mathrm{d}x_{_\mathrm{rtra}}\,e^{i\,k(\theta-\tilde{\theta})x_{_{\mathrm{rtra}}}} = \frac{2\pi}{k}\,\delta(\theta-\tilde{\theta}),
\end{eqnarray*}
and we can use it to eliminate the integral in $\tilde{\theta}$ to obtain
\begin{equation}
\int_{\-infty}^{\+infty}{\hspace{-0.25cm}}\mathrm{d}z_{_{\mathrm{rtra}}}\,\left|E_{_{\mathrm{rtra}}}^{^{\mathrm{[TE,TM]}}}\right|^{\2} = \left|t_{_{\mathrm{left}}}^{^{\mathrm{[TE,TM]}}}(\theta_{\0})\,t_{_{\mathrm{right}}}^{^{\mathrm{[TE,TM]}}}(\theta_{\0})\right|^{\2}\,\frac{2\pi}{k}\int_{\-infty}^{\+infty}{\hspace{-0.25cm}}\mathrm{d}\theta\,g^{\2}(\theta-\theta_{\0}).
\end{equation}

Since we are considering the Total Internal Reflection regime under the paraxial approximation, the reflection coefficient of the lower face of the prism only contributes with a complex phase, which is cancelled by its complex conjugate once the Dirac's Delta function is employed. Also, this condition allows us to change the integration limits of the angular integral, $\{-\pi/2, \pi/2\} \rightarrow \{-\infty, \infty\}$, since the Gaussian falls rapidly to zero. In the numerator of Eq. (\ref{eq:zmean}) the $x_{_{\mathrm{rtra}}}$ integral has the form
\begin{equation}
\label{eq:DiracDerivativeInt}
\int_{\-infty}^{\+infty}{\hspace{-0.25cm}}\mathrm{d}x_{_{\mathrm{rtra}}} \, x_{_{\mathrm{rtra}}}\,e^{i\,k(\theta-\tilde{\theta})x_{_{\mathrm{rtra}}}} = \frac{\pi}{i\,k^{\2}}\,\left[ \partial_{_{\theta}}\,\delta(\theta-\tilde{\theta}) - \partial_{_{\tilde{\theta}}}\,\delta(\theta-\tilde{\theta})\right].
\end{equation}
Integration by parts gives us then
\begin{eqnarray}
\label{eq:Hc}
\int_{\-infty}^{\+infty}{\hspace{-0.25cm}}\mathrm{d}x_{_{\mathrm{rtra}}}\,x_{_{\mathrm{rtra}}}\,\left|E_{_{\mathrm{rtra}}}^{^{\mathrm{[TE,TM]}}}\right|^{\2} & = & \left|t_{_{\mathrm{left}}}^{^{\mathrm{[TE,TM]}}}(\theta_{\0})\,t_{_{\mathrm{right}}}^{^{\mathrm{[TE,TM]}}}(\theta_{\0})\right|^{\2}\frac{\pi}{i\,k^{\2}}{\nonumber}
\\
& \times &\int_{\-infty}^{\+infty}{\hspace{-0.25cm}}\mathrm{d}\theta\,\begin{array}{ll}
g(\theta-\theta_{\0})e^{-i\,k\,z_{_{\mathrm{rtra}}}(\theta-\theta_{\0})^{\2}/\2+i\Phi_{_{\mathrm{GH}}}^{^{\mathrm{[TE,TM]}}}}
\\
\times\partial_{_{\theta}}\left[ g(\theta-\theta_{\0})e^{-i\,k\,z_{_{\mathrm{rtra}}}(\theta-\theta_{\0})^{\2}/\2+i\Phi_{_{\mathrm{GH}}}^{^{\mathrm{[TE,TM]}}}} \right]^{\ast}
\end{array}{\nonumber}
\\
&+&\mathrm{H.c.},
\end{eqnarray}
where H.c. stands for ``Hermitian conjugate''. \cor{Note} that, of the terms between brackets that are being derived, only complex terms will survive due to the summation with the Hermitian conjugate. Besides, the derivation of $e^{-i\,k\,z_{_{\mathrm{rtra}}}(\theta-\theta_{\0})^{\2}/\2}$ will give origin to an integration with an odd integrand and symmetrical integration limits, which returns zero. Finally, the derivation of $e^{i\Phi_{_{\mathrm{GH}}}^{^{\mathrm{[TE,TM]}}}}$ originates a term that is null before $\theta_{_{\mathrm{cri}}}$, and so, the mean Goos-H\"anchen shift is

\begin{equation}
\label{eq:dInt}
\langle d^{^{\mathrm{[TE,TM]}}}_{_{\mathrm{GH}}}\rangle = -\frac{1}{k}\frac{\displaystyle \int_{\theta_{_{\mathrm{cri}}}}^{\+infty}{\hspace{-0.25cm}}\mathrm{d}\theta\,g^{\2}(\theta-\theta_{\0})\frac{\partial\Phi_{_{\mathrm{GH}}}^{^{\mathrm{[TE,TM]}}}}{\partial\theta}} {\displaystyle \int_{\-infty}^{\+infty}{\hspace{-0.25cm}}\mathrm{d}\theta\,g^{\2}(\theta-\theta_{\0})}.
\end{equation}
We can see from the expression above that the difference between the maximum and the mean value calculations is the distribution $g(\theta-\theta_{\0})$, which is squared in the mean calculation case. For the TE-polarisation, carrying out the Goos-H\"anchen phase derivation and following the expansion (\ref{eq:expansion}), we obtain
\begin{eqnarray}
\langle d^{^{\mathrm{[TE]}}}_{_{\mathrm{GH}}}\rangle = \sqrt{2}\,\mathrm{w}_{\0}\sqrt{\frac{\tan\varphi_{\0}\cos\theta_{\0}}{2\,n\,\pi\,\cos\psi_{\0}}}\,\int_{{\mbox{\tiny $ \theta_{\mbox{\tiny $_0$}}+\sigma_{\mbox{\tiny $_0$}}$}}}^{\+infty}{\hspace{-0.25cm}}
\mbox{d}\theta\,\frac{\exp\left[\, -(k\mathrm{w}_{\0})^{\2}(\theta-\theta_{\0})^{\2}/\,2 \,\right]}{\sqrt{ \theta-\theta_{\0}-\sigma_{\0}}},
\end{eqnarray}
which is Eq. (\ref{eq:dghfI}) with $\mathrm{w}_{\0}\rightarrow \sqrt{2}\,\mathrm{w}_{\0}$. Consequently, we can adapt the result (\ref{eq:dghMAX}) to the mean value analysis:
\begin{equation}
\langle d_{_{\mathrm{GH}}}^{^{\mathrm{[TE]}}} \rangle= \sqrt{\frac{\,\,\pi \,\tan\varphi_{\0} \cos\theta_{\0}}{2\,\,n\,\cos\psi_{\0}}}\,\,\mathcal{S}\left[\,\frac{k\,{\mathrm{w}}_{\0}\sigma_{\0}}{2}\,\right]\,\,\sqrt{\frac{\mathrm{w}_{\0}}{k}}.
\end{equation}
Both results are related by
\begin{equation}
\langle\, d_{_{\mathrm{GH}}}^{^{\mathrm{[TE]}}} \rangle  = \, 2^{^{1/4}}\,\frac{\mathcal{S}\left(\,k\,{\mathrm{w}}_{\0}\sigma_{\0}/2\,\right)}{\mathcal{S}\left(\,k\,{\mathrm{w}}_{\0}\sigma_{\0}/2\,\sqrt{2}\,\right)}\,\, d_{_{\mathrm{GH}}}^{^{\mathrm{[TE]}}},
\end{equation}
and, at the critical angle,
\begin{equation}
\langle\, d_{_{\mathrm{GH(cri)}}}^{^{\mathrm{[TE]}}} \rangle  =  2^{^{1/4}}\,d_{_{\mathrm{GH(cri)}}}^{^{\mathrm{[TE]}}}.
\end{equation}
As before, the relation between TE and TM polarisations is

\begin{equation}
\langle d_{_{\mathrm{GH}}}^{^{\mathrm{[TM]}}} \rangle  =
\frac{\langle d_{_{\mathrm{GH}}}^{^{\mathrm{[TE]}}}\rangle}{n^{\2}\sin^{\2}\varphi_{\0}-\cos^{\2}\varphi_{\0}},
\end{equation}
and, for $k\mathrm{w}_{\0}\sigma_{\0}<-2\pi$, Artmann's result is reconstructed, that is, $\langle d^{^{\mathrm{[TE,TM]}}}_{_{\mathrm{GH}}(k\mathrm{w}_{\0}\sigma_{\0}<-2\pi)} \rangle \rightarrow d^{^{\mathrm{[TE,TM]}}}_{_{\mathrm{GH(Art)}}}$. This is an interesting result. The equivalence of the maximum and mean value analysis in the Artmann zone is due to the fact that away from the critical region the maximum intensity is at the center of the beam, while near the critical angle symmetry breaking effects occur. The nature of such effects will be discussed later. The curves for $\langle d_{_{\mathrm{GH}}}^{^{\mathrm{[TE,TM]}}} \rangle$ are plotted in Figure 11 against the incident angle $\theta_{\0}$, for the same set of parameters as Figure 10.

\WideFigureSideCaption{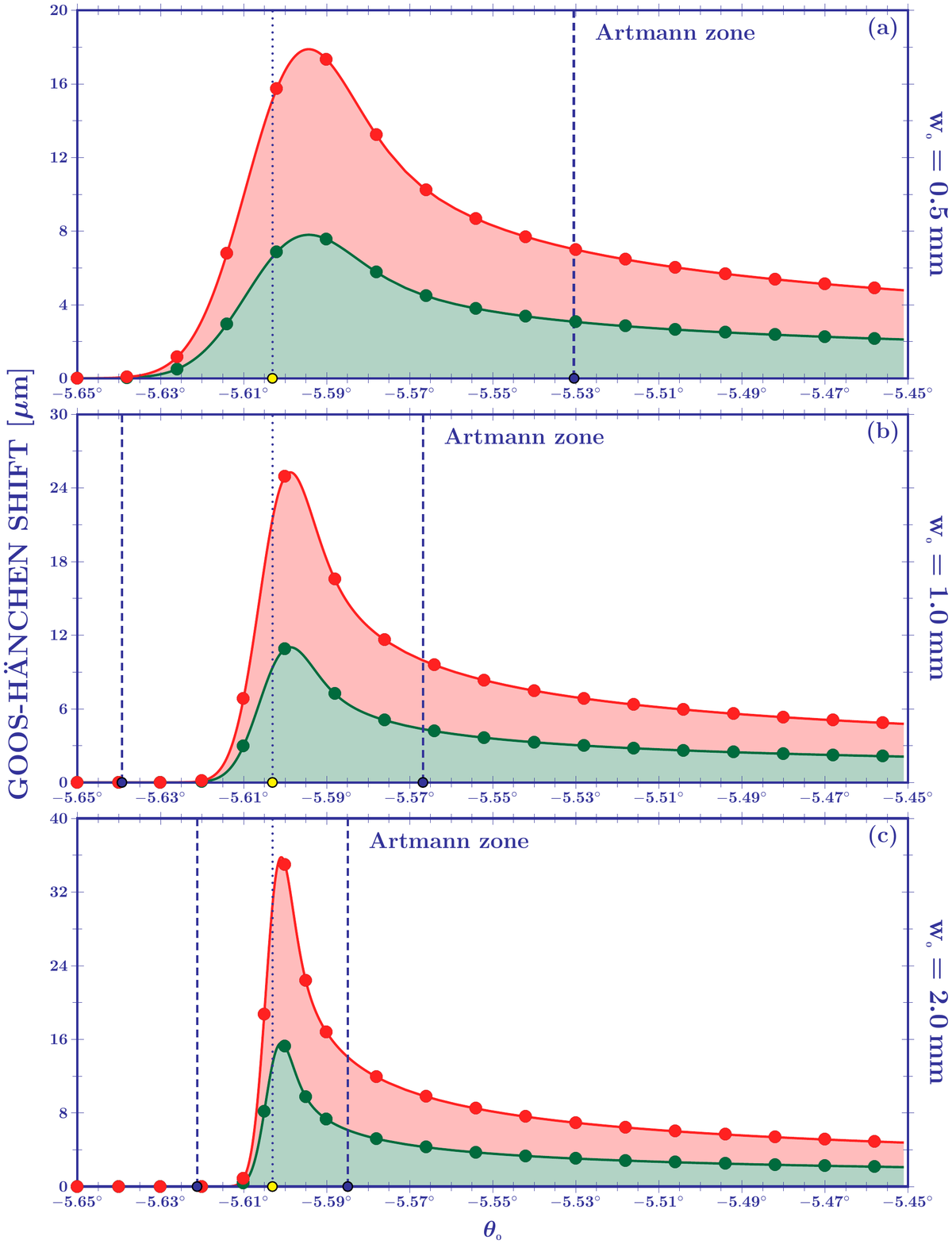}{The analytical curves for Goos-H\"anchen shift of the average intensity of a beam with $\lambda=0.633\,\mu\mathrm{m}$ and (a) $\mathrm{w}_{\0}=0.5\,\mathrm{mm}$, (b) $\mathrm{w}_{\0}=1\,\mathrm{mm}$, and (c) $\mathrm{w}_{\0}=2\,\mathrm{mm}$, as a function of the incidence angle $\theta_{\0}$. The prism considered is made of borosilicate $(n=1.515)$. The red curves denote the TM-polarisation and the green curves the TE-polarisation. The dots show numerical calculations, which are in good agreement with our analytical results.}

Another solution to this problem was proposed in 1970 by Horowitz and Tamir \cite{HoTa1970}. In their approach they used an approximated form of the Fresnel's reflection coefficient in the critical region that enabled them to calculate the integral of the reflected electric field, extracting thus, the information about its trajectory directly from its phase. Their formula reads
\begin{equation}
\delta_{_{\mathrm{GH(HoTa)}}}^{^{\mathrm{[TE,TM]}}} \approx \frac{A^{^{\mathrm{[TE,TM]}}}_{\0}}{2^{\5/\4}\cos\varphi_{\0}}\,\mathrm{Re}\left[ e^{i\pi/\4}D_{-\1/\2}(\gamma_{\0}) \right]e^{\gamma^{\2}/4}\sqrt{\frac{\mathrm{w}_{\0}}{nk}},
\end{equation}
with
\begin{equation}
\gamma_{\0} = i\,n\,k\,\mathrm{w}_{\0}\frac{\sin(\varphi_{\0}-\varphi_{_{\mathrm{cri}}})}{\sqrt{2}\cos\varphi_{\0}}
\end{equation}
and
\begin{equation}
\left\{ A^{^{\mathrm{[TE]}}}_{\0}, A^{^{\mathrm{[TM]}}}_{\0} \right\} = \frac{4\sin\varphi_{\0}}{\sqrt{(\sin\varphi_{\0}+\sin\varphi_{_{\mathrm{cri}}})\cos\varphi_{_{\mathrm{cri}}}}}\times \left\{ 1, \frac{n^{\2}\cos^{\2}\varphi_{_{\mathrm{cri}}}}{cos^{\2}\varphi_{\0}+n^{\4}(\sin^{\2}\varphi_{\0}-\sin^{\2}\varphi_{_{\mathrm{cri}}})} \right\},
\end{equation}
and where $D_{_{\alpha}}(\gamma_{\0})$ is the parabolic-cylinder (Weber) function. At the critical angle, $\gamma_{_{\mathrm{cri}}}=0$ and we have that
\begin{equation}
D_{_{-\1/\2}}(0) = \frac{\Gamma\left[\frac{1}{4}\right]}{2^{\3/\4}\sqrt{\pi}},
\end{equation}
and
\begin{equation}
\left\{ A^{^{\mathrm{[TE]}}}_{\0},A^{^{\mathrm{[TM]}}}_{\0} \right\} = \frac{2\sqrt{2}}{(n^{\2}-1)^{\1/\4}}\,\{ 1, n^{\2}\},
\end{equation}
and the Horowitz-Tamir formula, modified by our geometry factor, becomes
\begin{eqnarray}
\{ d^{^{\mathrm{[TE]}}}_{_{\mathrm{GH(HoTa\bullet cri)}}},d^{^{\mathrm{[TM]}}}_{_{\mathrm{GH(HoTa\bullet cri)}}} \} &=& \frac{\cos\theta_{_{\mathrm{cri}}}\cos\varphi_{_{\mathrm{cri}}}}{\cos\psi_{_{\mathrm{cri}}}}\,\{\delta_{_{\mathrm{GH(HoTa\bullet cri)}}}^{^{\mathrm{[TE]}}},\delta_{_{\mathrm{GH(HoTa\bullet cri)}}}^{^{\mathrm{[TM]}}} \}{\nonumber}
\\
&=&\frac{\Gamma\left[ \frac{1}{4} \right]}{2\sqrt{n\pi}(n^{\2}-1)^{\1/\4}}\,\sqrt{\frac{\mathrm{w}_{\0}}{k}}\,\{1,n^{\2}\},
\end{eqnarray}
displaying a perfect agreement with our results in Eqs. (\ref{eq:dghcriTE}) and (\ref{eq:dghcriTM}). The Horowitz-Tamir formula has met a relative success in comparison to experimental data, but it is important to \cor{note} that, not only it presents a cusplike structure near the critical angle for certain choices of parameters \cite{Cow1977}, \cor{as well} as its validity in the Artmann region is questionable due to mathematical inconsistencies in its derivation \cite{Lai1986}. Their result for incidence at the critical angle, however, is sound and in good agreement with experiments, which is why the comparison between our results and Horowitz and Tamir's was limited to this particular case. It is important to \cor{remember} that neither one of the three formulae presented in this section takes into consideration axial corrections. This means that the measurements have to be conducted as near as possible to the prism's face. The influence of such corrections may be the responsible for the small discrepancies found between experimental data and analytical formulae in \cite{Lai1986}.

In Figure 12, we present the comparison between the maximum and average intensity analyses. As stated before, both approaches agree with each other in the Artmann zone because in this zone the whole beam is being totally internally reflected and the mean and the maximum intensity points are coincident. In the critical region part of the beam is in partial reflection and such points become discordant, hence the shifts they undergo become different. In the next Section, the nature of this symmetry breaking effect will become clearer, since it is directly responsible for angular deviation phenomena.

\WideFigureSideCaption{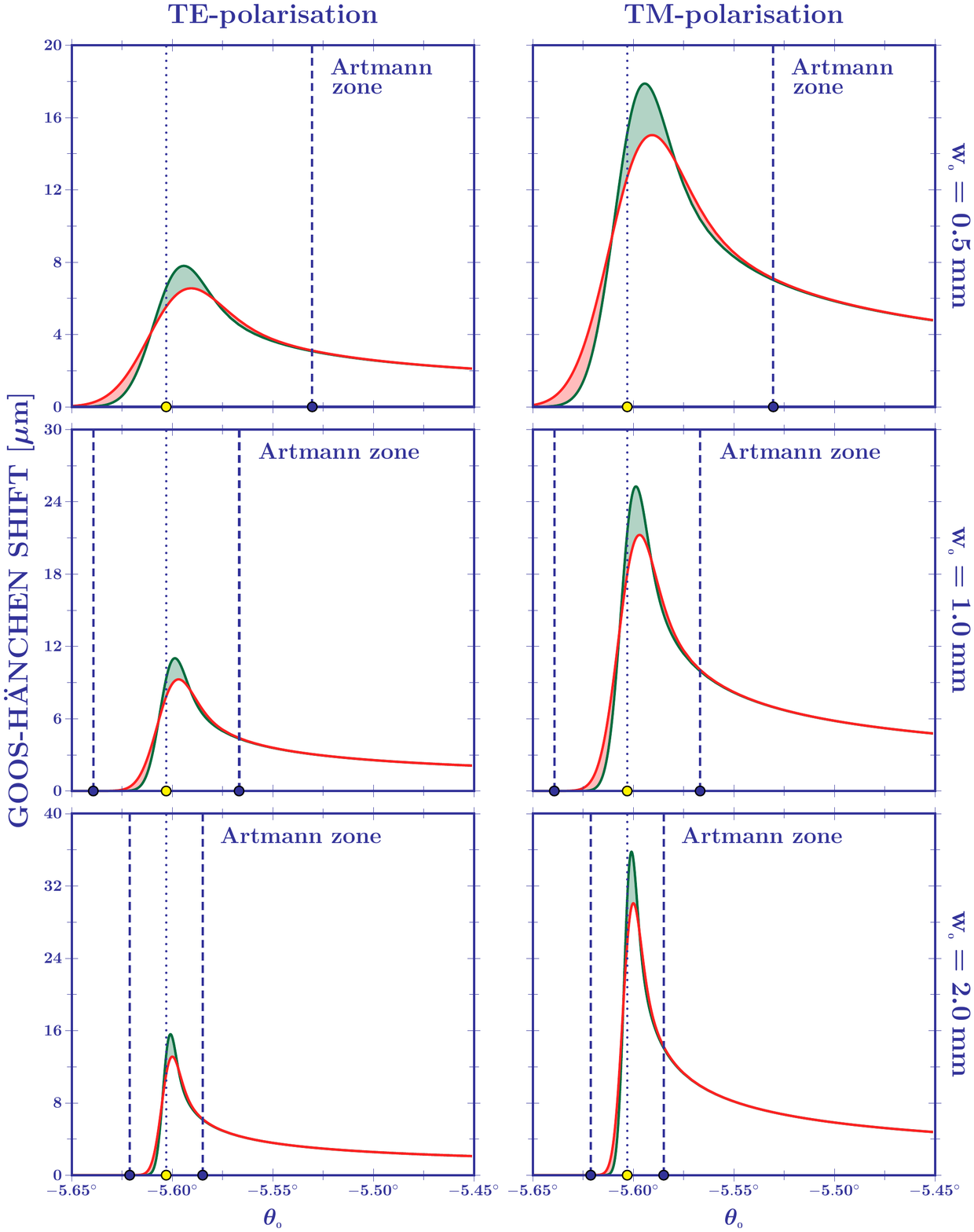}{Comparison between the Goos-H\"anchen shift of the maximum intensity point (curves in red) and of the average intensity point (curves in green) for a beam with $\lambda=0.633\,\mu\mathrm{m}$ and $\mathrm{w}_{\0}=0.5\,\mathrm{mm}$ (top row), $\mathrm{w}_{\0}=1\,\mathrm{mm}$ (middle row), and $\mathrm{w}_{\0}=2\,\mathrm{mm}$ (lower row), as a function of the incidence angle $\theta_{\0}$. The prism considered is made of borosilicate $(n=1.515)$. The left column displays the shift for the TE-polarisation and the right column for the TM. We can see that in the Artmann zone both shifts are in agreement. This is because in this zone the whole beam is totally internally reflected and the maximum and mean intensity points are very close. In the critical region the beam's symmetry is broken (part of it is under the Partial Reflection regime) and such points become discordant, yielding a different shift for each case.}

\section{Angular deviations}
\label{sec6}

As we saw in Section \ref{sec4}, the interaction of plane waves with dielectric interfaces is described by the Fresnel's coefficients, which are a function of the incidence angle. Plane waves have a well-defined wave vector, and, consequently, possess a unique, well-defined direction of incidence. For Gaussian beams, however, this is not the case. Beams have a finite angular distribution, centred around what is defined as their incidence angle ($\theta_{\0}$, in our notation), which generates a range of incoming wave vectors. The effect of this range is a symmetry breaking of the beam's structure induced by the Fresnel's coefficients. This can be thought of in the following terms: to every wave vector in the beam\cor{,} there is an associated set of Fresnel's coefficients. If all coefficients contribute with the same weight there is no preferred direction of transmission nor of reflection, and the laws of Geometrical Optics still hold. However, if the coefficients differ from angle to angle they will filter the beam, favouring the transmission or reflection of some waves and not of others. This effect is known as Fresnel Filtering \cite{TuSt2002} and is the same mechanism behind what is known as the angular Goos-H\"anchen shift \cite{GoShiHen2013}. Since the difference between both phenomena is essentially their name, we will not choose one of them, but rather refer to all symmetry breaking induced angular deviations from Geometrical Optics simply as \emph{angular deviations}.

In the angular interval where the Gaussian distribution is significant, see Eq. (\ref{eq:GaussianWidth}), the Fresnel's coefficients are, almost everywhere, a smooth function of $\theta$, and the structure of the beam is not altered appreciably, rendering small angular deviations, proportional to $1/(k\,\mathrm{w}_{\0})^{\2}$. As we will see, however, near the Brewster angles and near the critical angle the Fresnel's coefficients change more abruptly, increasing the magnitude of the angular deviations.

\subsection{Mean value calculation of the angular coefficient}

In Section \ref{sec4}, we saw that the mean path of a beam is calculated from a mean value integral of its transversal component, modulated by its intensity. For a free Gaussian beam, whose electric field is given by Eq. (\ref{eq:IncF}), we have that

\begin{equation}
\langle x_{_{\mathrm{inc}}} \rangle =  \frac{\displaystyle \int_{\-infty}^{\+infty}{\hspace{-0.25cm}} \mathrm{d} x_{_{\mathrm{inc}}}\, x_{_{\mathrm{inc}}} \, |E_{_{\mathrm{inc}}}|^{\2}}{\displaystyle \int_{\-infty}^{\+infty}{\hspace{-0.25cm}} \mathrm{d}x_{_{\mathrm{inc}}}\, |E_{_{\mathrm{inc}}}|^{\2}}\,.
\end{equation}
Using Eq. (\ref{eq:DiracD}), the denominator of the above expression is a straightforward calculation, and, with the aid of Eq. (\ref{eq:DiracDerivativeInt}), the numerator returns
\begin{eqnarray}
\int_{\-infty}^{\+infty}{\hspace{-0.25cm}}\mathrm{d}x_{_{\mathrm{inc}}}\,x_{_{\mathrm{inc}}}\,\left|E_{_{\mathrm{inc}}}\right|^{\2} & = & \frac{\pi}{i\,k^{\2}} \int_{\-infty}^{\+infty}{\hspace{-0.25cm}}\mathrm{d}\theta\,\begin{array}{ll}
g(\theta-\theta_{\0})e^{-i\,k\,z_{_{\mathrm{inc}}}(\theta-\theta_{\0})^{\2}/\2}
\\
\times\partial_{_{\theta}}\left[ g(\theta-\theta_{\0})e^{-i\,k\,z_{_{\mathrm{inc}}}(\theta-\theta_{\0})^{\2}/\2} \right]^{\ast}
\end{array}{\nonumber}
\\
&+&\mathrm{H.c.}\,.
\end{eqnarray}
The equation above can be evaluated through an integration by parts, transforming the spatial integral into its angular counterpart, see Eqs. (\ref{eq:Hc}) and (\ref{eq:dInt}). It can be written then as
\begin{eqnarray}
\langle\, x_{_{\mathrm{inc}}} \rangle & = & \frac{\displaystyle{
  \int_{\-infty}^{\+infty}{\hspace{-0.25cm}}\mathrm{d}\theta \, (\theta-\theta_{_{0}})\, g^{\2}(\theta-\theta_{_{0}}) }}{\displaystyle{\int_{\-infty}^{\+infty}{\hspace{-0.25cm}} {\mathrm{d}}
\theta\,g^{\2}(\theta-\theta_{_{0}}) }}\,\,\,z_{_{\mathrm{inc}}}\,.
\end{eqnarray}
The symmetry of the Gaussian angular distribution above implies a null value for the integral in the numerator. This means that the beam propagates along the $z_{_{\mathrm{inc}}}$ axis, i.e.
\begin{equation}
\label{eq:yINC}
\langle x_{_{\mathrm{inc}}} \rangle = 0\,.
\end{equation}
This result is expected, since the incident free beam has not yet interacted with anything, which makes its propagation direction a simple matter of coordinate system definition. It helps us \cor{to} illustrate, though, as will become clearer in the following sections, the importance of the beam's symmetry to the determination of its path.

\subsection{The external reflection}

Following the same steps as for the incident beam, the externally reflected beam, that is, the beam reflected by the left face of the prism, with an electric field given by Eq. (\ref{eq:ElrefFinal}), has its path \cor{determined}  by
\begin{eqnarray}
\label{meanref}
\langle\, x_{_{\mathrm{lref}}}^{^{\mathrm{[TE,TM]}}} \rangle & = &\frac{\displaystyle{
  \int_{\-infty}^{\+infty}{\hspace{-0.25cm}} \mathrm{d}
\theta\,\, (\theta-\theta_{_{0}})\, \left[\,g(\theta-\theta_{_{0}})\,r_{_{\mathrm{left}}}^{^{{\mathrm{[TE,TM]}}}}(\theta)\,\right]^{^{2}}}}{\displaystyle{\int_{\-infty}^{\+infty}{\hspace{-0.25cm}} \mathrm{d}
\theta\,\,\left[\, g(\theta-\theta_{_{0}})\,r_{_{\mathrm{left}}}^{^{{\mathrm{[TE,TM]}}}}(\theta)\,\right]^{^{2}} }}\,\,\,z_{_{\mathrm{lref}}}.
\end{eqnarray}
\cor{Note} that the symmetry of the incident beam's angular distribution is broken by the reflection coefficient of the left face of the prism, and so, the reflected beam's path is not parallel to $z_{_{\mathrm{lref}}}$. To obtain an analytical solution, let us develop the reflection coefficient up to the second order around the incidence angle $\theta_{\0}$. We have that
\[
\left[\,\frac{r_{_{\mathrm{left}}}^{^{{\mathrm{[TE,TM]}}}}(\theta)}{r_{_{\mathrm{left}}}^{^{{\mathrm{[TE,TM]}}}}(\theta_{\0})}\,\right]^{^{2}}
 =
 1\, +\,\,2\,\, \frac{r_{_{\mathrm{left}}}^{^{^{{\mathrm{[TE,TM]}}^\prime}}}(\theta_{\0})}{r_{_{\mathrm{left}}}^{^{^{\mathrm{[TE,TM]}}}}(\theta_{\0})}\,\,(\theta-\theta_{\0})\, \,+\]
 \[
  \left\{
  \,\left[\, \frac{r_{_{\mathrm{left}}}^{^{^{{\mathrm{[TE,TM]}}^\prime}}}(\theta_{\0})}{r_{_{\mathrm{left}}}^{^{^{\mathrm{[TE,TM]}}}}(\theta_{\0})}\,\right]^{^{2}} \,+\,\,
\frac{r_{_{\mathrm{left}}}^{^{^{{\mathrm{[TE,TM]}}^{\prime\prime}}}}(\theta_{\0})}{r_{_{\mathrm{left}}}^{^{^{\mathrm{[TE,TM]}}}}(\theta_{\0})}\,\right\}\,\,(\theta-\theta_{\0})^{^{2}}.
\]
Due to the symmetry of the integrands in the integrals of Eq. (\ref{meanref}), only the first order term of the above expansion contributes to the numerator, while the zeroth and second order terms contribute to the denominator. Regarding the second order term, we have that the squared fraction between brackets \cor{results}  much greater than the term with a second derivative, and we can neglect the effect of the \cor{latter}, writing then the expansion as
\begin{equation}
 \displaystyle{  \left[\,\frac{r_{_{\mathrm{left}}}^{^{{\mathrm{[TE,TM]}}}}(\theta)}{r_{_{\mathrm{left}}}^{^{{\mathrm{[TE,TM]}}}}(\theta_{\0})}\,\right]^{^{2}}}
 =
 1\, +\,\,2\,\,D_{_{\mathrm{lref}}}^{^{{\mathrm{[TE,TM]}}}}(\theta_{\0})\,\,(\theta-\theta_{\0})\, +
 \left[\,D_{_{\mathrm{lref}}}^{^{{\mathrm{[TE,TM]}}}}(\theta_{\0})\,\right]^{^{2}} \,(\theta-\theta_{\0})^{^{2}},
\end{equation}
where
\begin{eqnarray}
D_{_{\mathrm{lref}}}^{^{{\mathrm{[TE,TM]}}}}(\theta_{\0}) &=& \frac{\displaystyle r_{_{\mathrm{left}}}^{^{^{{\mathrm{[TE,TM]}}^\prime}}}(\theta_{\0})}{\displaystyle r_{_{\mathrm{left}}}^{^{^{\mathrm{[TE,TM]}}}}(\theta_{\0})}{\nonumber}\\
&=&\frac{2\,\sin\theta_{\0}}{n\,\cos\psi_{\0}}\,\left\{\, 1 \,,\, \frac{n^{\2}}{\sin^{^2}\theta_{\0} - n^{^{2}}\cos^{^{2}}\theta_{\0}}  \,\right\}\,\,.
\end{eqnarray}
Eq. (\ref{meanref}) can now be analytically integrated, giving
\begin{equation}
\label{adr}
\langle\, x_{_{\mathrm{lref}}}^{^{\mathrm{[TE,TM]}}} \rangle  =  \alpha_{_{\mathrm{lref}}}^{^{\mathrm{[TE,TM]}}}(\theta_{\0})\,\,\,z_{_{\mathrm{lref}}},
\end{equation}
where
\begin{equation}
\label{eq:alphalref}
 \alpha_{_{\mathrm{lref}}}^{^{\mathrm{[TE,TM]}}}(\theta_{\0})= \displaystyle{   \frac{2\, D_{_{\mathrm{lref}}}^{^{{\mathrm{[TE,TM]}}}}(\theta_{\0})}{(k\,{\mathrm{w}}_{\0})^{^{^2}}\,+\, \left[\,D_{_{\mathrm{lref}}}^{^{{\mathrm{[TE,TM]}}}}(\theta_{\0})\,\right]^{^{2}}}}\,\,\,z_{_{\mathrm{lref}}}\,\,.
 \end{equation}

For incidence at the Brewster angles, \cor{noting} that
\[  \left\{\,
D_{_{\mathrm{lref}}}^{^{{\mathrm{[TE]}}}}(\theta_{_{\mathrm{B(ext)}}}^{^{\pm}})
\,,\,D_{_{\mathrm{lref}}}^{^{{\mathrm{[TM]}}}}(\theta_{_{\mathrm{B(ext)}}}^{^{\pm}})
\, \right\}\,\,\to\,\,\left\{\, \pm\,\,2/n \,,\,\infty\,\right\}\,\,,
 \]
we have the angular coefficients
 \begin{equation}
 \left\{\,
 \alpha_{_{\mathrm{lref}}}^{^{\mathrm{[TE]}}}(\theta_{_{\mathrm{B(ext)}}}^{^{\pm}})\,,\,
 \alpha_{_{\mathrm{lref}}}^{^{\mathrm{[TM]}}}(\theta_{_{\mathrm{B(ext)}}}^{^{\pm}})
 \,\right\}\,=\,
 \left\{\,
 \pm\,\frac{4/n}{\,\,\,(\,k\,{\mathrm{w}}_{\0}\,)^{^{2}}}\,,\,
 0\,\right\}\,\,.
\end{equation}
It is interesting to note that, while $ \alpha_{_{\mathrm{lref}}}^{^{\mathrm{[TM]}}}(\theta_{_{\mathrm{B(ext)}}}^{^{\pm}})=0$ seems to imply the absence of angular deviations, no TM-polarised plane wave incident at this angle is actually reflected, see Figure 7(a). For a bounded beam however, this only means that the center of the beam's angular distribution is not reflected, though wave vectors with incidence angles around $(\theta_{_{\mathrm{B(ext)}}}^{^{\pm}})$ still are. This changes the incoming Gaussian distribution into a double-peaked structure which makes the concept of angular deviations hazy. The same effect happens for the reflection at the lower interface, and this can graphically \cor{be} seen in Figures 13(a-c). For this reason, it is more insightful to study the angular deviation of a TM-polarised beam in the close vicinity of the Brewster angle, rather than at the angle itself. Introducing a new parameter $\delta${\footnote{Not to be confused with the Dirac's Delta, which always displays an argument, \emph{e.g.} $\delta(x)$.}}, we have the incidence angle
$ \theta_{\0}= \theta_{_{\mathrm{B(ext)}}}^{^{\pm}} \,+\,\, \delta/k\,{\mathrm{w}}_{\0}\,\,. $
Observing that
\begin{eqnarray*}
\sin^{^2}\theta_{\0} - n^{^{2}}\cos^{^{2}}\theta_{\0} &\approx&
2\,\sin  \theta_{_{\mathrm{B(ext)}}}^{^{\pm}}
\,[\, \cos \theta_{_{\mathrm{B(ext)}}}^{^{\pm}}  \pm n\,\sin  \theta_{_{\mathrm{B(ext)}}}^{^{\pm}}  \,]\,\,\displaystyle{\frac{\delta}{k\,{\mathrm{w}}_{\0}}}
\\
&=&\pm\,\,2\,n\,\, \frac{\delta}{k\,{\mathrm{w}}_{\0}}\,\,,
\end{eqnarray*}
we have that
\[
D_{_{\mathrm{lref}}}^{^{{\mathrm{[TM]}}}}\left(\theta_{_{\mathrm{B(ext)}}}^{^{\pm}} \,+\,\,\displaystyle{\frac{\delta}{k\,{\mathrm{w}}_{\0}}}\right)
 \,= \,\frac{k\,{\mathrm{w}}_{\0}}{\delta},
\]
implying
\begin{equation}
\label{eq:TMbrewlref}
\alpha_{_{\mathrm{lref}}}^{^{\mathrm{[TM]}}} \left(\theta_{_{\mathrm{B(ext)}}}^{^{\pm}} \,+\,\,\displaystyle{\frac{\delta}{k\,{\mathrm{w}}_{\0}}}\right) \,=\,\frac{2\,\delta}{1+ \delta^{^{2}}}\,\frac{1}{k\,{\mathrm{w}}_{\0}}\,\,.
\end{equation}
From Eq. (\ref{eq:TMbrewlref}) we have the curious result that, in the vicinity of the Brewster angle, the angular shift does not depend on the relative refractive index. Besides, performing a derivation of $\alpha_{_{\mathrm{lref}}}^{^{\mathrm{[TM]}}}$ with respect to $\delta$ shows us that the maximum deviation occurs for $\delta=\pm\,1$.  These results are in agreement with Aiello and Woerdman's paper in reference \cite{AiWo2009}, where, by a different approach, they studied angular deviations in the Brewster region for an air/glass interface. In particular, their formula presents the same behaviour presented in Figure 14, where Eq. (\ref{eq:alphalref}) was plotted.

\WideFigureSideCaption{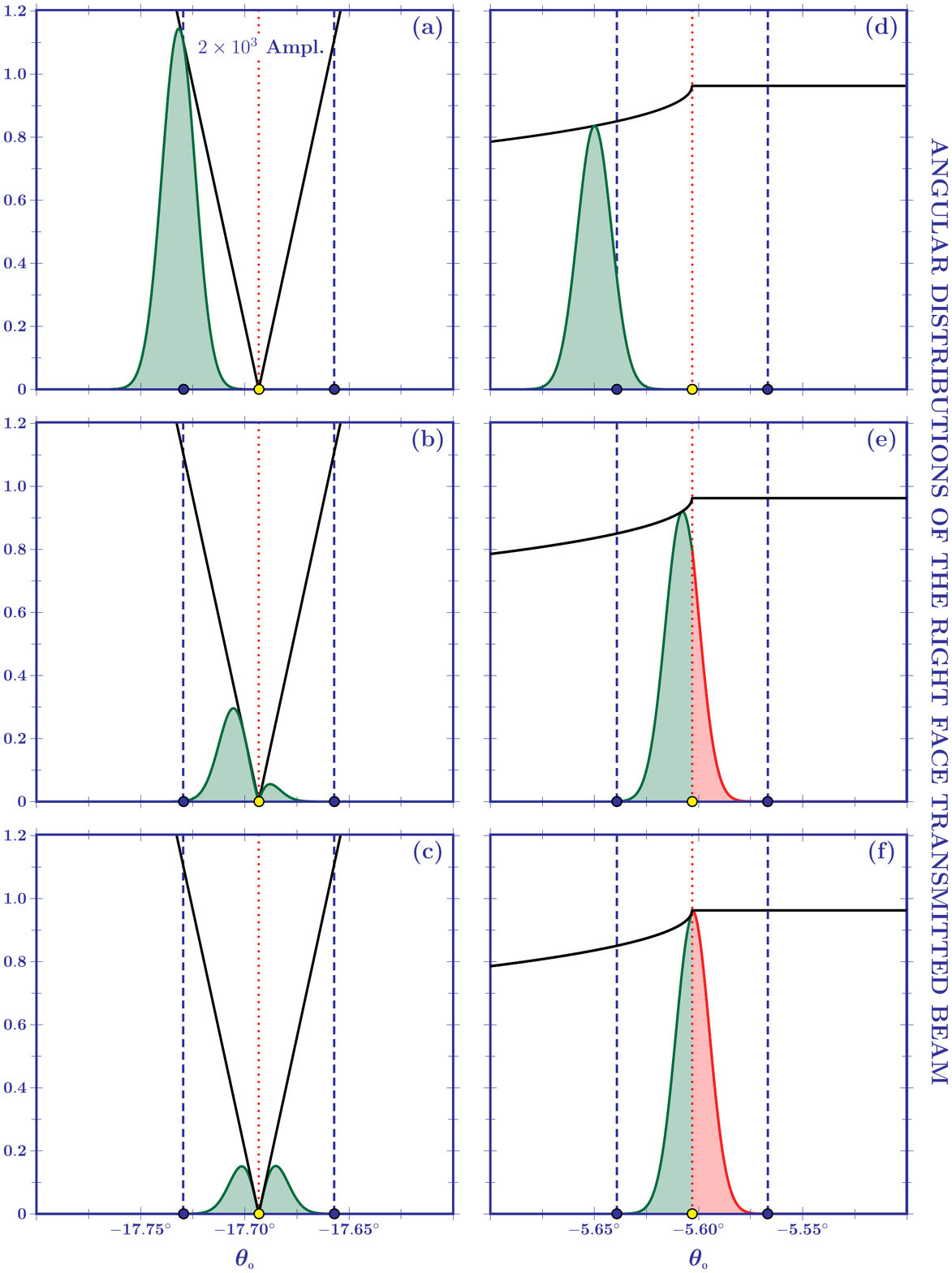}{(a-c) The angular distribution of a Gaussian beam approaching the internal Brewster angle. In (b) its symmetry is broken by the reflection coefficient (black lines) and in (c), exactly at the Brewster angle, it has become a double peaked structure, which makes the definition of angular deviations debatable. (d-f) A symmetry breaking of a different nature. The red portion of the distribution is totally internally reflected and there is a relative phase between both parts of the same distribution. The threshold is established by the critical angle. These plots consider a borosilicate prism ($n=1.515$) and a Gaussian distribution with $\mathrm{w}_{\0}=1\,\mathrm{mm}$.}

\WideFigureSideCaption{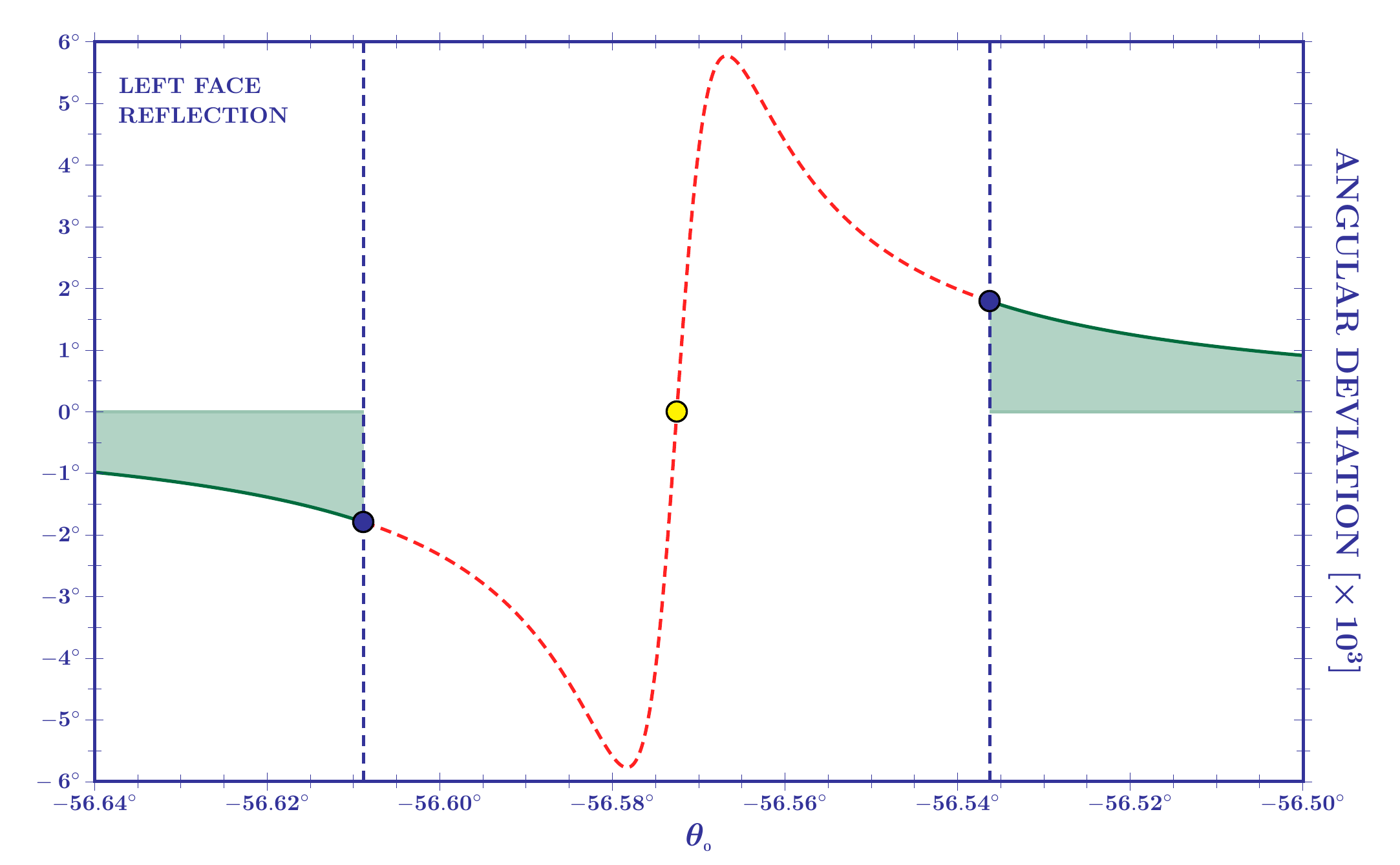}{Angular deviation for the reflection at the left face of the prism, \cor{an} external reflection. The dashed red portion of the curve is in the external Brewster region, where the concept of angular deviations is unclear due to the change in the incoming beam structure. The plot considers a borosilicate ($n=1.515$) prism and a TM-polarised beam with $\lambda=0.633\,\mu\mathrm{m}$ and $\mathrm{w}_{\0}=1\,\mathrm{mm}$.}

\subsection{The internal reflection}

The beam transmitted through the right face of the prism, with an electric field given by Eq. (\ref{eq:Ertrafinal}), has the symmetry of its angular distribution broken by the transmission coefficients of the left and right faces of the prism and by the lower face's reflection coefficient as well. Following the same steps carried out in the last Section, its path is then given by
\begin{equation}
\label{meantra}
\langle\, x_{_{\mathrm{rtra}}}^{^{\mathrm{[TE,TM]}}} \rangle
  = x_{_{\mathrm{rgeo}}} + \alpha_{_{\mathrm{rtra}}}^{^{\mathrm{[TE,TM]}}}(\theta_{\0})\,(z_{_{\mathrm{rtra}}}-z_{_{\mathrm{rgeo}}}),
\end{equation}
where
\begin{equation}
\label{eq:alphartra}
\alpha_{_{\mathrm{rtra}}}^{^{\mathrm{[TE,TM]}}}(\theta_{\0})\,=\,
\frac{2\, D_{_{\mathrm{rtra}}}^{^{{\mathrm{[TE,TM]}}}}(\theta_{\0})}{(k\,{\mathrm{w}}_{\0})^{^{^2}}\,+\, [\,D_{_{\mathrm{rtra}}}^{^{{\mathrm{[TE,TM]}}}}(\theta_{\0})\,]^{^{^{^2}}}},
\end{equation}
and
\begin{eqnarray}
D_{_{\mathrm{rtra}}}^{^{{\mathrm{[TE,TM]}}}} &=&  \frac{\displaystyle r_{_{\mathrm{lower}}}^{^{^{{\mathrm{[TE,TM]}}^\prime}}}(\theta_{\0})}{\displaystyle r_{_{\mathrm{lower}}}^{^{^{\mathrm{[TE,TM]}}}}(\theta_{\0})} {\nonumber}\\
&=&\frac{2\,\sin\varphi_{\0}\,\cos\theta_{\0}}{\cos\phi_{\0}\,\cos\psi_{\0}}
\displaystyle{ \left\{\, 1 \,,\, \frac{1}{n^{\2}\sin^{\2}\varphi_{\0} - \cos^{^{2}}\varphi_{\0}}  \,\right\}}\,\,.{\nonumber}
\end{eqnarray}
\cor{Note} that because the transmission coefficients are very smooth functions of the incidence angle, they can be factorised; the only contribution to the deviation coming then from the reflection coefficient, hence the section's title. At the internal Brewster angle, we have that
\begin{eqnarray*}
\left\{\,
D_{_{\mathrm{rtra}}}^{^{{\mathrm{[TE]}}}}(\theta_{_{\mathrm{B(int)}}})
\,,\,
D_{_{\mathrm{rtra}}}^{^{{\mathrm{[TM]}}}}(\theta_{_{\mathrm{B(int)}}})
\, \right\}\,\,&\to&\,\,
\frac{2\,\cos\theta_{_{\mathrm{B(int)}}} }{\cos\psi_{_{\mathrm{B(int)}}}}\,
\left\{\, 1 \, \,,\,\infty\,\right\}
\\
&=&\displaystyle{\left\{\, \frac{2\,\sqrt{n^{\2}-n^{\4}+2+2\,n^{\3}}}{n+1}\, \,,\,\infty\,\right\}},
\end{eqnarray*}
and, consequently,
 \begin{equation}
 \left\{\,
 \alpha_{_{\mathrm{rtra}}}^{^{\mathrm{[TE]}}}(\theta_{_{\mathrm{B(int)}}})\,,\,
 \alpha_{_{\mathrm{rtra}}}^{^{\mathrm{[TM]}}}(\theta_{_{\mathrm{B(int)}}})
 \,\right\}\,=\,
 \left\{\, \frac{4\,\sqrt{n^{\2}-n^{\4}+2+2\,n^{\3}}}{(n+1)\,\,(\,k\,{\mathrm{w}}_{\0}\,)^{^{2}}}
 \,,\,
 0\,\right\}.
\end{equation}
Here again, the fact that $\alpha_{_{\mathrm{rtra}}}^{^{\mathrm{[TM]}}}(\theta_{_{\mathrm{B(int)}}})=0$ does not actually imply that the beam's propagation direction is in \cor{accordance} with Geometrical Optics' predictions because, as for the externally reflected \cor{beam}, there is a structural change in the angular distribution, see Figure 13(a-c). For an incidence angle in the vicinity of the Brewster angle, however, we have that
\[
\theta_{\0}= \theta_{_{\mathrm{B(int)}}} \,+\,\, \frac{\delta}{k\,{\mathrm{w}}_{\0}}\,\,\,\,\,\,\,\Rightarrow\,\,\,\,\,\,\,\,
\varphi_{\0}= \varphi_{_{\mathrm{B(int)}}} \,+\,\, \frac{\cos\theta_{_{\mathrm{B(int)}}}}{n\,\cos\psi_{_{\mathrm{B(int)}}}}\,\,\frac{\delta}{k\,{\mathrm{w}}_{\0}}\,\,.
 \]
Observing then that
\begin{eqnarray*}
  n\,^{^2} \sin^{^2}\varphi_{\0} - \cos^{^{2}}\varphi_{\0}  &\approx &  2\,n\,\sin  \varphi_{_{\mathrm{B(int)}}}
\,[\, n\,\cos \varphi_{_{\mathrm{B(int)}}} + \, \sin  \theta_{_{\mathrm{B(int)}}}  \,]\,\,\frac{\cos\theta_{_{\mathrm{B(int)}}}}{n\,\cos\psi_{_{\mathrm{ B(int)}}}}\,\,\frac{\delta}{k\,{\mathrm{w}}_{\0}}\\
&=& \frac{2\,\cos\theta_{_{\mathrm{B(int)}}}}{\cos\psi_{_{\mathrm{B(int)}}}}\,\,\frac{\delta}{k\,{\mathrm{w}}_{\0}}\,\,,
 \end{eqnarray*}
we find
\begin{equation*}
D_{_{\mathrm{rtra}}}^{^{{\mathrm{[TM]}}}}\left(\theta_{_{\mathrm{B(int)}}}\,+\,\,\displaystyle{\frac{\delta}{k\,{\mathrm{w}}_{\0}}}\right)
\,= \,\frac{k\,{\mathrm{w}}_{\0}}{\delta}
\end{equation*}
and
\begin{equation}
\label{eq:alphaBintdev}
\alpha_{_{\mathrm{rtra}}}^{^{\mathrm{[TM]}}} \left(\theta_{_{\mathrm{B(int)}}} \,+\,\,\displaystyle{\frac{\delta}{k\,{\mathrm{w}}_{\0}}}\right) \,=\,\frac{2\,\delta}{1+ \delta^{^{2}}}\,\frac{1}{k\,{\mathrm{w}}_{\0}}\,\,,
\end{equation}
which are the same results obtained for the externally reflected beam. To \cor{have} an idea of the magnitude of the deviations, let us consider a borosilicate prism ($n=1.515$) and an incident Gaussian beam with a wavelength $\lambda=0.633\,\mu\mathrm{m}$ and a beam waist $\mathrm{w}_{\0} = 1\,\mathrm{mm}$. For incidence at $\theta_{_{\mathrm{B(int)}}}\,\pm\,1\,/\,k\,{\mathrm{w}}_{\0}$ we have
\begin{eqnarray*}
\alpha_{_{\mathrm{rtra}}}^{^{\mathrm{[TE]}}} & \approx & \frac{4}{(k\,{\mathrm{w}}_{\0})^{^{2}}}\,\approx \,2.3^{\circ}\,\times\,
10^{^{-6}}\,\,,\\
\alpha_{_{\mathrm{rtra}}}^{^{\mathrm{[TM]}}} & \approx & \pm\, \frac{1}{k\,{\mathrm{w}}_{\0}}\,\approx\,\pm\, \, 5.8^{\circ}\,\times\,
10^{^{-3}}\,\,.
\end{eqnarray*}
Figure 15 shows the angular deviations for a TM-polarised beam for incidence in the vicinity of the internal Brewster region.

For incidence in the critical region,
\[
\theta_{\0}= \theta_{_{\mathrm{cri}}} \,-\,\, \frac{|\delta|}{k\,{\mathrm{w}}_{\0}}\,\,\,\,\,\,\,\Rightarrow\,\,\,\,\,\,\,\,
\varphi_{\0}= \varphi_{_{\mathrm{cri}}} \,-\,\, \frac{\cos\theta_{_{\mathrm{cri}}}}{n\,\cos\psi_{_{\mathrm{cri}}}}\,\,\frac{|\delta|}{k\,{\mathrm{w}}_{\0}}\,\,,
 \]
we have
\begin{eqnarray}
\left\{\,
D_{_{\mathrm{rtra}}}^{^{{\mathrm{[TE]}}}}\left(\theta_{_{\mathrm{cri}}}\,-\,\,\displaystyle{\frac{|\delta|}{k\,{\mathrm{w}}_{\0}}}\right)
\,,\,
D_{_{\mathrm{rtra}}}^{^{{\mathrm{[TM]}}}}\left(\theta_{_{\mathrm{cri}}}\,-\,\,\displaystyle{\frac{|\delta|}{k\,{\mathrm{w}}_{\0}}}\right)
\,\right\} &=&
\sqrt{2}\,\left[\,\frac{2-n^{\2}+2\,\sqrt{n^{\2}-1}}{(n^{\2}-1)\,(n^{\2}+2\,\sqrt{n^{\2}-1})}\,\right]^{^{1/4}} {\nonumber} \\
&\times &\left\{\,1\,,\,n^{\2}\,\right\}\,\,\sqrt{\frac{k\,{\mathrm{w}}_{\0}}{|\delta|}},
\end{eqnarray}
which yields a refractive index-dependent angular deviation,
\begin{eqnarray}
\label{eq:alphacridev}
\left\{\alpha_{_{\mathrm{rtra}}}^{^{\mathrm{[TE]}}}\left(\theta_{_{\mathrm{cri}}} \,-\,\,\displaystyle{\frac{|\delta|}{k\,{\mathrm{w}}_{\0}}}\right),\,\alpha_{_{\mathrm{rtra}}}^{^{\mathrm{[TM]}}}\left(\theta_{_{\mathrm{cri}}} \,-\,\,\displaystyle{\frac{|\delta|}{k\,{\mathrm{w}}_{\0}}}\right)\right\}\,\,=\,\,
 \frac{f(n)}{\sqrt{|\delta|}}\,\left\{\,1\,,\,n^{\2}\,\right\}\,\,\frac{1}{\,\,\,(k\,{\mathrm{w}}_{\0})^{^{{3/2}}}},
\end{eqnarray}

\WideFigureSideCaption{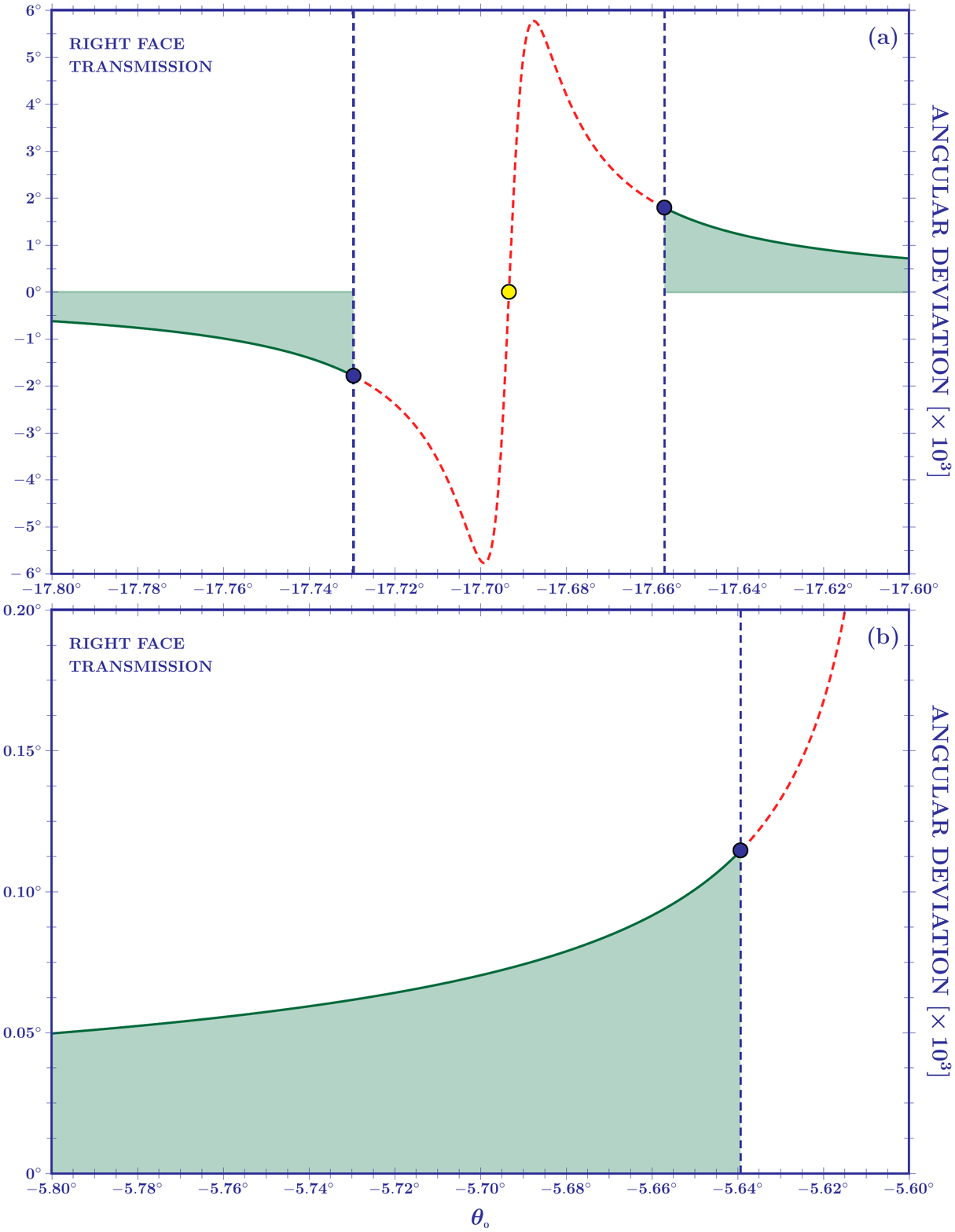}{(a) Angular deviation for the reflection at the lower face of the prism, a external reflection, in the vicinity of the internal Brewster angle. The dashed red portion of the curve is in the internal Brewster region, where the concept of angular deviations is unclear due to the change in the incoming beam structure. (b) Angular deviations in the vicinity of the critical angle. The dashed red curve marks the region where a relative phase exists between portions of the beam's angular distribution. The plots consider a borosilicate ($n=1.515$) prism and a TM-polarised beam with $\lambda=0.633\,\mu\mathrm{m}$ and $\mathrm{w}_{\0}=1\,\mathrm{mm}$.}
where
\begin{equation*}
 f(n)\,\,=\,\,2\,\sqrt{2}\,\left[\,\frac{2-n^{\2}+2\,\sqrt{n^{\2}-1}}{(n^{\2}-1)\,(n^{\2}+2\,\sqrt{n^{\2}-1})}\,
\right]^{^{1/4}}\,\,.
\end{equation*}
Figure 15 has the angular deviation of a TM-polarised beam near the critical region for the same parameters used before for the Brewster region. It is important to \cor{note} that, for incidence angles greater than $\theta_{_{\mathrm{cri}}} - \lambda/{\mathrm{w}}_{\0}$, the lower reflection coefficient becomes complex and the interference between real and imaginary parts has to be considered, which originates the so-called Composite Goos-H\"anchen shift \cite{Maia2017}. \cor{The} incidence at $\theta_{_{\mathrm{cri}}} -\lambda/{\mathrm{w}}_{\0}$, however, still has a real angular distribution, making angular deviations the only contribution to \cor{the} beam shift phenomena. This incidence angle also gives the maximal angular deviation in the critical region.

Let us compare angular deviations at the \cor{borders} of the region between the Brewster and the critical region, that is,
 \[  \theta_{_{\mathrm{B(int)}}} +  \,\displaystyle{\frac{\lambda}{\,{\mathrm{w}}_{\0}}}
     < \,\, \theta_{\0} <
    \theta_{_{\mathrm{cri}}} - \,\displaystyle{\frac{\lambda}{\,{\mathrm{w}}_{\0}}}\,\,.  \]
In this region angular deviations are a clear concept since the angular distribution does not present additional peaks nor complex terms. For a borosilicate prism and an incident Gaussian beam with  $\lambda=0.633\,\mu{\mathrm{m}}$ and ${\mathrm{w}}_{\0}=1\,{\mathrm{mm}}$, we have that,
 for incidence at $\theta_{_{\mathrm{B(int)}}} + \, \lambda\,/\, {\mathrm{w}}_{\0}$,
\[
\begin{array}{l}
\displaystyle{
\left\{\,\alpha_{_{\mathrm{rtra}}}^{^{\mathrm{[TE]}}}\,,\,\alpha_{_{\mathrm{rtra}}}^{^{\mathrm{[TM]}}}\,\right\} \,\approx  \, \left\{\,\frac{4}{(k\,{\mathrm{w}}_{\0})^{^{2}}}\,,\,\frac{1/\pi}{k\,{\mathrm{w}}_{\0}}\,\right\} \,\approx\,}\\
\left\{\,2.3^{\circ}\times\,10^{^{-3}}\,,\,1.8^{\circ}\,\right\}\,\times\,10^{^{-3}}\,\,,
\end{array}
\]
and, using $f(1.515)\approx 3/\sqrt{2}$, for incidence  at $\theta_{_{\mathrm{cri}}} - \, \lambda\,/\, {\mathrm{w}}_{\0}$,
\[
\begin{array}{l}
\displaystyle{
\left\{\,\alpha_{_{\mathrm{rtra}}}^{^{\mathrm{[TE]}}}\,,\,\alpha_{_{\mathrm{rtra}}}^{^{\mathrm{[TM]}}}\,\right\} \,\approx\,
\frac{3}{2\,\sqrt{\pi}}\,\{\,1\,,\,2.3\,\}\, \frac{1}{\,\,(k\,{\mathrm{w}}_{\0})^{^{3/2}}} \,\approx\,}\\
\displaystyle{
\left\{\,0.5^{\circ}\,,\,1.1^{\circ}\,\right\}\,\times\,10^{^{-4}}\,\,.}
\end{array}
\]
We see then that the critical region provides greater angular deviations than the Brewster region for the TE-polarisation while the opposite is true for the TM-polarised beams. The deviations for intermediary angles of incidence is given in Table 1.
\begin{table*}
{\large
{\color{navy}{
\begin{center}
\begin{tabular}{|c||c||c|}\hline
\rowcolor{bg!50} $\theta_{\0}$ & $10^{^{3}}\,\,\alpha_{_{\mathrm{TRA}}}^{^{\mathrm{[TE]}}}$ & $10^{^{3}}\,\,\alpha_{_{\mathrm{TRA}}}^{^{\mathrm{[TM]}}}$ \\ \hline
 \rowcolor[gray]{0.9}  $-\,17.65706^{\circ}$ & $0.0023^{\circ}$ & $1.7953^{\circ}$ \\ \hline
 \rowcolor[gray]{0.7}  $-\,17.65000^{\circ}$ & $0.0023^{\circ}$ & $1.5145^{\circ}$ \\ \hline
 \rowcolor[gray]{0.9}  $-\,16.65000^{\circ}$ & $0.0024^{\circ}$ & $0.0676^{\circ}$  \\ \hline
 \rowcolor[gray]{0.7}  $-\,15.65000^{\circ}$ & $0.0026^{\circ}$ & $0.0366^{\circ}$   \\ \hline
 \rowcolor[gray]{0.9}  $-\,14.65000^{\circ}$ & $0.0028^{\circ}$ & $0.0262^{\circ}$  \\ \hline
 \rowcolor[gray]{0.7}  $-\,13.65000^{\circ}$ & $0.0030^{\circ}$ & $0.0211^{\circ}$  \\ \hline
 \rowcolor[gray]{0.9}  $-\,12.65000^{\circ}$ & $0.0032^{\circ}$ & $0.0183^{\circ}$  \\ \hline
 \rowcolor[gray]{0.7}  $-\,11.65000^{\circ}$ & $0.0035^{\circ}$ & $0.0166^{\circ}$\\ \hline
 \rowcolor[gray]{0.9}  $-\,10.65000^{\circ}$ & $0.0039^{\circ}$ & $0.0158^{\circ}$  \\ \hline
 \rowcolor[gray]{0.7}  $-\,\,\,\,9.65000^{\circ}$ & $0.0045^{\circ}$ & $0.0156^{\circ}$  \\ \hline
\rowcolor[gray]{0.9}   $-\,\,\,\,8.65000^{\circ}$ & $0.0052^{\circ}$ & $0.0162^{\circ}$ \\ \hline
 \rowcolor[gray]{0.7}  $-\,\,\,\,7.65000^{\circ}$ & $0.0064^{\circ}$ & $0.0179^{\circ}$  \\ \hline
 \rowcolor[gray]{0.9}  $-\,\,\,\,6.65000^{\circ}$ & $0.0091^{\circ}$ & $0.0230^{\circ}$  \\ \hline
 \rowcolor[gray]{0.7}  $-\,\,\,\,5.65000^{\circ}$ & $0.0437^{\circ}$ & $0.1007^{\circ}$ \\ \hline
 \rowcolor[gray]{0.9}  $-\,\,\,\,5.63938^{\circ}$ & $0.0497^{\circ}$ & $0.1144^{\circ}$ \\ \hline
\end{tabular}
\end{center}
}}
}
\caption{\footnotesize{Table of angular deviation values for a borosilicate ($n=1.515$) prism and an incident beam with wavelength $\lambda=0.633\,\mu{\mathrm{m}}$ and minimal beam waist ${\mathrm{w}}_{\0}=1\,{\mathrm{mm}}$, for incidence angles (first column) ranging from the internal Brewster region to the critical region, and for TE- (second column) and TM-polarised (third column) waves.}}
\end{table*}


\subsection{The transmission through the lower face of the prism}

Analogous to the cases before, we \cor{see} that the path of the beam transmitted through the lower face of the prism is given by
\begin{equation}
\label{meantra}
\langle\, x_{_{\mathrm{ltra}}}^{^{\mathrm{[TE,TM]}}} \rangle
  = x_{_{\mathrm{lgeo}}} +\alpha_{_{\mathrm{ltra}}}^{^{\mathrm{[TE,TM]}}}(\theta_{\0})\,\,\,(z_{_{\mathrm{ltra}}}-z_{_{\mathrm{lgeo}}}),
\end{equation}
where
\begin{equation}
\alpha_{_{\mathrm{ltra}}}^{^{\mathrm{[TE,TM]}}}(\theta_{\0})\,=\,
\phi_{\0}^{\prime}\,\frac{2\, D_{_{\mathrm{ltra}}}^{^{{\mathrm{[TE,TM]}}}}(\theta_{\0})}{(k\,{\mathrm{w}}_{\0})^{^{^2}}\,+\, [\,D_{_{\mathrm{ltra}}}^{^{{\mathrm{[TE,TM]}}}}(\theta_{\0})\,]^{^{^{^2}}}}
\end{equation}
and
\vspace{-1cm}
\begin{eqnarray}
D_{_{\mathrm{ltra}}}^{^{{\mathrm{[TE,TM]}}}}(\theta_{\0})  &=& \frac{t_{_{\mathrm{lower}}}^{^{{\mathrm{[TE,TM]}}^{\prime}}}(\theta_{\0})}{t_{_{\mathrm{ lower}}}^{^{{\mathrm{[TE,TM]}}}}(\theta_{\0}) {\nonumber}}\\
&=&
\frac{n^{^2}-1}{n}\,\frac{\tan\varphi_{\0}\,\cos\theta_{\0}}{\cos\phi_{\0}\,\cos\psi_{\0}}\,\\
& \times & \,\left\{\, \frac{1}{n\,\cos\varphi{\0}+\cos\phi_{\0}} \,,\, \frac{n}{\cos\varphi{\0}+n\,\cos\phi_{\0}}  \,\right\}.
\end{eqnarray}
\cor{Note} that no reflection takes place in this case and so the dominant term in the symmetry breaking of beam's angular distribution is \cor{the} transmission coefficient through the lower face of the prism, \cor{with} the transmission coefficient of the left face being smoother. So, the Brewster angle is not present in this case and the only region of interest is the critical one. For incidence in this region we find
\begin{eqnarray}
\left\{\,
D_{_{\mathrm{ltra}}}^{^{{\mathrm{[TE]}}}}\left(\theta_{_{\mathrm{cri}}}\,-\,\,\displaystyle{\frac{|\delta|}{k\,{\mathrm{w}}_{\0}}}\right)
\,,\,
D_{_{\mathrm{ltra}}}^{^{{\mathrm{[TM]}}}}\left(\theta_{_{\mathrm{cri}}}\,-\,\,\displaystyle{\frac{|\delta|}{k\,{\mathrm{w}}_{\0}}}\right)
\,\right\} &=& \frac{f(n)}{4}\left\{\,1\,,\,n^{\2}\,\right\}\,\,\sqrt{\frac{k\,{\mathrm{w}}_{\0}}{|\delta|}},
\end{eqnarray}
and
\begin{equation}
\phi_{\0}^{\prime}\left(\theta_{_{\mathrm{cri}}}\,-\,\,\displaystyle{\frac{|\delta|}{k\,{\mathrm{w}}_{\0}}}\right) =
\displaystyle{\frac{f(n)}{4}\,\sqrt{n^{\2}-1}\,\sqrt{\frac{k\,{\mathrm{w}}_{\0}}{|\delta|}}},
\end{equation}
which provides an angular deviation of
\begin{equation}
\left\{ \alpha_{_{\mathrm{ltra}}}^{^{\mathrm{[TE]}}}\left(\theta_{_{\mathrm{cri}}}\,-\,\,\displaystyle{\frac{|\delta|}{k\,{\mathrm{w}}_{\0}}}\right), \alpha_{_{\mathrm{ltra}}}^{^{\mathrm{[TM]}}}\left(\theta_{_{\mathrm{cri}}}\,-\,\,\displaystyle{\frac{|\delta|}{k\,{\mathrm{w}}_{\0}}}\right) \right\} = \frac{f^{\2}(n)}{8}\,\frac{\sqrt{n^{\2}-1}}{k\mathrm{w}_{\0}|\delta|}\,\left\{1,\,n^{\2}\right\}.
\end{equation}

\WideFigureSideCaption{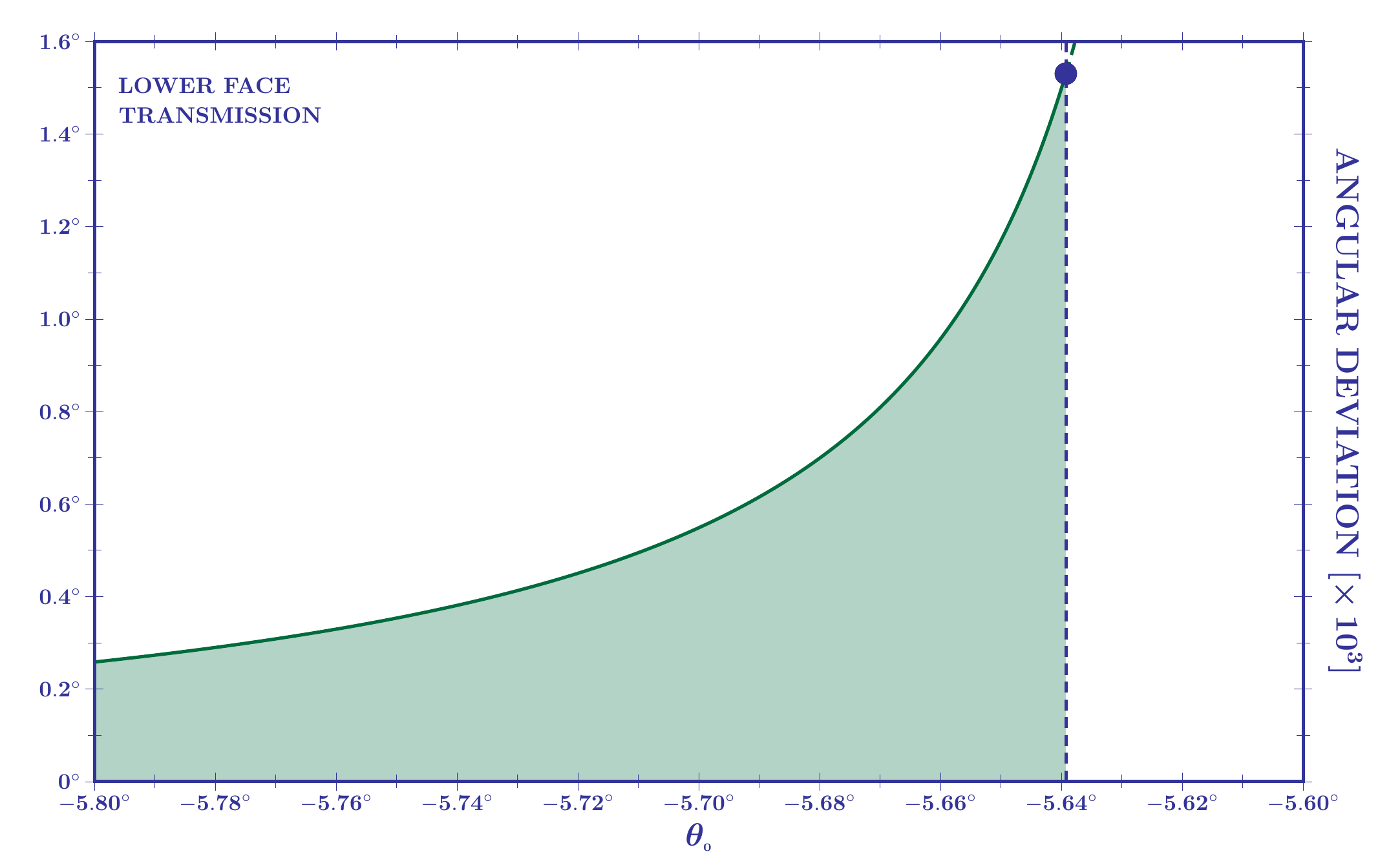}{Angular deviation for the transmission through the lower face of the prism in the vicinity of the critical angle. The dashed line marks the region where a relative phase exists between portions of the beam's angular distribution. The plot considers a borosilicate ($n=1.515$) prism and a TM-polarised beam with $\lambda=0.633\,\mu\mathrm{m}$ and $\mathrm{w}_{\0}=1\,\mathrm{mm}$.}

In the vicinity of the critical angle, considering an incident angle of $\theta_{\0} = \theta_{_{\mathrm{cri}}}-|\delta|/k\mathrm{w}_{\0}$, the angular deviations are given by
\begin{eqnarray*}
\alpha_{_{\mathrm{ltra}}}^{^{\mathrm{\{\,TE\,,\,TM\,\}}}} \left(\theta_{_{\mathrm{cri}}} \,-\,\,\displaystyle{\frac{|\delta|}{k\,{\mathrm{w}}_{\0}}}\right)\,\,=\,\,
 \frac{\sqrt{n^{\2}-1}\,f^{\2}(n)}{8}\,\left\{\,1\,,\,n^{\2}\,\right\}\,\,\frac{1}{\,\,\,|\delta|\,k\,{\mathrm{w}}_{\0}}\,\,.\end{eqnarray*}
At the border of the critical region, that is, for $\delta=2\pi$, the maximal angular deviation is achieved before part of the beam undergoes \cor{a} total internal reflection. At this angle we have
\begin{eqnarray}
\left\{\,\alpha_{_{\mathrm{ltra}}}^{^{\mathrm{[TE]}}}\left(\theta_{_{\mathrm{cri}}} \,-\,\,\displaystyle{\frac{\lambda}{{\mathrm{w}}_{\0}}}\right)\,,\,\alpha_{_{\mathrm{ltra}}}^{^{\mathrm{[TM]}}}\left(\theta_{_{\mathrm{cri}}} \,-\,\,\displaystyle{\frac{\lambda}{{\mathrm{w}}_{\0}}}\right)\,\right\} \,\approx\,\left\{\,0.6^{\circ}\,,\,1.3^{\circ}\,\right\}\,\times\,10^{^{-3}}\,\,.
\end{eqnarray}
The angular deviation of the beam transmitted through the lower face of the prism is plotted in Figure 16.

\section{The composite Goos-H\"anchen shift}
\label{sec7}

\cor{So far,}
we have seen that for incidence angles greater than $\theta_{_{\mathrm{cri}}}+\lambda/\mathrm{w}_{\0}$, a paraxial beam is entirely totally internally reflected, its reflection coefficient becoming complex, which yields a lateral displacement of its reflection point on the interface. On the other hand, for incidence angles before the threshold $\theta_{_{\mathrm{cri}}}-\lambda/\mathrm{w}_{\0}$, the reflection coefficient is real for the whole beam, but it stops being symmetrical and breaks the symmetry of the beam's angular distribution, which originates an angular deviation from the predictions of Geometrical Optics. This is what makes the critical region between both of these \cor{borders} so interesting: in this region part of the beam experiences total internal reflection and part of it suffers a change in the power balance between the plane waves that compose it. The interference between the real and complex parts of the same beam is the responsible for the so-called composite Goos-H\"anchen shift. This name \cor{means} that, in this region, angular and lateral Goos-H\"anchen shifts are verified simultaneously. In Section \ref{sec5}, we studied the Goos-H\"anchen shift in the critical region, but our approach considered only lateral shifts, disregarding angular effects. As we will see, taking into account both phenomena will not only lead to a lateral shift whose measurement is dependent upon the distance from where it is taken, as our geometrical intuition tells us, but also, \cor{it} will originate a new kind of oscillatory phenomena.

Let us consider the electric field of the beam being transmitted through the right face of the prism. Using Eqs. (\ref{eq:gaussianDist}) and (\ref{eq:Ertrafinal}) we can write the electric field of this beam as being proportional to
\begin{equation}
\label{eq:CGHErtra}
E_{_{\mathrm{rtra}}} \propto \int_{\-infty}^{\+infty}{\hspace{-0.25cm}}\mathrm{d}\theta \, r_{_{\mathrm{lower}}}^{^{\mathrm{[TE,TM]}}}(\theta) \exp\left\{ -\left[ \frac{\displaystyle k\,w(\tilde{z}_{_{\mathrm{rtra}}})}{2}\,(\theta-\theta_{\0}) \right]^{\2} + i\,k(\theta-\theta_{\0})\,\tilde{x}_{_\mathrm{rtra}} \right\},
\end{equation}
where $w(\tilde{z}_{_{\mathrm{rtra}}})=\mathrm{w}_{\0}\sqrt{1+i\zeta}$, not to be confused with $\mathrm{w}(\tilde{z})$, which is given by Eq. (\ref{eq:wz}). \cor{Note} that we have factored out the transmission coefficients since the Goos-H\"anchen shift is only associated to the reflection coefficient, being  also the main contribution to the beam's symmetry breaking. This reflection coefficient can be written as
\begin{equation}
\left\{r^{^{[\mathrm{TE}]}}_{_\mathrm{lower}}(\theta) , r^{^{[\mathrm{TM}]}}_{_\mathrm{lower}}(\theta) \right\} = \left\{ \frac{\displaystyle n\cos\varphi - \sqrt{1-n^{\2}\sin^{\2}\varphi}}{\displaystyle n\cos\varphi + \sqrt{1-n^{\2}\sin^{\2}\varphi}}{},  \frac{\displaystyle \cos\varphi - n\sqrt{1-n^{\2}\sin^{\2}\varphi}}{\displaystyle \cos\varphi + n\sqrt{1-n^{\2}\sin^{\2}\varphi}} \right\}.
\end{equation}
Now, by writing $\theta = \theta-\theta_{_\mathrm{cri}} + \theta_{_{\mathrm{cri}}} = \delta\theta + \theta_{_{\mathrm{cri}}}$, and by \cor{noting} that the Snell's law, $\sin\theta = n\,\sin\psi$, implies that
\begin{equation}
\delta\theta\,\cos\theta_{_\mathrm{cri}} = n\,\delta\psi\cos\psi_{_{\mathrm{cri}}},
\end{equation}
we have that
\begin{equation}
\delta\varphi = \frac{\displaystyle \cos\theta_{_{\mathrm{cri}}}}{\displaystyle \cos\psi_{_{\mathrm{cri}}}}\,\frac{\displaystyle \delta\theta}{\displaystyle n} \approx \frac{\displaystyle \delta\theta}{\displaystyle n},
\end{equation}
where we used the relation $\delta\psi = \delta\varphi$. Expanding now the elements in the reflection coefficients around critical incidence, we have that
\begin{subequations}
\begin{equation}
n\,\cos\varphi \approx n\,\cos\varphi_{_{\mathrm{cri}}} - n\,\sin\varphi_{_{\mathrm{cri}}}\,\delta\varphi = n\,\cos\varphi_{_{\mathrm{cri}}}-\delta\varphi,
\end{equation}
and that
\begin{eqnarray}
\sqrt{1-n^{\2}\,\sin^{\2}\varphi} &\approx& \sqrt{1 - n^{\2}\,\sin^{\2}\varphi_{_{\mathrm{cri}}} -2n^{\2}\,\sin\varphi_{_{\mathrm{cri}}}\cos\varphi_{_{\mathrm{cri}}}\delta\varphi }\\
&=&\sqrt{-2n\cos\varphi_{_{\mathrm{cri}}}\delta\varphi}.
\end{eqnarray}
\end{subequations}
Defining then $\delta\varepsilon = -\delta\varphi/n\,\cos\varphi_{_{\mathrm{cri}}}$, we can approximate the reflection coefficients in the critical region as
\begin{subequations}
\begin{equation}
r^{^{\mathrm{[TE]}}}_{_{\mathrm{lower}}}(\theta) = \frac{\displaystyle 1 + \delta\varepsilon - \sqrt{2\,\delta\varepsilon}}{\displaystyle 1 + \delta\varepsilon + \sqrt{2\,\delta\varepsilon}} \approx 1-2\sqrt{2\,\delta\varepsilon} + 4\,\delta\varepsilon,
\end{equation}
and
\begin{equation}
r^{^{\mathrm{[TM]}}}_{_{\mathrm{lower}}}(\theta) = \frac{\displaystyle 1 + \delta\varepsilon - n^{\2}\sqrt{2\,\delta\varepsilon}}{\displaystyle 1 + \delta\varepsilon + n^{\2}\sqrt{2\,\delta\varepsilon}} \approx 1-2\,n^{\2}\sqrt{2\,\delta\varepsilon} + 4\,n^{\4}\,\delta\varepsilon,
\end{equation}
\end{subequations}
which can be written collectively as
\begin{equation}
r^{^{\mathrm{[TE,TM]}}}_{_{\mathrm{lower}}} \approx 1 - \sqrt{2\,\gamma^{^{\mathrm{[TE,TM]}}}(\theta_{_{\mathrm{cri}}}-\theta)} + \gamma^{^{\mathrm{[TE,TM]}}}(\theta_{_{\mathrm{cri}}}-\theta),
\end{equation}
with $\{ \gamma^{^{\mathrm{[TE]}}},\gamma^{^{\mathrm{[TM]}}} \} = 4\{ 1,n^{\4} \}/n\sqrt{n^{\2}-1}$. Finally, defining the integration variable $\tau = k\,w(\tilde{z}_{_\mathrm{rtra}})(\theta_{_{\mathrm{cri}}}-\theta)/2$, the electric field of Eq. (\ref{eq:CGHErtra}) can be written as
\begin{equation}
\label{eq:ErtraApp}
E_{_{\mathrm{rtra}}}^{^{\mathrm{[TE,TM]}}} \propto \int_{\-infty}^{\+infty} {\hspace{-0.25cm}}\mathrm{d}\tau\,\left[ 1 - 2\,\sqrt{\frac{\displaystyle 2\,\gamma^{^{\mathrm{[TE,TM]}}}\,\tau}{\displaystyle k\,w(\tilde{z}_{_{\mathrm{rtra}}})}} + \frac{\displaystyle 2\,\gamma^{^{\mathrm{[TE,TM]}}}\,\tau}{\displaystyle k \, w(\tilde{z}_{_{\mathrm{rtra}}})} \right] \, e^{ -( \tau + \ell )^{\2}}\,\mathcal{G}(\tilde{x}_{_{\mathrm{rtra}}},\tilde{z}_{_{\mathrm{rtra}}}),
\end{equation}
where
\begin{equation}
\ell = \frac{k\,w(\tilde{z}_{_{\mathrm{rtra}}})}{2}\,(\theta_{\0}-\theta_{_{\mathrm{cri}}}) + i\frac{\tilde{x}_{_{\mathrm{rtra}}}}{w(\tilde{z}_{_{\mathrm{rtra}}})},
\end{equation}
and
\begin{equation}
\mathcal{G}(\tilde{x}_{_{\mathrm{rtra}}},\tilde{z}_{_{\mathrm{rtra}}}) = \exp\left[ -\frac{ \tilde{x}_{_{\mathrm{rtra}}}^{\2}}{w^{\2}(\tilde{z}_{_{\mathrm{rtra}}})} \right].
\end{equation}

Eq. (\ref{eq:ErtraApp}) presents three integrals to be solved. The first and third ones have straightforward solutions,
\begin{equation}
\int_{\-infty}^{\+infty} {\hspace{-0.25cm}} \mathrm{d}\tau\,e^{-(\tau+\ell)^{\2}} = \sqrt{\pi},
\end{equation}
and
\begin{equation}
\int_{\-infty}^{\+infty} {\hspace{-0.25cm}} \mathrm{d}\tau\,\tau\,e^{-(\tau+\ell)^{\2}} = -\sqrt{\pi}\,\ell,
\end{equation}
while the second one demands some more work. Following a procedure analogous to the one carried out in Section \ref{sec5}, see Eq. (\ref{eq:GHIntexp}), we can write the integrand of this integral as a series:
\begin{eqnarray}
\int_{\-infty}^{\+infty} {\hspace{-0.25cm}} \mathrm{d}\tau\,\sqrt{\tau}\,e^{-(\tau+d)^{\2}} &=& e^{-\ell^{\2}}\sum_{_{m=0}}^{\sinfty}\frac{\displaystyle (-2\,\ell)^{^m}}{\displaystyle m!} \int_{\-infty}^{\+infty} {\hspace{-0.25cm}}\mathrm{d}\tau \, \tau^{m+\frac{1}{2}}\,e^{-\tau^{\2}}{\nonumber}
\\
&=& e^{-\ell^{\2}}\sum_{_{m=0}}^{\sinfty} \frac{\displaystyle (2\,\ell)^{^m}}{\displaystyle m!}\,\frac{\displaystyle 1 + i\,(-1)^{^m}}{\displaystyle 2}\,\Gamma\left( \frac{\displaystyle 2m + 3}{\displaystyle 4} \right).
\end{eqnarray}
The integral returns then
\begin{equation}
\int_{\-infty}^{\+infty} \hspace{-0.25cm} \mathrm{d}\tau\,\sqrt{\tau}\,e^{-(\tau+\ell)^{\2}} = \frac{\displaystyle e^{i\frac{\pi}{\4}}}{2\,\sqrt{2}}\,\mathcal{F}(\tilde{x}_{_{\mathrm{rtra}}},\tilde{z}_{_{\mathrm{rtra}}};\theta_{\0}),
\end{equation}
being
\begin{equation}
\mathcal{F}(\tilde{x}_{_{\mathrm{rtra}}},\tilde{z}_{_{\mathrm{rtra}}};\theta_{\0}) = \left[2\Gamma\left(\frac{3}{4}\right)\,_{\1}F_{\1}\left(\frac{3}{4},\frac{1}{2},\ell^{\2}\right)+i\,\ell\,\Gamma\left(\frac{1}{4}\right)\,_{\1}F_{\1}\left(\frac{5}{4},\frac{3}{2},\ell^{\2}\right)\right]\,e^{-\ell^{\2}},
\end{equation}
where $_{\1}F_{\1}(x)$ is the Kummer confluent hypergeometric function. Finally, the electric field coming out of the right face of the prism has the form
\begin{equation}
\label{eq:IntegratedErtra}
E_{_{\mathrm{rtra}}} \propto \left[ 1-\sqrt{\frac{\displaystyle \gamma^{^{\mathrm{[TE,TM]}}}}{\displaystyle 2\pi k\,w(\tilde{z}_{_{\mathrm{rtra}}})}}\,e^{i\frac{\pi}{\4}}\mathcal{F}(\tilde{x}_{_{\mathrm{rtra}}},\tilde{z}_{_{\mathrm{rtra}}}\;\theta_{\0})-\frac{\displaystyle 2\,\ell\,\gamma^{^{\mathrm{[TE,TM]}}}}{\displaystyle k\,w(\tilde{z}_{_{\mathrm{rtra}}})} \right]\,\mathcal{G}(\tilde{x}_{_{\mathrm{rtra}}},\tilde{z}_{_{\mathrm{rtra}}}).
\end{equation}
The study of the intensity of this approximated field, $I_{_{\mathrm{rtra}}} = |E_{_{\mathrm{rtra}}}|^{^{2}}$, is what holds the information on the composite Goos-H\"anchen shift. We are interested in the maximum value of the above equation for a given set of parameters, or, more specifically, how the position of the maximum changes, \cor{for a given wavelength and beam width,} when the incidence angle and the position of the camera that makes the measurements change.

Firstly, in order to check the validity of our approximations we must consider measurements being carried out extremely close to the right face of the prism. The reason for this is to remove from the picture the angular deviations, in which case we should obtain the known curves for the Goos-H\"anchen shift. In Figures 17 and 18 we show the results of this numerical analysis for TE- and TM-polarised beams,

\WideFigureSideCaption{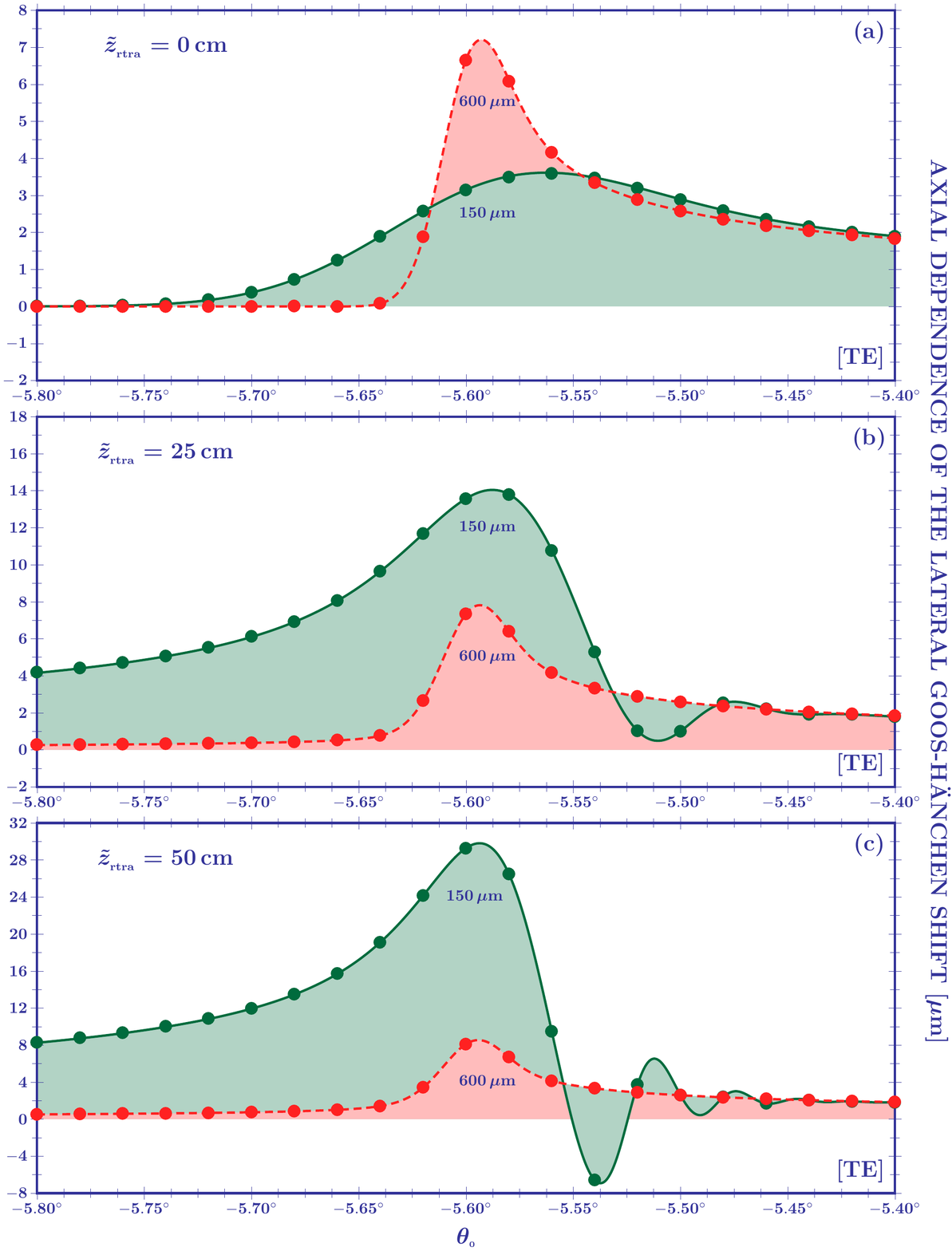}{The Composite Goos-H\"anchen shift for a TE-polarised beam with $\lambda=0.633\,\mu\mathrm{m}$ interacting with a borosilicate ($n=1.515$) prism. The solid green curves represent a minimum beam waist of $150\,\mu\mathrm{m}$ and the dashed red ones of $600\,\mu\mathrm{m}$. For a camera distance of 0 cm (a) the curves reproduce the known Goos-H\"anchen shift. As the camera moves to 25 cm (b) and then 50 cm (c) we can see an amplification of the shift as well as an oscillatory behaviour. Both effects are due to the symmetry breaking at the critical angle. The dots \cor{represent} the numerical data.}
respectively, with $\lambda=0.633\,\mu\mathrm{m}$ and beam waists of 150, 300, and 600 $\mu$m, transmitted through a borosilicate ($n=1.515$) prism. The numerical results obtained from using Eq. (\ref{eq:CGHErtra}) directly are in good agreement with the numerical analysis of maxima of intensity using the approximated expression in Eq. (\ref{eq:IntegratedErtra}). We can see in these figures how the behaviour of the shift changes when the camera is moved away from the transmitting interface. These axial effects depend upon the ratio $\tilde{z}_{_\mathrm{rtra}}/z_{_{R}}$, and, as expected, the farther away the measurement is taken the larger the amplification of the shift. For two measurements made at the same distance, as $\mathrm{w}_{\0}$ increases, the shift becomes less pronounced. This is due to the fact that a narrow angular distribution is more strongly centred at the incidence angle, and less susceptible to the symmetry breaking of the reflection coefficient.

In Figures 17 and 18 we can also see an oscillatory behaviour of the shift. This can be explained by the arguments of the Kummer confluent hypergeometric function, since this function is real for real-valued arguments and imaginary for imaginary-valued ones. For $\theta_{_{\0}}<\theta_{_{\mathrm{cri}}}$ the real part of the parameter $\ell$ is dominant and no oscillations are verified. As the incidence angle moves further into the critical zone, for $\theta_{_{\mathrm{cri}}} < \theta_{\0} < \theta_{_{\mathrm{cri}}} + \lambda/\mathrm{w}_{\0}$ and for an appropriate ratio $\tilde{z}_{_{\mathrm{rtra}}}/z_{_{\mathrm{R}}}$, real and imaginary parts become comparable, which generates the oscillations of the shift. In the Artmann zone the real part of $\ell$ becomes the main contribution once again, the beam recovers its symmetry, and the composite Goos-H\"anchen shift goes back to being the standard Goos-H\"anchen shift, which does not depend on the beam's waist. 

The composite Goos-H\"anchen shift has recently \cor{been} measured experimentally by Santana \emph{et al.} \cite{San2016}, using weak measurement techniques. The amplification of the Goos-H\"anchen shift was observed, but not the oscillatory behaviour, due to their choice of parameters. Figure 17(a-b) shows us that the oscillations start, for a beam width of $150\,\mu$m, at a camera distance of 25 cm. In the experiment, however, a $170\,\mu$m wide beam was employed, for a camera 20 to 25 cm away from the face of the prism, meaning that the camera was not far away enough for the size of the beam used.

\WideFigureSideCaption{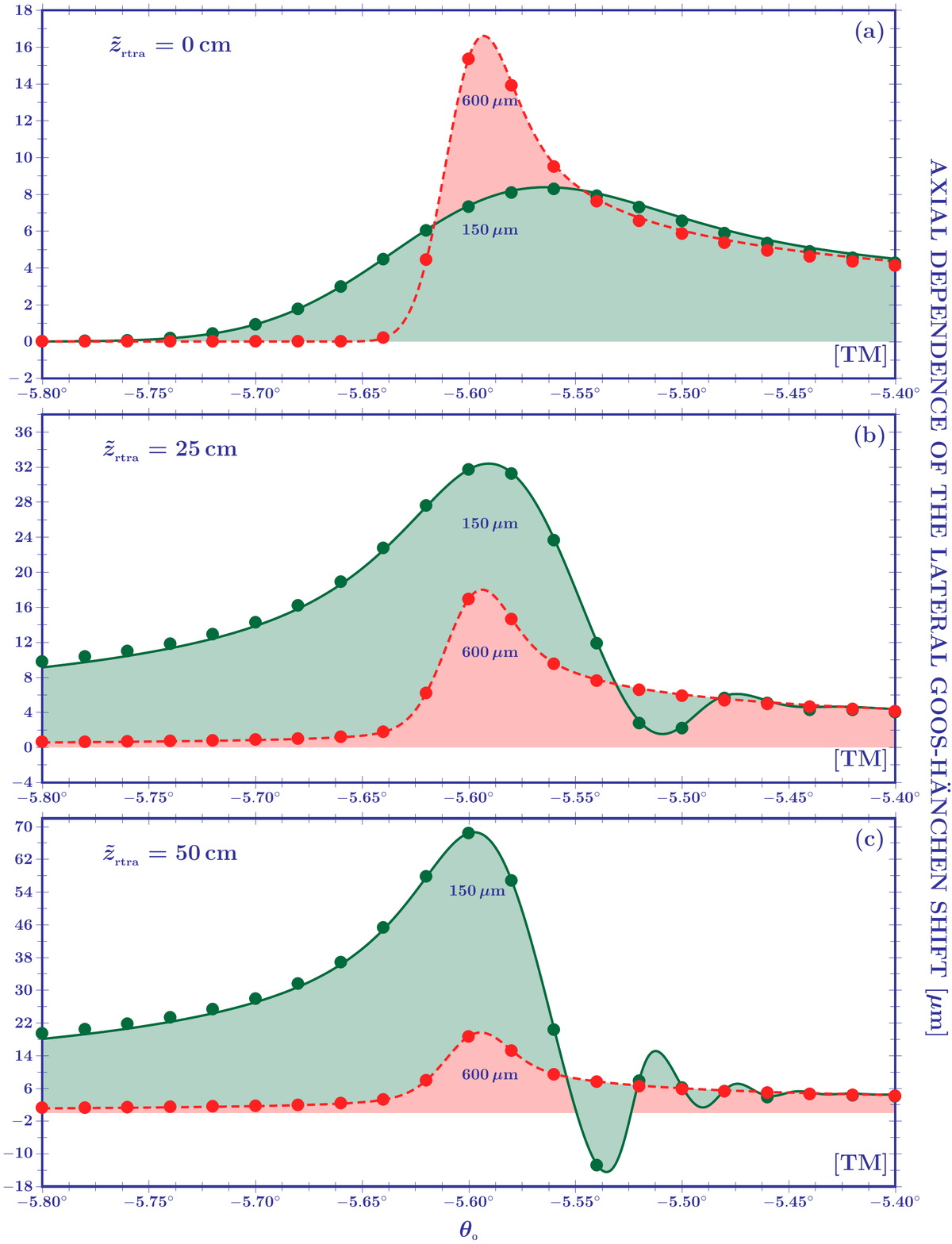}{The Composite Goos-H\"anchen shift for a TM-polarised beam with $\lambda=0.633\,\mu\mathrm{m}$ interacting with a borosilicate ($n=1.515$) prism. The solid green curves represent a minimum beam waist of $150\,\mu\mathrm{m}$ and the dashed red ones of $600\,\mu\mathrm{m}$. For a camera distance of 0 cm (a) the curves reproduce the known Goos-H\"anchen shift. As the camera moves to 25 cm (b) and then 50 cm (c) we can see an amplification of the shift as well as an oscillatory behaviour. Both effects are due to the symmetry breaking at the critical angle. The dots are the numerical data.}

\section{Optical weak measurements}
\label{sec8}

\subsection{Mathematical description}


Optical weak measurements are an indirect approach to enhance and measure small signal phenomena, taking advantage of the different responses an optical system has for different polarisation states. The mechanism behind it is analogous to the ones employed in PSA ellipsometry{\footnote{Polariser-Sample-Analyser ellipsometry}}\cite{Roth1945, Azz1987, Azz1991, Tom2005}, only, instead of focusing on interference patterns, optical weak measurements focus on intensity profiles. The experimental set-up can be seen in Figure 19: An optical beam out of a laser source passes through a polariser, becoming diagonally polarised. After this, it interacts with a dielectric structure where each of its components will be modified in a different way (since Fresnel's coefficients are polarisation dependent). Leaving the dielectric, light passes through a set of waveplates in order to remove any relative phase it may have acquired, meeting then a second polariser (analyser) which mixes both polarised fields' amplitudes and produces an intensity profile depending on the relative shift between the centres of the TE and TM beams. After following this path, light is finally collected by a camera. As we change the analyser's angle, the relative contribution of each polarisation component to the mixture also changes, and, consequently, so does the intensity profile. This change is observed in the relative position of intensity peaks, and the measurement of their relative distance allows an indirect measurement of the relative shift. Under appropriate conditions, this distance can overcome the real shifts, hence the status of this technique as an amplification technique.

\WideFigureSideCaption{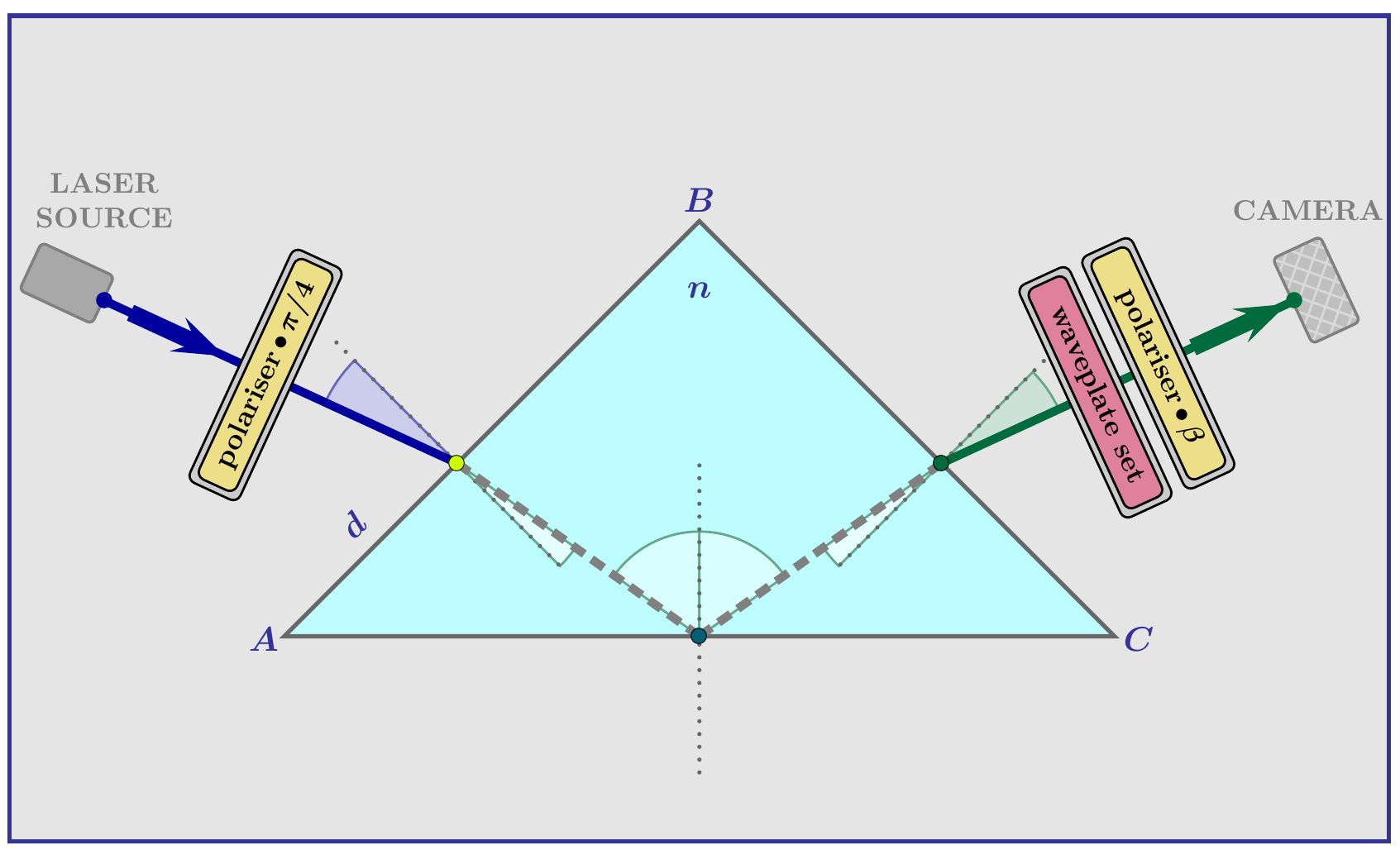}{The optical weak measurement system. Light comes out of a laser source and passes through a polariser which \cor{turn it into} a mixture of TE- and TM-polarised beams. It then interacts with the dielectric prism, where each of its components undergoes a different Goos-H\"anchen shift. Leaving the structure, light passes through a set of waveplates in order to have any relative phase it might have acquired removed, and then passes through a second polariser (analyser) which will \cor{mix} its components' amplitudes, creating a double peaked intensity profile. The beam is then finally collected by a camera.}

Let us begin our mathematical description of an optical weak measurement system. When a beam with an electric field $\mathbf{E}(x,y,z) = E_{_{x}}(x,y,z)\,\mathbf{\hat{x}} \,+\, E_{_{y}}(x,y,z)\,\mathbf{\hat{y}}$ passes through a polariser at an angle $\alpha$, the outgoing beam is modified by the Jones matrix associated to the polariser, which is given by
\begin{equation}
M_{_{\alpha}} = \left[\begin{array}{cc}
\cos^{\2}\alpha & \cos\alpha\,\sin\alpha \\
\cos\alpha\,\sin\alpha & \sin^{\2}\alpha
\end{array}\right],
\end{equation}
resulting in a field
\begin{eqnarray}
\mathbf{E_{_{\alpha}}}(x,y,z) &=& M_{_{\alpha}} \, E(x,y,z) {\nonumber}\\
&=& \left[ E_{_{x}}(x,y,z)\cos\alpha + E_{_{y}}(x,y,z)\sin\alpha\right]\,\left(\cos\alpha\,\mathbf{\hat{x}} + \sin\alpha\,\mathbf{\hat{y}}\right).
\end{eqnarray}
In the expression above, the $x-z$ plane is implied as being the plane of incidence, and the term $ E_{_{x}}(x,y,z)\cos\alpha + E_{_{y}}(x,y,z)\sin\alpha$ can be written as $E_{_{\alpha}}$. Upon interaction with the dielectric structure, however, we have that $E_{_{\mathrm{\alpha}}}\cos\alpha \rightarrow E_{_{\mathrm{tra}}}^{^{\mathrm{[TM]}}}$ and $E_{_{\mathrm{\alpha}}}\sin\alpha \rightarrow E_{_{\mathrm{tra}}}^{^{\mathrm{[TE]}}}$, where ``tra'' stands for ``transmitted''. The transmitted electric field becomes then
\begin{equation}
\mathbf{E}_{_{\mathrm{tra}}}(x,y,z)=E^{^{\mathrm{[TM]}}}_{_{\mathrm{tra}}}\,(x,y,z)\,\boldsymbol{\hat{x}}+E^{^{\mathrm{[TE]}}}_{_{\mathrm{tra}}}\,(x,y,z)\,\boldsymbol{\hat{y}}.
\end{equation}
The amplitudes $E_{_{\mathrm{tra}}}^{^{\mathrm{[TE,TM]}}}$ depend on the angle $\alpha$ of the first polariser. For simplicity, $\alpha$ is chosen to be $\pi/4$, so the incoming beam is an equal mixture of TE and TM-polarisation states. After the beam leaves the dielectric a set of waveplates removes any relative phase between $E_{_{\mathrm{tra}}}^{^{\mathrm{[TE,TM]}}}$ and it then passes through a second polariser, at an angle $\beta$, where these components are mixed. The electric field captured by the camera is then
\begin{eqnarray}
\label{eq:MbetaE}
\mathbf{E}_{_{\mathrm{cam}}}(x,y,z) &=& M_{_{\beta}}\,\mathbf{E}_{_{\mathrm{tra}}}(x,y,z) {\nonumber}\\
&=& \left[ E^{^{\mathrm{[TM]}}}_{_{\mathrm{tra}}}(x,y,z)\cos\beta + E^{^{\mathrm{[TE]}}}_{_{\mathrm{tra}}}(x,y,z)\sin\beta\right]\,\left(\cos\beta\,\mathbf{\hat{x}} + \sin\beta\,\mathbf{\hat{y}}\right).
\end{eqnarray}

In the previous Sections we have not considered diagonally polarised beams, which is why the vector form of their electric fields could be neglected, and we could focus only on the amplitudes $E^{^{\mathrm{[TE,TM]}}}_{_{\mathrm{tra}}}$. Besides, since all the phenomena we've studied were confined to the plane of incidence, we could also neglect the $y$ component of such fields. This spatial variable was displayed only for the completeness of the discussion, which had a general character, but, since it is also irrelevant for the treatment of optical weak measurements we are going to carry out, we will \cor{omit it}. Adapting Eq. (\ref{eq:MbetaE}) to the notation used in previous Sections, and considering that the electric field transmitted through the right face of the prism is our subject of interest, we have that
\begin{eqnarray}
\label{eq:OWMEcam}
\mathbf{E}_{_{\mathrm{cam}}}(\tilde{x}_{_{\mathrm{rtra}}},\tilde{z}_{_{\mathrm{rtra}}})=\left[\,E^{^{\mathrm{[TM]}}}_{_{\mathrm{rtra}}}(\tilde{x}_{_{\mathrm{rtra}}},\tilde{z}_{_{\mathrm{rtra}}})\cos\beta + \,E^{^{\mathrm{[TE]}}}_{_{\mathrm{rtra}}}(\tilde{x}_{_{\mathrm{rtra}}},\tilde{z}_{_{\mathrm{rtra}}})\sin\beta\,\right]\nonumber
\\
\times\left(\cos\beta\,\boldsymbol{\hat{\tilde{x}}_{_{\mathrm{rtra}}}}+\sin\beta\,\boldsymbol{\hat{\tilde{y}}_{_{\mathrm{rtra}}}}\right)\,\,,
\end{eqnarray}
where, in the Total Internal Reflection Regime, $E^{^{\mathrm{[TE,TM]}}}_{_{\mathrm{rtra}}}(\tilde{x}_{_{\mathrm{rtra}}},\tilde{z}_{_{\mathrm{rtra}}})$ can be approximated by the incident field function as
\begin{eqnarray}{\label{eq:Et}}
E^{^{\mathrm{[TE,TM]}}}_{_{\mathrm{rtra}}}(\tilde{x}_{_{\mathrm{rtra}}},\tilde{z}_{_{\mathrm{rtra}}}) \approx \left|t^{^{\mathrm{[TE,TM]}}}(\theta_{\0})\right|E_{_{\mathrm{inc}}}\left(\tilde{x}_{_{\mathrm{rtra}}} - d^{^{\mathrm{[TE,TM]}}}_{_{\mathrm{GH}}},\tilde{z}_{_{\mathrm{rtra}}} \right),
\end{eqnarray}
being $t^{^{\mathrm{[TE,TM]}}}(\theta_{\0}) = t^{^{\mathrm{[TE,TM]}}}_{_{\mathrm{left}}}(\theta_{\0})\,t^{^{\mathrm{[TE,TM]}}}_{_{\mathrm{right}}}(\theta_{\0}) $ and $d^{^{\mathrm{[TE,TM]}}}_{_{\mathrm{GH}}}$ the Goos-H\"anchen shift. \cor{Note} that, ordinarily, this field should have a phase
\[e^{i \left( \Phi_{_{\mathrm{rgeo}}}+\Phi^{^{\mathrm{[TE,TM]}}}_{_{\mathrm{GH}}} \right)},\]
but we are considering the use of a set of waveplates before the second polariser to remove relative phases, and so, this term is suppressed. The intensity associated to the electric field in Eq. (\ref{eq:OWMEcam}) is
\begin{equation}
{\label{eq:OWMIw}}
\mathcal{I}(\tilde{x}_{_{\mathrm{rtra}}},\tilde{z}_{_{\mathrm{rtra}}}) = \left|E^{^{\mathrm{[TM]}}}_{_\mathrm{rtra}}(\tilde{x}_{_{\mathrm{rtra}}},\tilde{z}_{_{\mathrm{rtra}}})\cos\beta + E^{^{\mathrm{[TE]}}}_{_\mathrm{rtra}}(\tilde{x}_{_{\mathrm{rtra}}},\tilde{z}_{_{\mathrm{rtra}}})\sin\beta\right|^{^2}.
\end{equation}
\cor{Note} that the amplitudes $E^{^{\mathrm{[TE,TM]}}}_{_{\mathrm{rtra}}}(\tilde{x}_{_{\mathrm{rtra}}},\tilde{z}_{_{\mathrm{rtra}}})$ are Gaussian functions that do not share their centres, which generates a double peaked curve, being the peaks' contributions to the intensity controlled by the second polariser's angle.

Let us define the following new variables:
\begin{subequations}
\begin{equation}
X = \tilde{x}_{_{\mathrm{rtra}}} - \frac{\displaystyle d^{^{\mathrm{[TE]}}}_{_{\mathrm{GH}}} + d^{^{\mathrm{[TM]}}}_{_{\mathrm{GH}}} }{\displaystyle 2},
\end{equation}
\begin{equation}
\Delta d_{_{\mathrm{GH}}} = d^{^{\mathrm{[TM]}}}_{_{\mathrm{GH}}} - d^{^{\mathrm{[TE]}}}_{_{\mathrm{GH}}},
\end{equation}
and
\begin{equation}
\tau = \frac{\displaystyle \left| t^{^{\mathrm{[TE]}}}(\theta_{\0}) \right|}{\displaystyle\left| t^{^{\mathrm{[TM]}}}(\theta_{\0}) \right|}.
\end{equation}
\end{subequations}
The relative Goos-H\"achen shift $\Delta d_{_{\mathrm{GH}}}$ (evaluated numerically) is plotted in Figure 20. The field intensity (\ref{eq:OWMIw}) can now be written as being proportional to
\begin{eqnarray}{\label{eq:Iwapp}}
\mathcal{I}  \propto \bigg{|}  \tau\,\tan\beta\,\exp\left[ - \left( \frac{\displaystyle X + \Delta d_{_{\mathrm{GH}}}/2}{\displaystyle \mathrm{w}(\tilde{z}_{_{\mathrm{rtra}}})} \right)^{\2}\right] + \exp\left[ - \left( \frac{\displaystyle X - \Delta d_{_{\mathrm{GH}}}/2}{\displaystyle \mathrm{w}(\tilde{z}_{_{\mathrm{rtra}}})} \right)^{\2} \right] \bigg{|}^{\2}.
\end{eqnarray}
A few approximations can be made to the equation above. \cor{First}, in the vicinity of the critical angle, the region we are interested \cor{in}, the transmission coefficients of the prism for the TE and TM polarisations are close enough for \cor{us} to consider $\tau \approx 1$. \cor{Second}, by choosing the angle of the second polariser as
\begin{equation}
\beta = \frac{3\,\pi}{4} + \Delta\epsilon,
\end{equation}
and considering a very small perturbation $\Delta\epsilon$ about the fixed angle $3\pi/4$, that is, $\Delta\epsilon \ll 1$, which is, incidentally, a crucial step in optical weak measurements, it can be shown that
\begin{equation}
\tan\beta \, \approx \, 2\Delta\epsilon-1\,\,.
\end{equation}
These approximations allow us to express the electric field intensity collected by the camera as
\begin{eqnarray}
\mathcal{I} &\propto& \bigg{|}\,(2\Delta\epsilon-1)\,\left(1 - \frac{\displaystyle X\,\Delta d_{_{\mathrm{GH}}}}{\mathrm{w}^{\2}(\tilde{z}_{_{\mathrm{rtra}}})} \right) + \left(1 + \frac{\displaystyle X\,\Delta d_{_{\mathrm{GH}}}}{\displaystyle \mathrm{w}^{\2}(\tilde{z}_{_{\mathrm{rtra}}})} \right) \bigg{|}^{\2} \,  \mathrm{exp}\left[-\frac{2\,X^{\2}}{|\mathrm{w}(\tilde{z}_{_{\mathrm{rtra}}})|^{\2}} \right]\,,
\end{eqnarray}

\WideFigureSideCaption{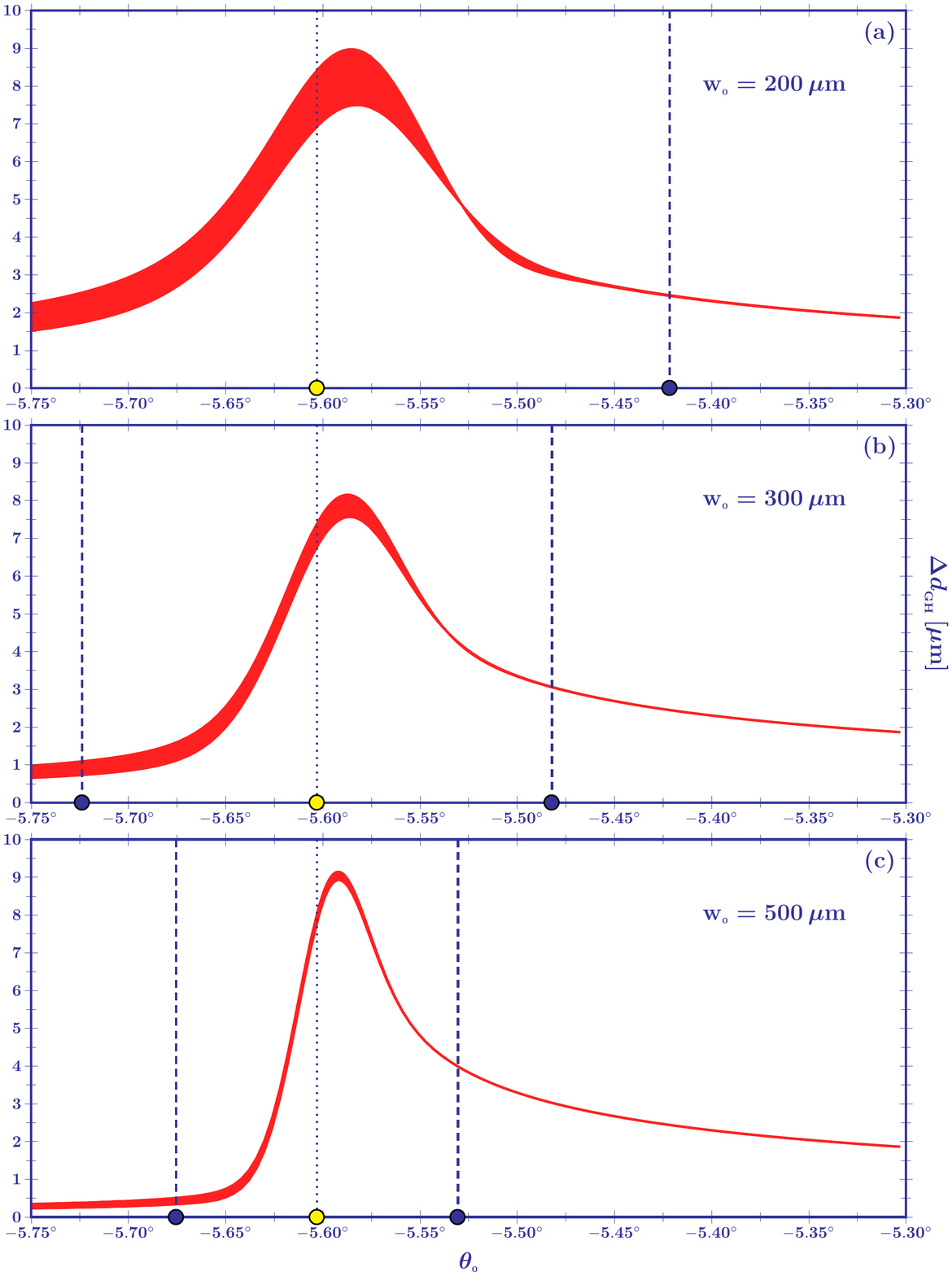}{The relative Goos-H\"anchen shift for a beam with $\lambda = 0.633\, \mu\mathrm{m}$ and minimum beam waist (a) $200\,\mu\mathrm{m}$, (b) $300\,\mu\mathrm{m}$, and (c) $600\,\mu\mathrm{m}$. The interaction is considered to be with a borosilicate ($n=1.515$) prism, and the camera distance from the right face of the prism is considered to \cor{range} between 10 and 15 cm. \cor{These limits correspond} to the lower and top sides, respectively, of the each curve's width. This data is the numerical evaluation of the shift of the maximum of intensity of the beam.}
which can be further simplified to
\begin{eqnarray}
\label{eq:finalIwnp}
\mathcal{I}(\Delta\epsilon,X) \propto \left[ \Delta\epsilon +  \frac{\displaystyle \Delta d_{_{\mathrm{GH}}}}{\displaystyle |\mathrm{w}(\tilde{z}_{_{\mathrm{rtra}}})|^{^2}}\,X\right]^{^{2}}\,\mathrm{exp}\left[-\frac{\displaystyle 2\,X^{\2}}{\displaystyle |\mathrm{w}(\tilde{z}_{_{\mathrm{rtra}}})|^{^{2}}} \right]\,.
\end{eqnarray}

\subsection{The axial dependence of measurements}

The peaks' positions of Eq. (\ref{eq:finalIwnp}) are controlled by the polariser parameter $\Delta \epsilon$. For $\Delta\epsilon=0$, the intensity profile has a minimum at $X_{_{\mathrm{min}}}=0$ and two symmetric peaks at $X_{_{\mathrm{max}}} = \pm |\mathrm{w}(\tilde{z}_{_{\mathrm{rtra}}})|/\sqrt{2}$. For a non-zero $\Delta\epsilon$, the minimum is at

\begin{equation}
X_{_\mathrm{min}}(\Delta\epsilon) = -\frac{\displaystyle \Delta\epsilon}{\displaystyle \Delta d_{_{\mathrm{GH}}}}\,|\mathrm{w}(\tilde{z}_{_{\mathrm{rtra}}})|^{^2}
\end{equation}
and the two maximums are at
\begin{equation}
X^{^{\pm}}_{_{\mathrm{max}}}(\Delta\epsilon) = \frac{\displaystyle -\Delta\epsilon \pm \sqrt{(\Delta\epsilon)^{^2}+ 2 \left[\,\Delta  d_{_\mathrm{GH}}/|\mathrm{w}(\tilde{z}_{_{\mathrm{rtra}}})| \,\right]^{^{2}}}}{\displaystyle 2\Delta d_{_\mathrm{GH}}}\,|\mathrm{w}(\tilde{z}_{_{\mathrm{rtra}}})|^{\2}.
\end{equation}
For a positive $\Delta\epsilon$ (which corresponds to an anticlockwise rotation), the main peak is centred at $X^{^{+}}_{_{\mathrm{max}}}(|\Delta\epsilon|)$, while for a negative $\Delta\epsilon$ (a clockwise rotation), the main peak's centre is at $X^{^{-}}_{_{\mathrm{max}}}(-|\Delta\epsilon|)$. The distance between such peaks is then
\begin{eqnarray}
\label{eq:DXmax}
\Delta X_{_{\mathrm{max}}} &=& X_{_{\mathrm{max}}}^{^{+}}(|\Delta\epsilon|) - X_{_{\mathrm{max}}}^{^{-}}(-|\Delta\epsilon|)\nonumber
\\
&=& \frac{\displaystyle -|\Delta\epsilon| + \sqrt{|\Delta\epsilon|^{^2} + 2 \left[\,\Delta d_{_\mathrm{GH}}/|\mathrm{w}(\tilde{z}_{_{\mathrm{rtra}}})| \,\right]^{^2}}}{\displaystyle \Delta d_{_\mathrm{GH}}}\,|\mathrm{w}(\tilde{z}_{_{\mathrm{rtra}}})|^{^2}.
\end{eqnarray}
Figure 21 displays this distance for the same parameters of Figure 20 and for different values of $\Delta\epsilon$. We can see, in comparison to Figure 20, that the scale of the peaks' distance is multiplied b a factor 50, and that the smaller the $\Delta\epsilon$ parameter, the better the amplification. Besides, in the critical region a strong axial dependence is still verified.

\WideFigureSideCaption{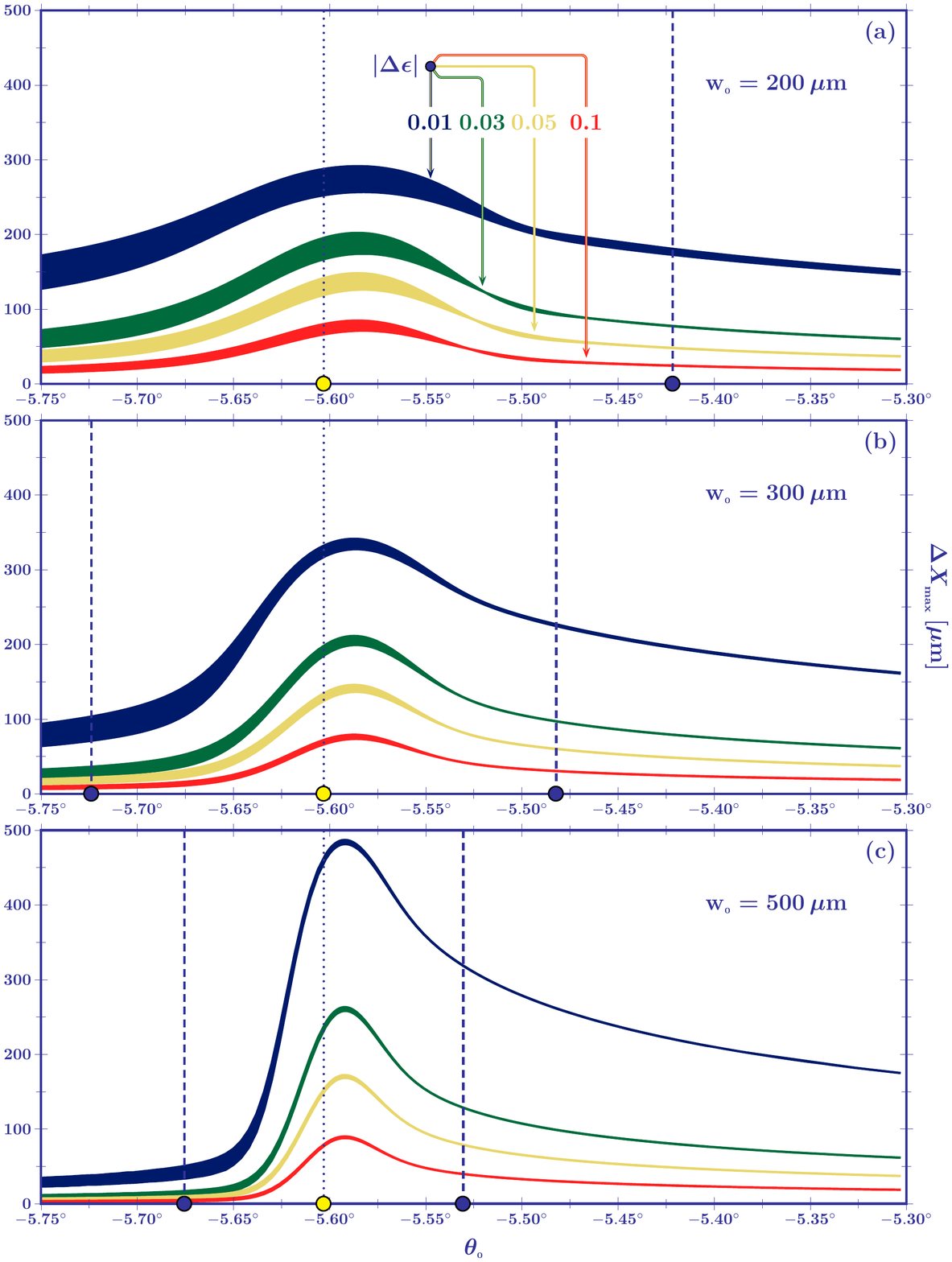}{The optical weak measurement amplification of the relative Goos-H\"anchen shift for a beam with $\lambda = 0.633\, \mu\mathrm{m}$ and minimum beam waist (a) $200\,\mu\mathrm{m}$, (b) $300\,\mu\mathrm{m}$, and (c) $600\,\mu\mathrm{m}$. The interaction is considered to be with a borosilicate ($n=1.515$) prism, and the camera distance from the right face of the prism is considered to \cor{range} between 10 and 15 cm. Different values of the polariser parameter $\Delta\epsilon$ were used. We can see that the scale of this amplification, in comparison to the direct measurement of the Goos-H\"anchen shift, is a factor \cor{of} 50.}
We can see that for $0\leq \Delta\epsilon \leq \Delta d_{_{\mathrm{GH}}}/|\mathrm{w}(\tilde{z}_{_{\mathrm{rtra}}})|$, the peaks' distance fall on the range

\[ \sqrt{2}\,|\mathrm{w}(\tilde{z}_{_{\mathrm{rtra}}})| \leq \Delta X_{_{\mathrm{max}}} \leq (\sqrt{3}-1) \,|\mathrm{w}(\tilde{z}_{_{\mathrm{rtra}}})|, \]
showing that as we increase $\Delta\epsilon$ the distance between the peaks decreases. For
\begin{equation}
\label{eq:Decond}
|\Delta\epsilon| \gg \Delta d_{_{\mathrm{max}}} / |\mathrm{w}(\tilde{z}_{_{\mathrm{rtra}}})|,
\end{equation}
we have that
\begin{equation}
\label{eq:DX}
\Delta X_{_{\mathrm{max}}} \approx \frac{\Delta d_{_{\mathrm{GH}}}}{|\Delta\epsilon|},
\end{equation}
\cor{meaning}, the relative Goos-H\"anchen shift is amplified by a factor $|\Delta\epsilon|^{^{-1}}$. For incidence angles far from the critical angle, the condition (\ref{eq:Decond}) is easily satisfied. In the critical region, however, the Goos-H\"anchen shift is naturally amplified, the condition (\ref{eq:Decond}) is not satisfied, and, consequently, the weak measurement amplification is no longer proportional to $|\Delta\epsilon|^{^{-1}}$.  Increasing the beam waist $\mathrm{w}_{\0}$ removes the axial dependence of the amplification because it makes the beam's collimation around the incidence angle stronger. To the right of the critical region this axial dependence is no longer verified because the beam's symmetry is recovered.

From an experimental \cor{point of view}, it makes more sense to invert Eq. (\ref{eq:DXmax}) to express the relative Goos-H\"anchen shift as a function of the distance between intensity peaks, which is what the experimentalist will actually measure. In doing so we obtain
\begin{equation}
\Delta d_{_{\mathrm{GH}}} = \frac{2\,|\Delta \epsilon| \,|\mathrm{w}(\tilde{z}_{_{\mathrm{rtra}}})|^{^{\2}}}{2\,|\mathrm{w}(\tilde{z}_{_{\mathrm{rtra}}})|^{^2} - \Delta X^{^{2}}_{_{\mathrm{max}}}}\,\Delta X_{_{\mathrm{max}}}.
\end{equation}

\subsection{The effect of the Goos-H\"anchen phase}

In the last Sections we studied the amplification of the Goos-H\"anchen shift via an optical weak measurement set-up. \cor{When} establishing such a set-up, a crucial step was the removing of the relative phase between the electric field's components of the beam. This relative phase is the relative Goos-H\"anchen phase, acquired in the interaction with the dielectric structure. As we will see, this phase has a destructive effect on measurements, and the study of such \cor{an} effect is interesting not only for the completeness of the theory, but it is also important from a pragmatic \cor{point of view} as an efficiency test for phase removal techniques. A waveplate, for instance, that leaves a residual phase may compromise experimental results, and the quantitative knowledge of how Goos-H\"anchen phases affect weak measurements may help us to verify when they are completely removed.

Each internal reflection prompts a new shift of the beam, in such way that, for $N_{_\mathrm{r}}$ reflections the total shift becomes $N_{_{\mathrm{r}}}\,d^{^{\mathrm{[TE,TM]}}}_{_{\mathrm{GH}}}$. As the Goos-H\"anchen shift and the Goos-H\"anchen phase are related through Eq. (\ref{eq:PhsShft}), we see that the same factor is present in the final phase acquired by the beam, that is, Eq. (\ref{eq:GHphs}) becomes
\begin{equation}
\label{eq:NGHphs}
\left\{ \Phi_{_{\mathrm{GH}}}^{^{[\mathrm{TE}]}}, \Phi_{_{\mathrm{GH}}}^{^{[\mathrm{TM}]}} \right\} = -2 \, N_{_{\mathrm{r}}}\, \left\{ \arctan\left[\frac{\displaystyle \sqrt{n^{\2}\sin^{\2}\varphi-1}}{n\cos\varphi}\right],\arctan\left[\frac{\displaystyle n\,\sqrt{n^{\2}\sin^{\2}\varphi-1}}{\cos\varphi}\right] \right\}.
\end{equation}
Since this phase is a function of the incidence angle and of the number of reflections, it can be controlled by changing this angle and by changing the length of the dielectric, which is why we will in this Section consider a multiple-reflection structure, as the one presented in Figure 8. For simplicity, we will consider a structure composed of an even number of right angle triangular prisms, in such way that the incoming and outgoing beams are parallel. In this set-up, Eq. (\ref{eq:rgeophs}) for the geometrical phase of the beam transmitted through the right face of the structure becomes
\begin{equation}
\label{eq:Nphirgeo}
 \Phi_{_{\mathrm{rgeo}}} =\frac{N_{_{\mathrm{r}}}}{2} \,k\, \left(\sqrt{2}\,n\,\cos\varphi + n\cos\psi-\cos\theta\,\right) \overline{AB},
 \end{equation}
and the geometrical shift given by Eq. (\ref{eq:xrgeo})
\begin{eqnarray}
\label{eq:Nxrgeo}
x_{_{\mathrm{rgeo}}} = \frac{N_{_{\mathrm{rgeo}}}}{2}\,\Big{(}\,\cos\theta-\sin\theta+2\,\tan\psi\,\cos\theta\,\Big{)}\,\overline{AB}\,.
\end{eqnarray}

The beam intensity (\ref{eq:Iwapp}) can then be written as
\begin{eqnarray}
\label{eq:Iphs}
\mathcal{I}  \propto \bigg{|}  \tau\,\tan\beta\,\exp\left[ - \left( \frac{\displaystyle X + \Delta d_{_{\mathrm{GH}}}/2}{\displaystyle \mathrm{w}(\tilde{z}_{_{\mathrm{rtra}}})} \right)^{\2} + i\Delta\Phi_{_{\mathrm{GH}}}\right] + \exp\left[ - \left( \frac{\displaystyle X - \Delta d_{_{\mathrm{GH}}}/2}{\displaystyle \mathrm{w}(\tilde{z}_{_{\mathrm{rtra}}})} \right)^{\2} \right] \bigg{|}^{\2},
\end{eqnarray}
where
\begin{equation}
\Delta\Phi_{_{\mathrm{GH}}}= \Phi^{^{\mathrm{[TE]}}}-\Phi^{^{\mathrm{[TM]}}} = 2 \, N_{_{\mathrm{r}}} \arctan\left[\frac{\displaystyle \sqrt{n^{\2} \, \sin^{\2}\varphi-1}}{\displaystyle n \, \sin\varphi\,\tan\varphi}\right] \,
\end{equation}
is the relative Goos-H\"anchen phase. \cor{Note} that, as said before, we are considering an even number of prisms in our dielectric chain, see Figure 18, and so the right and left faces of our structure are parallel, and so, consequently, are \cor{also} the incoming and outgoing beams. So the definition of the system $\tilde{x}_{_{\mathrm{rtra}}}-\tilde{z}_{_{\mathrm{rtra}}}$, given by Eq. (\ref{eq:ytildertra}), must be updated to
\begin{subequations}
\label{eq:ytildertraN}
\begin{equation}
\tilde{x}_{_{\mathrm{rtra}}} = x_{_{\mathrm{inc}}} + \Phi_{_{\mathrm{rgeo}}}^{\prime}(\theta_{\0})/k = x_{_{\mathrm{rtra}}} - x_{_{\mathrm{rgeo}}},
\end{equation}
and
\begin{equation}
\tilde{z}_{_{\mathrm{rtra}}} = z_{_{\mathrm{inc}}} + \Phi_{_{\mathrm{rgeo}}}^{\prime\prime}({\theta_{\0}})/k = z_{_{\mathrm{rtra}}} - z_{_{\mathrm{rgeo}}},
\end{equation}
\end{subequations}
according to Eqs. (\ref{eq:Nphirgeo}) and (\ref{eq:Nxrgeo}). Following then the approximations made for the case without phase, the intensity (\ref{eq:Iphs}) can be written as
\begin{eqnarray}
\label{eq:finalIw}
\mathcal{I}(\Delta\epsilon,X) \propto \left[ \left(\,\Delta\epsilon  +  \frac{\displaystyle \Delta d_{_{\mathrm{GH}}}}{\displaystyle |\mathrm{w}(\tilde{z}_{_{\mathrm{rtra}}})|^{^2}}\,X\right)^{^2}  +  \sin^{\2}\left( \frac{\Delta\Phi_{_{\mathrm{GH}}}}{2}\right)\right]\,\mathrm{exp}\left[-\frac{\displaystyle 2\,X^{\2}}{\displaystyle |\mathrm{w}(\tilde{z}_{_{\mathrm{rtra}}})|^{^{2}}} \right]\,.
\end{eqnarray}
This intensity profile has its peaks given by the equation
\begin{equation}
\label{eq:inPhs}
\left(\Delta\epsilon+\frac{\Delta d_{_{\mathrm{GH}}} \, X}{\mathrm{w}^{\2}(\tilde{z}_{_\mathrm{rtra}})}\right)\left(\Delta d_{_{\mathrm{GH}}} - 2\Delta\epsilon \, X - \frac{2\Delta d_{_{\mathrm{GH}}} \, X^{\2}}{\mathrm{w}^{2}(\tilde{z}_{\mathrm{rtra}})}\right) - 2\,\sin^{\2}\left(\frac{\Delta\Phi_{_{\mathrm{GH}}}}{\displaystyle 2}\right)\, X  =  0.
\end{equation}
If $\Delta\epsilon\,=\,0$ the equation above returns
\begin{equation}
X\,\left[\frac{\Delta d^{\2}_{_{\mathrm{GH}}}}{\mathrm{w}^{\2}(\tilde{z}_{\mathrm{rtra}})}- \frac{2\,\Delta d^{\2}_{_{\mathrm{GH}}} \, X^{\2}}{\mathrm{w}^{\4}(\tilde{z}_{_{\mathrm{rtra}}})}-2\,\sin^{\2}\left(\frac{\displaystyle \Delta\Phi_{_{\mathrm{GH}}}}{\displaystyle 2}\right)\right] = 0,
\end{equation}
providing an intensity's minimum at $X=0$ and two symmetrical maxima at
\begin{equation}
X_{_{\mathrm{max}}} = \pm\frac{\displaystyle \mathrm{w}(\tilde{z}_{_{\mathrm{rtra}}})}{\displaystyle \sqrt{2}}\,\sqrt{1-\left[\frac{\displaystyle 2\,\mathrm{w}(\tilde{z}_{_{\mathrm{rtra}}})}{\displaystyle \Delta d_{_{\mathrm{GH}}}}\,\sin\left(\frac{\displaystyle \Delta\Phi_{_{\mathrm{GH}}}}{\displaystyle 2}\right)\right]^{^{2}}},
\end{equation}
which returns the expected value if the phase is made zero. For a non-zero $\Delta\epsilon$, Eq. (\ref{eq:inPhs}) can be approximated to
\begin{equation}
2\,\Delta\epsilon\, \frac{\Delta d_{_{\mathrm{GH}}} \, X^{\2}}{\mathrm{w}^{\2}(\tilde{z}_{_{\mathrm{rtra}}})} + \left[\Delta\epsilon^{\2}-\frac{\Delta d^{\2}_{_{\mathrm{GH}}}}{2\,\mathrm{w}^{\2}(\tilde{z}_{_{\mathrm{rtra}}})}+\sin^{\2}\left(\frac{\displaystyle \Delta\Phi_{_{\mathrm{GH}}}}{\displaystyle 2}\right)\right]\,X-\frac{\displaystyle 1}{\displaystyle 2}\,\Delta\epsilon\,\Delta d_{_{\mathrm{GH}}} = 0
\end{equation}
with maxima at
\begin{equation}
\begin{split}
X_{_{\mathrm{max}}}^{\pm} = \frac{\displaystyle \Delta\epsilon\,\mathrm{w}^{\2}(\tilde{z}_{_\mathrm{rtra}})}{\displaystyle 4\,\Delta d_{_{\mathrm{GH}}}}\left[-\left(1-\frac{\displaystyle \Delta d^{\2}_{_{\mathrm{GH}}}}{\displaystyle 2\,\mathrm{w}^{\2}(\tilde{z}_{_{\mathrm{rtra}}})\Delta\epsilon^{\2}}+\frac{\displaystyle \sin^{\2}\left( \Delta\Phi_{_{\mathrm{GH}}}/2\right)}{\displaystyle \Delta\epsilon^{\2}}\right)\right.\pm \\
\left.\sqrt{\left(1-\frac{\displaystyle \Delta d^{\2}_{_{\mathrm{GH}}}}{\displaystyle 2\,\mathrm{w}^{\2}(\tilde{z}_{_{\mathrm{rtra}}})\Delta\epsilon^{\2}}+\frac{\displaystyle \sin^{\2}\left( \Delta\Phi_{_{\mathrm{GH}}}/2\right)}{\Delta\epsilon^{\2}}\right)^{2}+\frac{\displaystyle 4\,\Delta d_{_{\mathrm{GH}}}^{\2}}{\displaystyle \mathrm{w}^{\2}(\tilde{z}_{_{\mathrm{rtra}}})\,\Delta\epsilon^{\2}}}\right].
\end{split}
\end{equation}
Now, let us consider that the incidence angle is far greater than the critical one, that is, the approximation $|\Delta\epsilon|>>\Delta d_{_\mathrm{GH}}/\mathrm{w}(\tilde{z}_{_{\mathrm{rtra}}})$ is valid, as we did for the case without phase. This gives us the distance between peaks as
\begin{equation}
\label{eq:DXphsz}
\Delta X_{_{\mathrm{max}}} \,\approx\, \frac{\displaystyle \Delta d_{_{\mathrm{GH}}}}{\displaystyle |\Delta\epsilon|}\frac{\displaystyle 1}{\displaystyle 1+\left[\frac{\displaystyle \sin(\Delta\Phi_{_{\mathrm{GH}}}/2)}{\displaystyle |\Delta\epsilon|}\right]^{\2}}.
\end{equation}
Note that removing the phase the result (\ref{eq:DX}) is reconstructed and the shift is amplified by a factor $|\Delta\epsilon|$. It is, however, possible to obtain this amplification by controlling the phase, without removing it. We can see that for
\begin{equation}
\Delta\Phi_{_{\mathrm{GH}}} \, = \, 2\,m\,\pi \,\,\,\,\,\,\,\,\,\,\,\,\,\,\,\,\,\,\,\, \mathrm{for} \,\,\,\,\,\,\,\,\,\,\,\,\,\,\,\,\,\,\,\, m \,=\, 0,1,2...
\end{equation}
this aim is achieved, but this happens only for particular combinations of $N_{_{\mathrm{r}}}$ and $\theta_{\0}$, as can be seen in Figure 22. In this Figure the sinusoidal function is plotted as a function of the incidence angle for different numbers of internal reflections.

\WideFigureSideCaption{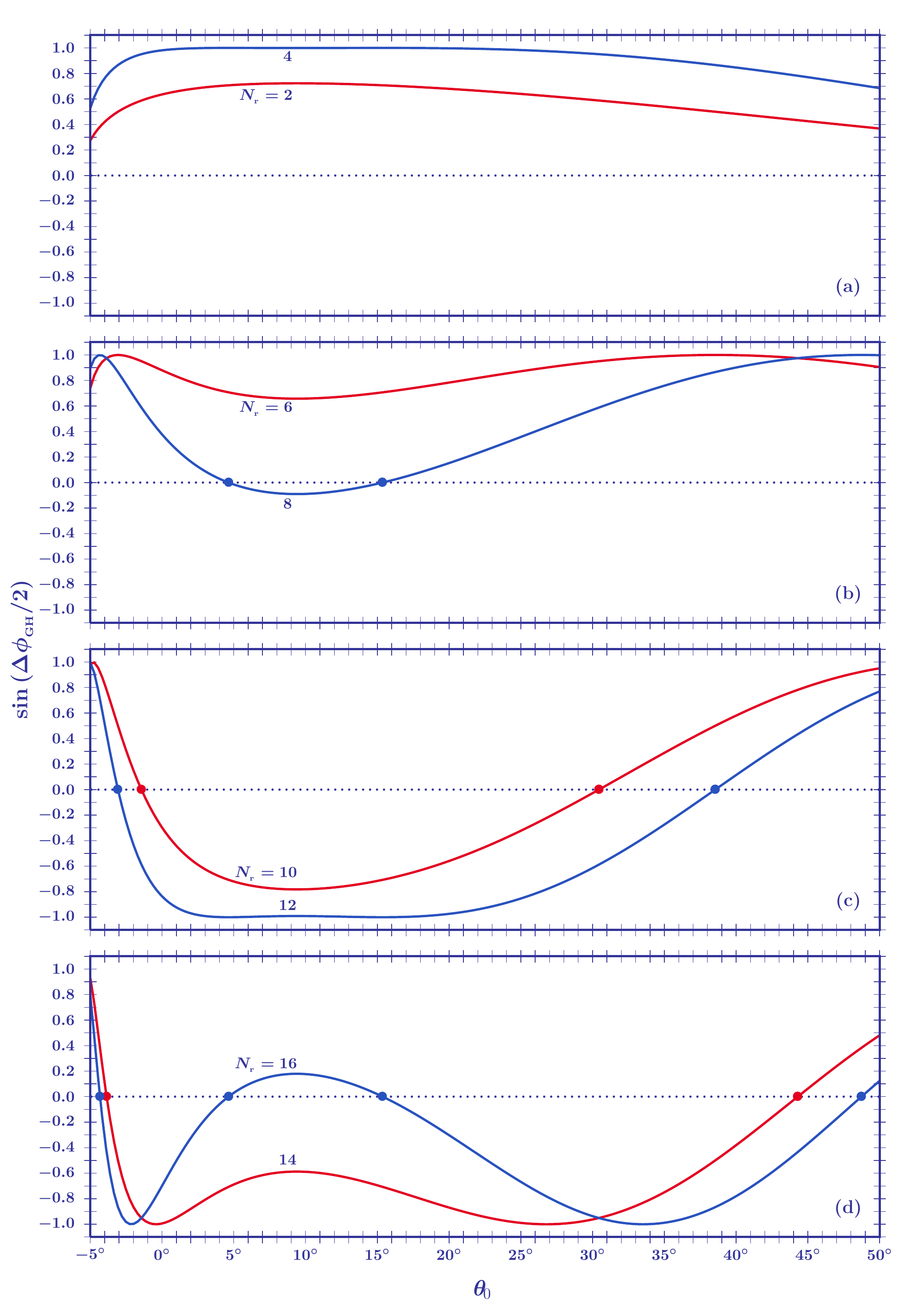}{The sine of half the relative Goos-H\"anchen phase for a borosilicate ($n=1.515$) dielectric structure, for different numbers of total internal reflections $N_{_{\mathrm{r}}}$. Starting at $N_{_{\mathrm{r}}}=8$ it is always possible to find incidence angles for which this sine function is zero and the results of optical weak measurements without phase can be reconstructed. As $N_{_{\mathrm{r}}}$ increases so does the number of available zeros.}
\noindent
We can see that there is a minimum number of internal reflections needed to trigger the reconstruction of Eq. (\ref{eq:DX}). For a borosilicate prism, for instance, for $N_{_{\mathrm{r}}}<8$, $\Delta\Phi_{_{\mathrm{GH}}}$ is never an integer multiple of $2\pi$, but as the number of reflections increases, this result becomes accessible more often, starting with two angles for $N_{_{\mathrm{r}}}=8$.

The effect of the Goos-H\"{a}nchen phase on $\Delta X_{_{\mathrm{max}}}$ is studied for $N_{_{\mathrm{r}}}=8$ and $N_{_{\mathrm{r}}}=16$ in Figure 23, where the dashed lines represents the weak measurement amplification without phase. We can see that Eq. (\ref{eq:DXphsz}) is virtually null for every incidence angle except for the ones around which $\sin(\Delta\Phi_{_{\mathrm{GH}}}/2)=0$. For incidence precisely at the angles for which this result is obtained the weak measurement amplification without phase is obtained. These results are of a pragmatic interest for the experimentalist. As Figure 21 shows us, the separation between intensity peaks may be nearly 0.5 mm long, which in comparison to the direct measurement of the Goos-H\"anchen shift is a huge amplification. A Goos-H\"anchen phase that is not completely removed, however, will give the appearance of only one peak \cor{existing} in a fixed position, or, at least, in a scenario not as dramatic, will provide a non-optimal amplification.

\WideFigureSideCaption{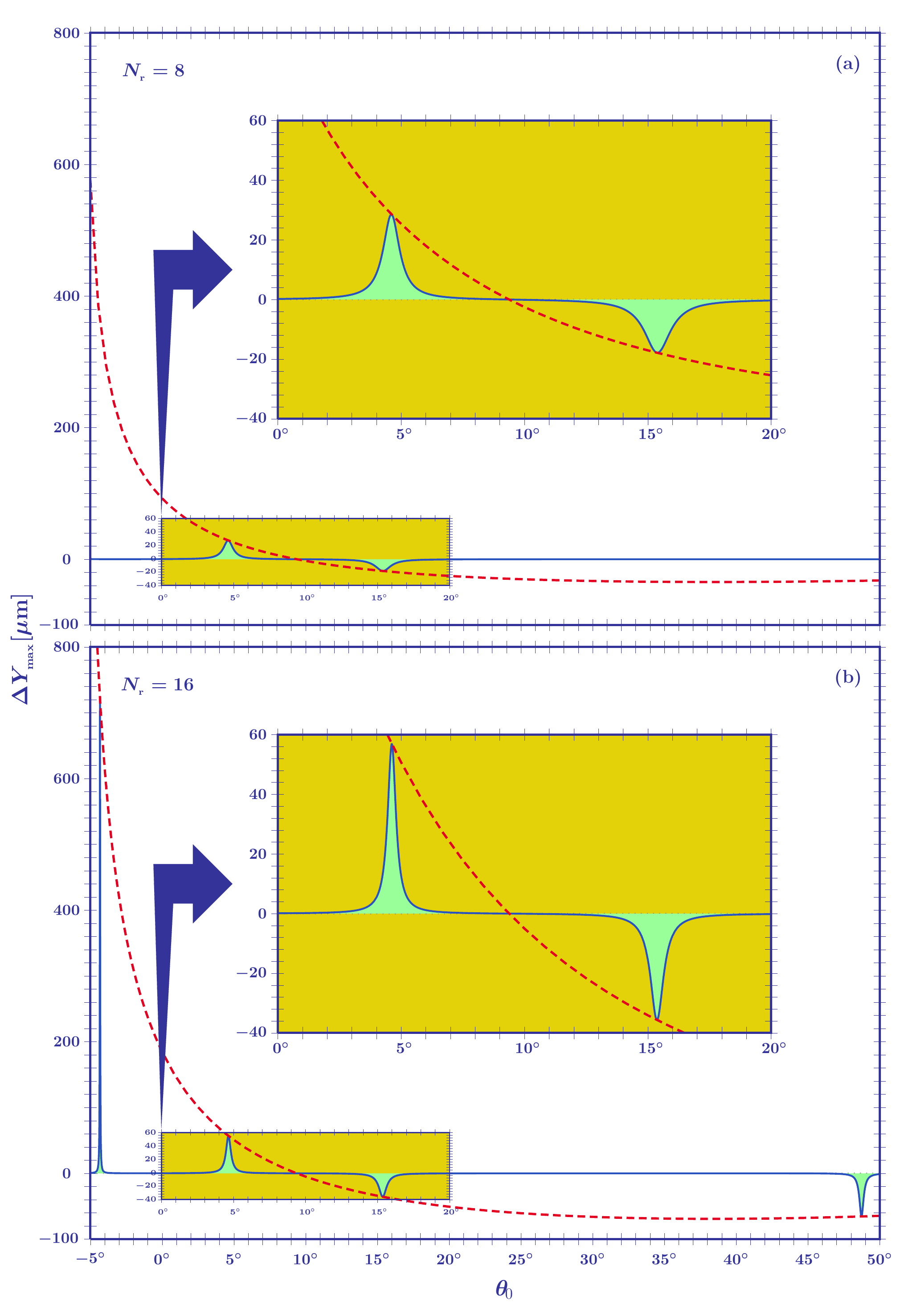}
{The optical weak measurement amplification of the Goos-H\"anchen shift without the removal of the relative Goos-H\"anchen phase for a borosilicate ($n=1.515$) structure allowing (a) $8$ and (b) 16 total internal reflections. The curves represent a beam with $\lambda = 0.633\,\mu$m and $\mathrm{w}_{\0}= 1$ mm. The phase renders the distance between intensity peaks virtually null, except around the angles for which $\sin(\Delta\Phi_{_{\mathrm{GH}}}/2) = 0$. Precisely for such angles the amplification obtained is the same given by weak measurements without phase, that is, $\Delta X_{_{\mathrm{max}}} \approx \Delta d_{_{\mathrm{GH}}}/|\Delta\epsilon|$. The $|\Delta\epsilon|$ factor used was $1^{\circ}$.}

\section{Weak measurements vs. direct measurements}
\label{sec9}

In Section \ref{sec6}, we studied the angular deviations a light beam suffers when it interacts with a dielectric structure outside of critical incidence. Even though the angular deviations from the path predicted by Geometrical Optics acquired larger values in the vicinities of the Brewster and critical angles (for the beam transmitted through the right face of the right angle triangular prism we have been considering) even in these regions the effect was fairly small. Optical weak measurements \cor{present} themselves then as a possible amplification solution, as was the case for the lateral Goos-H\"anchen shift. In this Section, we will develop the Optical Weak Measurement theory for angular deviations, which requires simple modifications from what we did in the last Section, and \cor{we} will compare the results obtained with the direct measurements. In fact, in a way, angular deviations, as we will see, present a more direct problem to both the theorist and the experimentalist, since no phase is acquired by the beam outside of the Total Internal Reflection regime.

The optical system we will consider is the same presented in Figure 19, with the exclusion of the waveplates set. The electric field intensity collected by the camera is still given by Eq. (\ref{eq:OWMIw}),
\begin{equation*}
\mathcal{I}(\tilde{x}_{_{\mathrm{rtra}}},\tilde{z}_{_{\mathrm{rtra}}}) = \left|E^{^{\mathrm{[TM]}}}_{_\mathrm{rtra}}(\tilde{x}_{_{\mathrm{rtra}}},\tilde{z}_{_{\mathrm{rtra}}})\cos\beta + E^{^{\mathrm{[TE]}}}_{_\mathrm{rtra}}(\tilde{x}_{_{\mathrm{rtra}}},\tilde{z}_{_{\mathrm{rtra}}})\sin\beta\right|^{^2},
\end{equation*}
but the electric fields expressions are now approximated by
\begin{equation*}
E_{_{\mathrm{rtra}}}^{^{\mathrm{[TE,TM]}}}\,
\propto \,\,\,  r^{^{\mathrm{[TE,TM]}}}_{_{\mathrm{lower}}}(\theta_{\0}) \,\,
\mathrm{exp}\left\{\, -\,\frac{\displaystyle \left[\,\tilde{x}_{_{\mathrm{rtra}}} -\,\, \alpha_{_{\mathrm{rtra}}}^{^{\mathrm{[TE,TM]}}}(\theta_{\0}) \,\,
\tilde{z}_{_{\mathrm{rtra}}}\right]^{^2}}{\displaystyle {\mathrm{w}}^{\2}(\widetilde{z}_{_{\mathrm{rtra}}})} \right\},
\end{equation*}
where we can see that the displacement of the centre of the Gaussian function is determined by the angular coefficient $\alpha_{_{\mathrm{rtra}}}^{^{\mathrm{[TE,TM]}}}$. Defining then the new variables
\begin{subequations}
\begin{equation}
X \, = \, \widetilde{x}_{_{\mathrm{rtra}}}- \,\frac{  \alpha_{_{\mathrm{rtra}}}^{^{\mathrm{[TE]}}}(\theta_{\0}) + \, \alpha_{_{\mathrm{rtra}}}^{^{\mathrm {[TM]}}}(\theta_{\0}) }{2} \,\,\, \tilde{z}_{_{\mathrm{rtra}}} \,\,,
\end{equation}
\begin{equation}
\Delta\alpha_{_{\mathrm{rtra}}}(\theta_{\0}) \, = \,\alpha_{_{\mathrm{rtra}}}^{^{\mathrm{[TM]}}}(\theta_{\0}) - \, \alpha_{_{\mathrm{rtra}}}^{^{\mathrm {[TE]}}} (\theta_{\0})\,\,,
\end{equation}
and
\begin{equation}
\tau(\theta_{\0}) = r^{^{\mathrm{[TM]}}}_{_{\mathrm{lower}}}(\theta_{\0})\,/\,r^{^{\mathrm{[TE]}}}_{_{\mathrm{lower}}}(\theta_{\0}) \,\,,
\end{equation}
\end{subequations}
we can write the intensity as being proportional to
\begin{eqnarray}
\mathcal{I}_{_{\mathrm{rtra}}} \, \propto \, \left\{ \,\tan\beta \, \exp\left[\,-\,\left( X + \,\frac{\Delta\alpha_{_{\mathrm{rtra}}}(\theta_{\0})}{2} \,\, \,\tilde{z}_{_{\mathrm{rtra}}}\, \right)^{^2}\,\bigg/\mathrm{w}^{\2}(\tilde{z}_{_\mathrm{rtra}}) \right] +\,\right.\nonumber\\
 \left.\tau(\theta_{\0})\, \exp\left[\,-\,\left( X - \,\frac{\Delta\alpha_{_{\mathrm{rtra}}}(\theta_{\0})}{2} \,\, \,\tilde{z}_{_{\mathrm{rtra}}}\, \right)^{^2} \bigg/\mathrm{w}^{\2}(\tilde{z}_{_\mathrm{rtra}})\,\right]
 \,\right\}^{^2}\,\,.
\end{eqnarray}

Setting the angle of the second polariser to $\beta  \, = \, -\,\arctan[\,\tau(\theta_{\0})\,] \,+\,\Delta\epsilon$ and considering that $|\Delta \epsilon| \ll 1$, we have that
\[
\tan \beta \,\approx \,-\,\tau(\theta_{\0})\,+\,\left[\,1 \,+\,\tau^{^{2}}(\theta_{\0})  \,\right]\,\Delta \epsilon\,\,.
\]
Using this result and by \cor{noting} that $\Delta\alpha_{_{\mathrm{rtra}}}(\theta_{\0}) \ll 1$, the transmitted intensity can be simplified to
\begin{eqnarray}
\label{eq:IntWM}
\mathcal{I}_{_{\mathrm{rtra}}} & \propto & \left\{ \,[\,-\,\tau(\theta_{\0})+(1+\tau^{\2}(\theta_{\0}))\,\Delta \epsilon\,] \, \left(\,1- \Delta\alpha(\theta_{\0})\,\frac{X\,\tilde{z}_{_{\mathrm{rtra}}}}{\mathrm{w}^{\2}(\tilde{z}_{_{\mathrm{rtra}}})}\,\right) \,+\, \right. \nonumber\\ & & \left. \tau(\theta_{\0})\,\left(\,1+ \Delta\alpha(\theta_{\0})\,\frac{X\,}{\mathrm{w}^{\2}(\tilde{z}_{_{\mathrm{rtra}}})}\,\right)\,\right\}^{^{2}}\,\exp\left[\,-\,2\,\frac{X^{^{2}}}{\mathrm{w}^{\2}(\tilde{z}_{_{\mathrm{rtra}}})}\,\right]\nonumber \\ \nonumber \\
 & \propto &  \left[ \,\frac{1+\tau^{\2}(\theta_{\0})}{2\, \tau(\theta_{\0})}\,\Delta \epsilon\, +\,\Delta\alpha(\theta_{\0})\,\frac{X\,\tilde{z}_{_\mathrm{rtra}}}{\mathrm{w}^{\2}(\tilde{z}_{_{\mathrm{rtra}}})}\,\right]^{^{2}} \,\exp\left[\,-\,2\,\frac{X^{^{2}}}{\mathrm{w}^{\2}(\tilde{z}_{_{\mathrm{rtra}}})}\,\right]\nonumber \\ \nonumber \\
 & = &  \left[ \,\frac{\Delta\epsilon}{A(\theta_{\0})}\, +\,\Delta\alpha(\theta_{\0})\,\frac{X\,\tilde{z}_{_{\mathrm{rtra}}}}{\mathrm{w}^{\2}(\tilde{z}_{_{\mathrm{rtra}}})}\,\right]^{^{2}} \,\exp\left[\,-\,2\,\frac{X^{^{2}}}{\mathrm{w}^{\2}(\tilde{z}_{_{\mathrm{rtra}}})}\,\right]\,\,,
\end{eqnarray}
where we have defined
\begin{equation}
{A(\theta_{\0})} = \frac{2\,\tau(\theta_{\0})}{1+\tau^{\2}(\theta_{\0})}.
\end{equation}
The intensity (\ref{eq:IntWM}) has peaks at
\begin{equation}
X_{_{\mathrm{max}}}^{^{\pm}}(\Delta \epsilon) = \frac{-\,\Delta\epsilon\,\pm\,\sqrt{(\Delta\epsilon)^{^{2}} + 2\,[\,A(\theta_{\0})\,\Delta\alpha(\theta_{\0})\,\tilde{z}_{_{\mathrm{rtra}}}/\mathrm{w}(\tilde{z}_{_{\mathrm{rtra}}})\,\,]^{^{2}}}}{2\,A(\theta_{\0})\,\Delta\alpha(\theta_{\0})\,\tilde{z}_{_{\mathrm{rtra}}}}\,\mathrm{w}^{\2}(\tilde{z}_{_{\mathrm{rtra}}})\,,
\end{equation}
and for $|\Delta\epsilon|\,\gg\,A(\theta_{\0})\,\Delta\alpha(\theta_{\0})\,{\tilde{z}_{_{\mathrm{rtra}}}}/{\mathrm{w}(\tilde{z}_{_{\mathrm{rtra}}})}$
we can approximate the square root above as
\[|\Delta\epsilon| \,\,+\,\,\frac{[\,A(\theta_{\0})\,\Delta\alpha(\theta_{\0})\,\tilde{z}_{_{\mathrm{rtra}}}/\mathrm{w}(\tilde{z}_{_{\mathrm{rtra}}})\,\,]^{^{2}}}{|\Delta \epsilon|}\,\,. \]

Now, a positive rotation of the second polariser, $\Delta \epsilon = |\Delta \epsilon |$, yields, by using the previous approximation,
\begin{eqnarray}
\left\{\,X_{_{\mathrm{max}}}^{^{-}}(|\Delta \epsilon|)\,,\, X_{_{\mathrm{max}}}^{^{+}}(|\Delta \epsilon|)\,\right\}\,=\,
\left\{\,-\,\frac{|\Delta \epsilon|\,\mathrm{w}^{\2}(\tilde{z}_{_\mathrm{rtra}})}{A(\theta_{\0})\,\Delta\alpha(\theta_{\0})\,\tilde{z}_{_{\mathrm{rtra}}}}\,,\,  \frac{A(\theta_{\0})\,\Delta\alpha(\theta_{\0})\,\tilde{z}_{_{\mathrm{rtra}}}}{2\,|\Delta \epsilon|}  \,\right\}\,\,,
\end{eqnarray}
which shows that for a positive rotation the main peak is at $X_{_{\mathrm{max}}}^{^{+}}(|\Delta \epsilon|)$. For a negative rotation, $\Delta \epsilon = -\,|\Delta \epsilon |$, we have then that
\begin{eqnarray}
\left\{\,X_{_{\mathrm{max}}}^{^{-}}(-\,|\Delta \epsilon|)\,,\, X_{_{\mathrm{max}}}^{^{+}}(-\,|\Delta \epsilon|)\,\right\}\,=\,
\left\{\,  \frac{|\Delta \epsilon|\,\mathrm{w}^{\2}(\tilde{z}_{_{\mathrm{rtra}}})}{A(\theta_{\0})\,\Delta\alpha(\theta_{\0})\,\tilde{z}_{_{\mathrm{rtra}}}}\,,\, -\, \frac{A(\theta_{\0})\,\Delta\alpha(\theta_{\0})\,\tilde{z}_{_{\mathrm{rtra}}}}{2\,|\Delta \epsilon|} \,\right\}\,\,,
\end{eqnarray}
being then the main peak at $X_{_{\mathrm{max}}}^{^{+}}(-\,|\Delta \epsilon|)$. The distance between these peaks is
\begin{eqnarray}
\label{eq:AngX}
\Delta X_{_{\mathrm{max}}} &=&  X_{_{\mathrm{max}}}^{^{+}}(|\Delta \epsilon|) - X_{_{\mathrm{max}}}^{^{+}}(-\,|\Delta \epsilon|) \nonumber \\ &=&\frac{ A(\theta_{\0})}{|\Delta \epsilon\,|}\,\,\Delta\alpha_{_{\mathrm{rtra}}}(\theta_{\0})\,\,{\tilde{z}_{_{\mathrm{rtra}}}} = \Delta\alpha_{_{\mathrm{rtra}}}^{^{\mathrm{WM}}}(\theta_{\0})\,\,{\tilde{z}_{_{\mathrm{rtra}}}}\,\,.
\end{eqnarray}
The angular coefficient $\Delta \alpha^{^{\mathrm{WM}}}_{_{\mathrm{rtra}}}(\theta_{\0})$ \cor{has}, in contrast to the coefficient measured by a direct procedure, $\Delta\alpha_{_{\mathrm{rtra}}}(\theta_{\0})$, an amplification factor of $1\,/\,|\Delta \epsilon|$.  Near the critical angle we have that $A_{_{\mathrm{cri}}}(\theta_{\0})\approx 1$, and so, a weak measurement in this region is
\begin{equation}
\Delta\alpha_{_{\mathrm{ rtra, cri}}}^{^{\mathrm{WM}}}(\theta_{\0})\,\, \approx\,\,
\frac{\Delta\alpha_{_{\mathrm{rtra, cri}}}(\theta_{\0})}{|\Delta \epsilon|}\,\,\propto\,\, \frac{1}{|\Delta \epsilon|\,\,(k\,{\mathrm{w}}_{\0})^{^{3/2}}}\,\,.
\end{equation}
In the vicinity of the Brewster region we have that $A_{_{\mathrm{B(int)}}}(\theta_{\0})\propto 1/\,k\,{\mathrm{w}}_{\0}$, providing
\begin{equation}
\Delta\alpha_{_{\mathrm{rtra, B(int)}}}^{^{\mathrm{WM}}}(\theta_{\0})\,\, \propto\,\,
\frac{\Delta\alpha_{_{\mathrm{rtra, B(int)}}}(\theta_{\0})}{k\,{\mathrm{w}}_{\0}\,|\Delta \epsilon|}\,\,\propto\,\, \frac{1}{|\Delta \epsilon|\,\,(k\,{\mathrm{w}}_{\0})^{^{2}}}\,\,.
\end{equation}

Figure 24 shows the weak measurement amplification of angular deviations for different values of the parameter $|\Delta \epsilon|$. We can see in Figure 24(a) that as one approaches the critical incidence the weak measurements approach becomes increasingly better. Figures 24(a) and (b) are zoom-ins in the regions of interest. At the boarder of the internal Brewster region, for instance, which is located at $\theta_{\0} = \theta_{_{\mathrm{B(int)}}} = \lambda/\mathrm{w}_{\0}$, for $|\Delta \epsilon|=0.1^{\circ}$, we have that
\begin{eqnarray}
\left\{\,\Delta\alpha_{_{\mathrm{rtra}}}\,,\,\Delta\alpha_{_{\mathrm{rtra}}}^{^{\mathrm{WM}}}  \,\right\} &=& \left\{\,1.8^{\circ} \,,\,\,\,\,3.0^{\circ}\,\right\}\,\times\,10^{^{-3}}
\end{eqnarray}
while at the boarder of the critical region at $\theta_{_{\mathrm{cri}}} - \, \lambda\,/\, {\mathrm{w}}_{\0}$, we have
\begin{eqnarray}
\left\{\,\Delta\alpha_{_{\mathrm{rtra}}}\,,\,\Delta\alpha_{_{\mathrm{rtra}}}^{^{\mathrm{WM}}}  \,\right\} &=& \left\{\,6.5^{\circ}\,\times\,10^{^{-2}} \,,\,37.0^{\circ}\,\right\}\,\times\,10^{^{-3}},
\end{eqnarray}
which \cor{demonstrates} the power of amplification of \cor{this} technique. For intermediate angles, the comparison between direct and weak measuring procedures is given in Table 2.

\WideFigureSideCaption{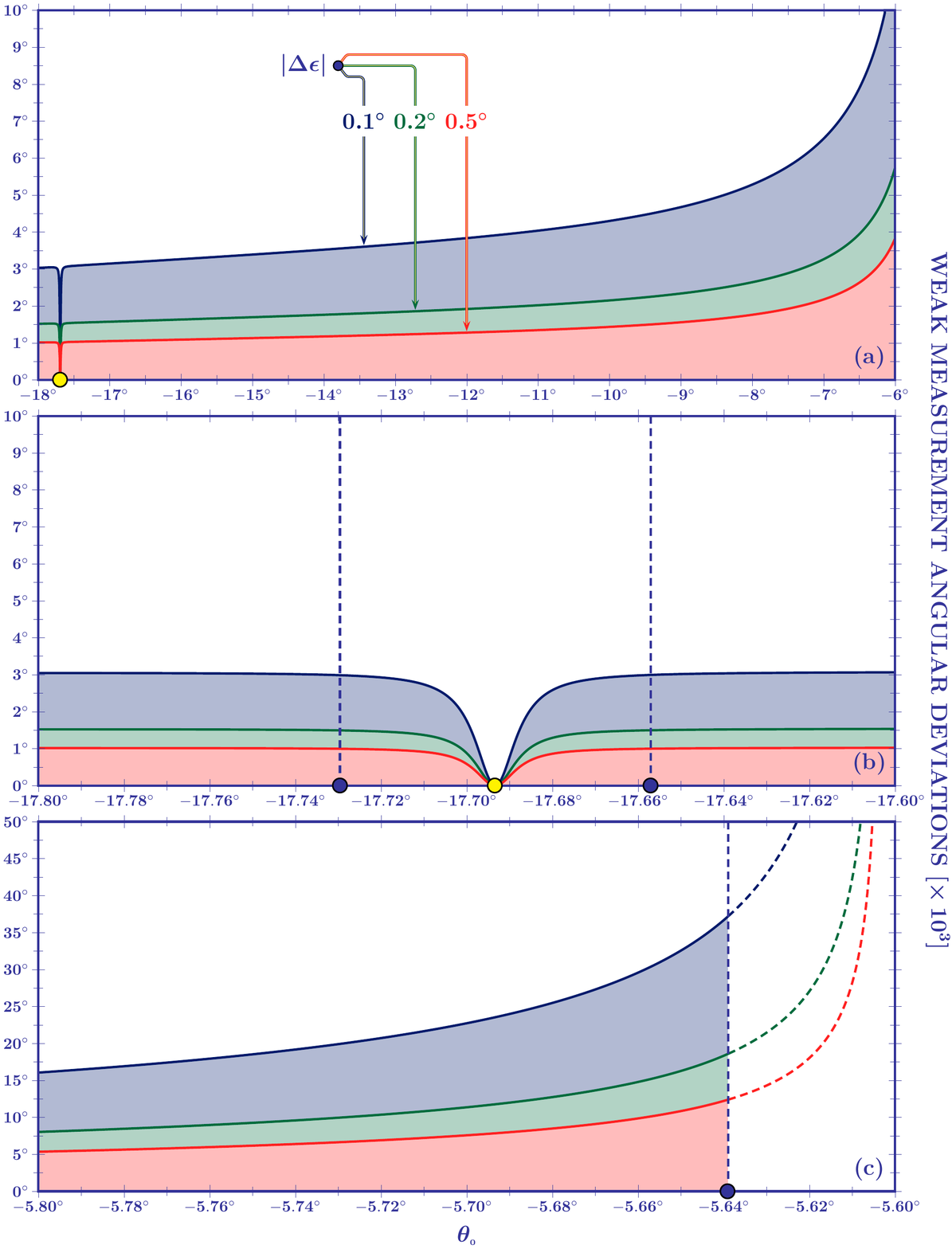}{The optical weak measurement amplification of angular deviations for a borosilicate ($n=1.515$). The curves represent a beam with $\lambda = 0.633\,\mu$m and $\mathrm{w}_{\0}= 1$ mm, and $|\Delta\epsilon|$ $0.1^{\circ}$, $0.2^{\circ}$, and $0.5^{\circ}$. (a) As the incidence angle approaches the critical incidence we can see the increasing efficiency of the weak measurement approach. (b) Amplification on the internal Brewster region. The region between dashed vertical lines in (b) represent the range of incidence angles for which the concept of angular deviations is obscure. (c) The vicinity of the critical region. The dashed lines in (c) represent the region were the Composite Goos-H\"anchen shift is present.}

\begin{table*}
{\large
{\color{navy}{
\begin{center}
\begin{tabular}{|c||c|c||c||c||r|}\hline
\rowcolor{bg} $\theta_{\0}$ & $10^{^{3}}\,\,\alpha_{_{\mathrm{TRA}}}^{^{\mathrm{[TE]}}}$ & $10^{^{3}}\,\,\alpha_{_{\mathrm{TRA}}}^{^{\mathrm{[TM]}}}$ & $10^{^{3}}\,\,\Delta\alpha_{_{\mathrm{TRA}}}$   &  $10^{^{3}}\,\,\Delta\alpha_{_{\mathrm{TRA}}}^{^{\mathrm{WM^{\color{bg}x}}}}$  & {\textsc{Ampl.}}    \\ \hline
 \rowcolor[gray]{0.9}  $-\,17.65706^{\circ}$ & $0.0023^{\circ}$ & $1.7953^{\circ}$ & $1.7930^{\circ}$
 & $2.9938^{\circ}$  & $1.6697$  \\ \hline
 \rowcolor[gray]{0.7}  $-\,17.65000^{\circ}$ & $0.0023^{\circ}$ & $1.5145^{\circ}$ & $1.5122^{\circ}$
 & $3.0169^{\circ}$  & $1.9950$   \\ \hline
 \rowcolor[gray]{0.9}  $-\,16.65000^{\circ}$ & $0.0024^{\circ}$ & $0.0676^{\circ}$ & $0.0652^{\circ}$
  & $3.1877^{\circ}$    & $48.8911$\\ \hline
 \rowcolor[gray]{0.7}  $-\,15.65000^{\circ}$ & $0.0026^{\circ}$ & $0.0366^{\circ}$ & $0.0340^{\circ}$
  & $3.3086^{\circ}$ & $97.3118$  \\ \hline
 \rowcolor[gray]{0.9}  $-\,14.65000^{\circ}$ & $0.0028^{\circ}$ & $0.0262^{\circ}$ & $0.0234^{\circ}$
  & $3.4354^{\circ}$  & $146.8077$  \\ \hline
 \rowcolor[gray]{0.7}  $-\,13.65000^{\circ}$ & $0.0030^{\circ}$ & $0.0211^{\circ}$ & $0.0181^{\circ}$
  & $3.5717^{\circ}$  & $197.3315$\\ \hline
 \rowcolor[gray]{0.9}  $-\,12.65000^{\circ}$ & $0.0032^{\circ}$ & $0.0183^{\circ}$ & $0.0151^{\circ}$
  & $3.7234^{\circ}$ & $246.7099$  \\ \hline
 \rowcolor[gray]{0.7}  $-\,11.65000^{\circ}$ & $0.0035^{\circ}$ & $0.0166^{\circ}$ & $0.0131^{\circ}$
  & $3.9000^{\circ}$ & $297.7099$\\ \hline
 \rowcolor[gray]{0.9}  $-\,10.65000^{\circ}$ & $0.0039^{\circ}$ & $0.0158^{\circ}$ & $0.0119^{\circ}$
  & $4.1196^{\circ}$  & $346.1849$ \\ \hline
 \rowcolor[gray]{0.7}  $-\,\,\,\,9.65000^{\circ}$ & $0.0045^{\circ}$ & $0.0156^{\circ}$ & $0.0111^{\circ}$
  & $4.4133^{\circ}$ & $397.5946$\\ \hline
\rowcolor[gray]{0.9}   $-\,\,\,\,8.65000^{\circ}$ & $0.0052^{\circ}$ & $0.0162^{\circ}$ & $0.0110^{\circ}$
  & $4.8493^{\circ}$  & $440.8455$ \\ \hline
 \rowcolor[gray]{0.7}  $-\,\,\,\,7.65000^{\circ}$ & $0.0064^{\circ}$ & $0.0179^{\circ}$ & $0.0115^{\circ}$
  & $5.6049^{\circ}$ & $487.3826$\\ \hline
 \rowcolor[gray]{0.9}  $-\,\,\,\,6.65000^{\circ}$ & $0.0091^{\circ}$ & $0.0230^{\circ}$ & $0.0139^{\circ}$
  & $7.3749^{\circ}$  & $530.5683$ \\ \hline
 \rowcolor[gray]{0.7}  $-\,\,\,\,5.65000^{\circ}$ & $0.0437^{\circ}$ & $0.1007^{\circ}$ & $0.0570^{\circ}$
  & $32.5569^{\circ}\,\,$ & $571.1737$\\ \hline
 \rowcolor[gray]{0.9}  $-\,\,\,\,5.63938^{\circ}$ & $0.0497^{\circ}$ & $0.1144^{\circ}$ & $0.0647^{\circ}$
  & $36.9881^{\circ}\,\,$ & $571.6862$\\ \hline
\end{tabular}
\end{center}
}}
}
\caption{\footnotesize{Table of angular deviation values for a borosilicate ($n=1.515$) prism and an incident beam with wavelength $\lambda=0.633\,\mu{\mathrm{m}}$ and minimal beam waist ${\mathrm{w}}_{\0}=1\,{\mathrm{mm}}$, for incidence angles (first column) ranging from the internal Brewster region to the critical region, and for TE- (second column) and TM-polarised (third column) waves. The fourth column displays the relative angular deviation and the fifth the relative angular deviation in a weak measurement system. The last column has the amplification factors between direct and weak measurements of the relative angular deviation.}}
\end{table*}

Let us come back to the constraint
\begin{equation*}
|\Delta\epsilon|\,\gg\,A(\theta_{\0})\,\Delta\alpha(\theta_{\0})\,\frac{\tilde{z}_{_{\mathrm{rtra}}}}{\mathrm{w}(\tilde{z}_{_{\mathrm{rtra}}})},
\end{equation*}
in order to determine the condition on $|\Delta\epsilon|$ which validates the analysis we carried out so far. In the region between the internal Brewster and the critical regions, that is, for $\theta_{_{\mathrm{B(int)}}}+\lambda/{\mathrm{w}}_{\0}<\theta_{\0}< \theta_{_{\mathrm{cri}}}-\lambda/{\mathrm{w}}_{\0}$, the main restriction comes from the critical region where $A_{_{\mathrm{cri}}}(\theta_{\0})\approx 1$ and angular deviations are proportional to $(k\,{\mathrm{w}}_{\0})^{^{-\,3/2}}$. Consequently, we have the condition
\begin{equation}
|\Delta\epsilon|\,\gg\,  \frac{\tilde{z}_{_{\mathrm{rtra}}}}{\,\,\,(k\,{\mathrm{w}}_{\0})^{^{3/2}}\,\mathrm{w}(\tilde{z}_{_{\mathrm{rtra}}})}\,\,.
\end{equation}
For a beam with ${\mathrm{w}}_{\0}=1\,{\mathrm{mm}}$, $\lambda=0.633\,\mu{\mathrm{m}}$ and for a camera positioned at $\tilde{z}_{_{\mathrm{rtra}}}=25\,cm$, $\tilde{z}_{_{\mathrm{rtra}}}/\mathrm{w}(\tilde{z}_{_{\mathrm{rtra}}}) \approx 250$, \cor{thus} we have that $ |\Delta \epsilon|\,\gg\, 0.014^{\circ}\,\,$.

Finally, let us \cor{note} that the structure of the beam is important for weak measurements as well as for direct measurements. Eq. (\ref{eq:alphartra}) has a $(k \,\mathrm{w}_{\0})^{\2}$ dependency in its denominator and so, as $\mathrm{w}_{\0}$ becomes wider the less pronounced becomes the angular deviation. This happens because increasing the minimal beam waist collimates the beam and restricts the wave vectors of its component plane waves to angles closer to the centre of its angular distribution. The same effect occurs for the relative angular coefficients in weak measurements, because they depend on the relative coefficients of direct measurements. Let us define an efficiency factor given by the polarisation-relative distance that must be measured from the line along which Geometrical Optics predicts the maximum electric field intensity \cor{to} be  to where it actually is, divided by the beam width where the measurement is made, that is
\begin{equation}
\rho = \frac{\Delta\alpha_{_{\mathrm{rtra}}}\,\tilde{z}_{_{\mathrm{rtra}}}}{\mathrm{w}(\tilde{z}_{_{\mathrm{rtra}}})}.
\end{equation}
Using Eqs. (\ref{eq:alphaBintdev}) and (\ref{eq:alphacridev}) we have that, for a direct measurement at a distance of $25$ cm,
\begin{eqnarray}
\{\rho_{_{\mathrm{B(int)}}},\rho_{_{\mathrm{cri}}}\}\,& = & \left\{1,\,\frac{1}{\sqrt{k\,{\mathrm{w}}_{\0}}}\right\}\frac{\tilde{z}_{_{\mathrm{rtra}}}}{k\,\mathrm{w}_{\0}\,\mathrm{w}(\tilde{z}_{_{\mathrm{rtra}}})} \approx  \left\{2.515\%,\,0.025\% \right\},
\end{eqnarray}
while for a weak measurement
\begin{eqnarray}
\{\rho_{_{\mathrm{B(int)}}},\rho_{_{\mathrm{cri}}}\}_{_{\mathrm{WM}}} & = & \left\{\frac{1}{k\,\mathrm{w}_{\0}},\,\frac{1}{\sqrt{k\,{\mathrm{w}}_{\0}}}\right\}\frac{\tilde{z}_{_{\mathrm{rtra}}}}{|\Delta\epsilon|\,k\,\mathrm{w}_{\0}\,\mathrm{w}(\tilde{z}_{_{\mathrm{rtra}}})} \nonumber \\
&\approx & \left\{0.145\%,\,14.466\% \right\}.
\end{eqnarray}
By comparing the efficiency factors for direct and weak measurements we see that near the Brewster angle weak measurements have the opposite of the desired effect, ``breaking'' the measurement. Direct measurements, in contrast, are more efficient near the Brewster angle than near the critical one.

\section{Conclusions}
\label{sec10}

\cor{More than 70 years have passed} since the original publication of Goos and H\"anchen, and the \cor{research} field they started is still prolific and full of interesting questions. In the present, work we have addressed \cor{same} of such questions, limiting ourselves to planar beam shifts. The phenomena analysed have different manifestations, but share in their core the same nature, \cor{i.e.} the fact that the Fresnel coefficients, \cor{describing} the interaction of light with dielectric structures, are not in \cor{accordance} with Geometrical Optics \cor{and present} corrections to it. The way the Fresnel's coefficients change the expected path of light in an optical system depends on the system itself through its refractive index, but, more importantly, on the incidence angle, which determines if the light beam is in the Partial or Total Internal Reflection regime.

For a totally internally reflected beam the reflection coefficient becomes complex and a new phase is acquired by its electric field. Because optical phases and optical paths are intrinsically related, this generates a shift of the reflection point in relation to the incoming point: for a partial reflection these points are the same, but not in the total internal reflection case. This is the Goos-H\"anchen shift, experimentally verified for the first time in 1947. Since then, its analytical description was an important investigation point. Artmann described it correctly \cor{at a distance} from the critical angle and Horowitz and Tamir successfully calculated the shift precisely at it. In Section \ref{sec5}, in accordance with both these results, we obtained an analytical expression for the Goos-H\"anchen shift everywhere for a Gaussian beam. The approach employed followed Artmann's original idea of using the Stationary Phase Method to find the optical path, but while his results are only valid for a plane wave description of the problem, we considered the beam's structure, integrating the shift weighted by the angular distribution of the beam. This result is in agreement with Artmann's \cor{calculation at a distance}  from the critical angle because in this region the shift is nearly constant and can be factored out of the integral.

Implicit in the Stationary Phase Method is the choice to look at the shift of the maximum intensity point of the reflected beam. A different approach would be to look at the shift of the mean intensity point, which was also done in Section \ref{sec5} by calculating the centroid of the totally internally reflected beam. Both results are not the same everywhere, agreeing with each other only in the Artmann zone (the zone where Artmann's results are valid). This is expected and is due to the structure of the beam. An angular Gaussian distribution has an appreciable value in the interval $\theta_{\0}-\lambda/\mathrm{w}_{\0}<\theta<\theta_{\0}+\lambda/\mathrm{w}_{\0}$. If $\theta_{\0}>\theta_{_{\mathrm{cri}}}+\lambda/\mathrm{w}_{\0}$ the whole beam can be considered reflected and because the reflection coefficient is complex, with magnitude 1, the beam is symmetric. As a result its mean intensity point is coincident with its maximum intensity point. If, however, $\theta_{_{\mathrm{cri}}}-\lambda/\mathrm{w}_{\0}<\theta_{\0}<\theta_{_{\mathrm{cri}}}+\lambda/\mathrm{w}_{\0}$, part of the beam is outside the Total Internal Reflection regime and its symmetry is broken. As we saw, this can be \cor{observed} in the difference between results obtained by the mean and maximum intensity approaches to the Goos-H\"anchen shift in the critical zone. The analytical formulae for both cases were compared to numerical calculations, achieving an excellent agreement.

This symmetry breaking effect of the beam was the subject analysed in Section \ref{sec6}, where we showed it to be the responsible for angular shifts. We calculated  analytical expressions for such shifts in the region between the Brewster and the critical regions. An optical beam is a packet of plane waves, each with an incidence angle following the beam's angular distribution. The propagation direction of the beam is given by the incidence angle of the distribution's centre, which is the main contribution to the packet. When light interacts with a dielectric interface the Fresnel's coefficients break the symmetry of this distribution, favouring the transmission (reflection) of plane waves other than the one originally at the centre of the beam. This shifts the mean intensity point of the beam, which now propagates in a slight different direction than the incident angle would have it propagate, according to Geometrical Optics. This is the trigger behind angular deviations. All the effects we have discussed are dependent on polarisation states because the Fresnel coefficients discriminate between such states. For angular deviations, in particular, an important result arises from this fact. The Brewster angle does not reflect TM-polarised light and, consequently, since beams have an angular aperture around their centre, a TM-polarised beam has its symmetry more strongly broken at this point, yielding a greater angular deviation. How to interpret such deviations in the Brewster region, however, is still an open topic. Incidence at the Brewster angle turns a Gaussian distribution into a double-peaked structure, which makes the concept of angular deviations hazy. In the literature on the subject it is possible to find analyses of the deviations underwent by each peak separately as well as by the mean intensity of the whole structure, which, similarly to the centre of mass of a boomerang, is outside the portion of space that contains the bulk of the electric field.

Moving the incidence angle closer to the critical angle a new, interesting effect was verified. In the critical region the angular distribution can be divided in two portions. One of them is totally internally reflected and is laterally shifted (Goos-H\"anchen shift) and one of them is partially reflected and angularly shifted. The beam, however, is still a single entity, and not a multiple-peaked structure, and the conjunction of both these phenomena yields the effect known as the Composite Goos-H\"anchen shift, which was numerically studied in 
Section \ref{sec7}. The hallmark of this shift is the axial dependence of the measurements. The farther you position the camera that collects the beam the greater will be the measured lateral shift, because the presence of angular deviations makes the actual beam not parallel to the optical path predicted by Geometrical Optics. Away from the critical angle, where no symmetry breaking effects occur, this parallelism is restored and the axial position of the camera ceases to be relevant. Also, since in the critical region there is a relative phase between both portions of the beam, an oscillatory behaviour of the shift was verified, the amplitude of which decreases as one moves away from the critical angle. Finally, it is interesting to \cor{note} the role played by the minimal beam waist in the Composite Goos-H\"anchen shift. The axial element of the measurement becomes smoother the wider \cor{ $\mathrm{w}_{\0}$ becomes}, meaning that for axial amplifications to occur, one must position the camera much farther away. For a TE-polarised beam, for instance, we saw that either at a distance of 0 or 50 cm the maximum shift was about 7 $\mu$m for $\mathrm{w}_{\0}=600\,\mu$m while for $\mathrm{w}_{\0}=150\,\mu$m it goes from 3 $\mu$m at 0 cm to nearly 30 $\mu$m at 50 cm. The reason for this behaviour is the collimation of beam. The $\mathrm{w}_{\0}=600\,\mu$m-beam is strongly centred around its incidence angle and its symmetry breaking is not as \cor{pronounced}.

All these phenomena are but minute corrections to Geometrical Optics. As we saw, in the Artmann zone the Goos-H\"anchen shift is proportional to $\lambda$, while in the critical region \cor{is proportional} to $\sqrt{\mathrm{w}_{\0}\,\lambda}$. Angular deviations are proportional to $\lambda/\mathrm{w}_{\0}$ in the Brewster region and to $\sqrt{\mathrm{w}_{\0}/\lambda}$ near the critical region. Even for TM-polarised beams, which have a $n^{\2}$ factor in relation to TE-polarised beams, these effects are small. In this work we have considered a wavelength of $0.633\,\mu$m, leaving measurements in the micrometer scale. This, however, does not diminish the importance of such corrections, which are relevant not only for the sake of a better understanding of light's behaviour, but also, from a pragmatic \cor{point of view}, for the precision design of optical systems such as optical resonators and ellipsometric probes. Amplification techniques are, nevertheless, available to address the problem of such precise measurements. Multiple-reflection systems and large wavelengths have been common solutions to this issue, and in Sections \ref{sec8} and \ref{sec9} we discussed the indirect approach of Optical Weak Measurements. This powerful method uses polarisers to mix components of a elliptically polarised beam, creating a particular double-peaked intensity profile with peaks' positions that change according to a polarising parameter $\Delta\epsilon$. We saw that for two consecutive measurements with a clockwise and an anticlockwise rotation of the polariser by the same amount $|\Delta\epsilon|$, the position of the main intensity peak changes, and the distance between such positions yields an indirect measurement of the relative shift. For the Goos-H\"anchen shift, weak measurements amplified the direct measurements by a factor as big as 30, allowing a relative shift of more than 0.5 mm for a beam with $\mathrm{w}_{\0}=500\,\mu$m. The axial effect of the critical region was still present in the weak measurements, and, just as in the direct measurements case, wider minimal beam waists minimised this effect. In the Artmann zone this amplification is constant, with an amplification factor $1/|\Delta\epsilon|$, showing that the technique's power is limited to the precision with which a polariser's angle can be set. Also, we have found an analytical description of the effect of the relative Goos-H\"anchen phase on weak measurements. This phase is usually removed after light leaves the dielectric prism it is interacting with, but a description of its effects was still lacking in the literature. We found it poses a destructive influence on weak measurements, generating an intensity profile with virtually fixed peaks. For incidence angles for which $\Delta\Phi_{_{\mathrm{GH}}}=2\,m\,\pi$, however, this phase is naturally removed and the results of measurements without phase are reconstructed. This is of practical interest since a phase that is not completely removed will render experimental data discrepant from theoretical expectations, and can be used to evaluate the efficiency of the phase-removal technique employed.

For angular shifts, on the other hand, our analysis showed that the Optical Weak Measurement technique is not as effective throughout the whole incidence angle spectrum. Direct measurements of the relative angular shift are nearly 30 times greater at the \cor{border} of the Brewster region than at the \cor{border} of the critical region, but while weak measurements offered an amplification factor of nearly 600 in the \cor{latter}, it was only nearly 2 in the \cor{previous}. By defining an efficiency factor as the relative angular coefficient multiplied by the ratio between the distance of the camera which carries \cor{out} the measurement and the aperture of the beam at this distance, weak measurements near the Brewster region become even more inefficient. This factor is the ratio between the transversal distance the experimentalists have to measure and the size of the object they are measuring. If the beam is wider than the distance it is shifted the efficiency factor is lower than \cor{the} opposite case and the measurement is harder. For a camera at 25 cm from the optical system and a beam with a minimal beam waist of $1$ mm, there is an efficiency of 2.5$\%$ for a direct measurement versus 0.1$\%$ for a weak measurement near the Brewster region. In comparison, near the critical region these factors are of 0.03$\%$ for a direct measurement and 14.5$\%$ for a weak measurement.

Finally, \cor{regarding}  outlooks and possible future lines of research, it is possible to extend everything that was done in the present work for different beam profiles, such as Laguerre- and Hermite-Gaussian beams, Bessel beams, Airy beams, and so forth. In particular, the extension of optical weak measurements to such profiles presents itself as an interesting topic, since such investigations have been strongly centred around Gaussian beams. The Goos-H\"anchen shifts as well as the angular shifts depend on the structure of the beam, and so, beams with different properties are bound to produce new, interesting, results. The possibility of extension to other areas of Physics is also promising. The interaction of particles with potential barriers in Quantum Mechanics has in its description coefficients analogous to Fresnel's, as does the interaction of seismic waves with geological interfaces, and the propagation of acoustic waves\cite{DK2018}. \cor{It is also important to recall that the relative Goos-H\"anchen phase plays a fundamental role in the power oscillation of lasers transmitted through dielectric blocks. Theoretical investigations of the power oscillation presented in \cite{POT1,POT2} have recently found an experimental confirmation\cite{POExp}. This oscillation is caused by the zero order term in the Taylor expansion  of the Fresnel (Goos-H\"anchen) phase and represent a different phenomenon with respect to the oscillatory behaviour of light predicted in \cite{Maia2017} and observed in \cite{OBoLExp}.}

\subsection*{Acknowledgements}

The authors gratefully acknowledge the many helpful suggestions of  Dr. Manoel P. Araújo and  Prof. Silvânia A. Carvalho during the preparation of the paper. One of the author (S.D.L) wishes to thank the University of Salento
for the hospitality, and the CNPq (grant 2018/303911) and Fapesp (grant 2019/06382-9) for financial support.


\end{document}